\documentclass[11pt, letterpaper]{article}
\usepackage{epsfig,amssymb,amsmath, multirow, url, tcolorbox,booktabs, enumitem}

\usepackage{hyperref}
\usepackage{graphicx}
\usepackage{enumitem}
\usepackage{natbib}
\usepackage[utf8]{inputenc}
\usepackage{authblk}
\usepackage{lscape}
\usepackage{adjustbox}
\usepackage{setspace}
\usepackage[ruled,vlined]{algorithm2e}
\SetKwRepeat{Do}{do}{while}%
\SetKwInput{KwInit}{Initialization}%
\SetKwIF{If}{ElseIf}{Else}{if}{then}{else if}{else}{endif}%

\usepackage{dsfont}
\usepackage{bm}
\usepackage{amssymb}
\usepackage{amsthm}
\usepackage{mleftright}
\usepackage{multibib}
\newcites{supp}{Supplementary References}

\bibpunct{(}{)}{,}{a}{,}{,}
\newtheorem{@assumption}{\bf Assumption}[section]

 \newtheorem{@remark}{\bf Remark}[section]
 \newenvironment{remark}{\begin{@remark}\rm}{\end{@remark}}

\newtheorem{theorem}{Theorem}[section]
\newtheorem{proposition}{Proposition}[section]
\newtheorem{lemma}{Lemma}[section]
\newtheorem{corollary}{Corollary}[section]

\newtheorem{Lassump}{Assumption}

\newtheorem{Cassump}{Assumption}

\newtheorem{Fassump}{Assumption}

\newtheorem{Eassump}{Assumption}


\newcommand{\mfh}{\mathfrak{H}}

\newcommand{\bbE}{{\mathbb E}}

\newcommand{\bbP}{{\mathbb P}}

\newcommand{\E}{\bbE}

\newcommand {\corr}{\qopname\relax n{\textrm{Corr}}}

\newcommand{\argmin}{\mathop{\rm argmin}}

\def\RR{\mathbb R}
\def\ZZ{\mathbb Z}

\def\NN{\mathbb N}

\DeclareFontFamily{U}{mathx}{\hyphenchar\font45}
\DeclareFontShape{U}{mathx}{m}{n}{<-> mathx10}{}
\DeclareSymbolFont{mathx}{U}{mathx}{m}{n}
\DeclareMathAccent{\widebar}{0}{mathx}{"73}

\newcommand{\vecop}{\operatorname{vec}}

\newcommand{\diag}{\operatorname{diag}}

\newcommand{\Prob}{\mathbb{P}} 

\newcommand{\rig}{\operatorname{r}} 
\newcommand{\aaa}{\widetilde{q}} 

\usepackage{bm}

\newcommand{\prob}{\operatorname{p}} 
\newcommand{\Dim}{d} 
\newcommand{\Lag}{p} 
\newcommand{\VO}{p} 
\newcommand{\TM}{\Psi} 
\newcommand{\maxF}{\max} 
\newcommand{\conZ}{\bm{c}_{Z}}
\newcommand{\bmS}{\bm{\Sigma}_{X}}
\newcommand{\bmhS}{\bm{\widehat{\Sigma}}_{X}}
\newcommand{\bmR}{\bm{R}_{Z}}
\newcommand{\bmhR}{\bm{\widehat{R}}_{Z}}
\newcommand{\bmSY}{\bm{\Sigma}_{Y}}
\newcommand{\bmL}{\bm{\Lambda}}

\newcommand{\vertiii}[1]{{\vert\kern-0.25ex\vert\kern-0.25ex\vert #1 
    \vert\kern-0.25ex\vert\kern-0.25ex\vert}}
\newcommand{\BIGvertiii}[1]{{\left\vert\kern-0.25ex\left\vert\kern-0.25ex\left\vert #1 
    \right\vert\kern-0.25ex\right\vert\kern-0.25ex\right\vert}}

\usepackage{mleftright}
\usepackage{xparse}

\NewDocumentCommand{\evaluat}{sO{\big}mm}{%
  \IfBooleanTF{#1}
   {\mleft. #3 \mright|_{#4}}
   {#3#2|_{#4}}%
}

\usepackage{multirow}
\providecommand{\keywords}[1]
{
{  \small	
  {\textit{Keywords---}} #1}
}
\providecommand{\keywordsmsc}[1]
{
{  \small	
  {\textit{MSC2020---}} #1}
}
\usepackage{cleveref}
\crefname{equation}{equation}{equations}
\Crefname{equation}{Equation}{Equations}
\crefrangelabelformat{equation}{(#3#1#4--#5#2#6)}
\crefmultiformat{equation}{equations (#2#1#3}{, #2#1#3)}{#2#1#3}{#2#1#3}
\Crefmultiformat{equation}{Equations (#2#1#3}{, #2#1#3)}{#2#1#3}{#2#1#3}
\usepackage{caption}
\usepackage{subcaption}
\usepackage{graphicx}
\usepackage{rotating}
\usepackage{tikz}

\numberwithin{equation}{section}

\allowdisplaybreaks

\begin{document}
\title{Latent Gaussian dynamic factor modeling and forecasting \\ 
for multivariate count time series} 

\author[1]{Younghoon Kim}
\author[2]{Marie-Christine D\"uker}
\author[3]{Zachary F. Fisher}
\author[4]{Vladas Pipiras$^{*}$} 

\affil[1]{Cornell University}
\affil[2]{Friedrich-Alexander-Universit\"at Erlangen-N\"urnberg}
\affil[3]{The Pennsylvania State University}
\affil[4]{University of North Carolina at Chapel Hill}

\def\thefootnote{$*$}\footnotetext{Corresponding author. Email: pipiras@email.unc.edu}

\date{\today}
\date{}
\maketitle

\begin{abstract}
    This work considers estimation and forecasting in a multivariate, possibly high-dimensional count time series model constructed from a transformation of a latent Gaussian dynamic factor series. The estimation of the latent model parameters is based on second-order properties of the count and underlying Gaussian time series, yielding estimators of the underlying covariance matrices for which standard principal component analysis applies. Theoretical consistency results are established for the proposed estimation, building on certain concentration results for the models of the type considered. They also involve the memory of the latent Gaussian process, quantified through a spectral gap, shown to be suitably bounded as the model dimension increases, which is of independent interest. In addition, novel cross-validation schemes are suggested for model selection. The forecasting is carried out through a particle-based sequential Monte Carlo, leveraging Kalman filtering techniques. A simulation study and an application are also considered.
\end{abstract}


\keywordsmsc{Primary 62M10, 62H12; secondary 62H20.}

\section{Introduction}
\label{se:introduction}

This work develops theory, estimation, and forecasting methods for dynamic factor modeling of discrete-valued multivariate, possibly high-dimensional time series. Count time series are widespread in the natural, health, social and other sciences, for example, monthly counts of earthquakes or the amount of rainfall above a certain magnitude, daily counts of virus infections over spatial locations, item responses on surveys, and number of followers in social network over time. Mathematically, we consider a $d-$vector time series $\{X_{t}\}_{t\in\mathbb{Z}}=\{(X_{i,t})_{i=1,\ldots,d}\}_{t\in\mathbb{Z}}$, where $X_{i,t}\in\mathbb{N}_0=\{0,1,2,\ldots\}$ and $t\in\mathbb{Z}$ represents time. While the model can handle the range $\mathbb{Z}$, discrete-valued models are commonly applied to counts and the range $\mathbb{N}_0$. Any set of finite discrete values can be represented as a subset of $\mathbb{N}_0$. The primary focus will be on stationary models, though we also discuss the inclusion of covariates and differencing.

In general, modeling time series with discrete (count) values is delicate. In the continuous case, the class of autoregressive moving average (ARMA) models parsimoniously spans all non-deterministic stationary series (by the classical Wold decomposition). In the count setting, the landscape is much less established, with no single class of models dominating in popularity. In fact, researchers have developed numerous methods for constructing stationary count time series. The majority of work on count time series has been devoted to the univariate case. Popular approaches include those based on thinning operators \cite[e.g.][]{mckenzie1985some,alzaid1993some} and the generalized state-space models \cite[e.g.][]{davis2016handbook}, including hidden Markov models (HMMs) \cite[e.g.,][]{macdonald1997hidden}, Bayesian dynamic models \cite[e.g.][]{gamerman2015dynamic}. Integer-valued autoregressive conditional heteroskedasticity modeling \cite[e.g.][]{ferland2006integer,fokianos2009poisson,zhu2011negative} is another popular observation-driven approach. Recent reviews of this research area are given in \cite{weiss2018introduction,davis2021count} and \cite{fokianos2021multivariate}.

Multivariate, and potentially high-dimensional, count time series have received considerably less attention. A recent popular approach uses generalized linear model (GLM) constructions in high-dimensional settings with component series means, conditionally on the past, depending on their past values or those of the counts themselves akin to vector autoregression (VAR) models; see \cite{chen2017multivariate,hall2018learning,mark2018network,mark2019estimating}, and \cite{fokianos2020multivariate}. A variation of the approach is to use dynamic factor model (DFM) constructions instead; see \cite{jung2011dynamic,cui2014generalized,wang2018modeling}, and \cite{brauning2020dynamic}. This approach posits conditional distributions of the counts and is convenient for likelihood estimation procedures. In a collection of articles on count time series, \cite{karlis2016models} surveys relatively recent works on multivariate discrete-valued time series models.

In a recent paper, \cite{jia2023latent} proposed a new count time series model driven by latent Gaussian time series. The model offers a flexible and most general correlation structure that can accommodate any count marginal distribution. The marginal distribution can exhibit over- or under-dispersion, or have zero-inflation. The autocovariance function (ACVF) of the model is as general as possible for a given marginal distribution, in particular, capable of achieving the most negative pairwise correlations. \cite{kong2023seasonal} further extended the model to incorporate periodic and seasonal features by replacing the vanilla ARMA with periodic autoregressive moving average (PARMA) and seasonal autoregressive moving average (SARMA) models. The model bins the latent Gaussian series into discrete values and is particularly suitable for data that can be thought of as a discretization of an underlying continuous-valued signal (e.g., when a discrete scale is used for response over a continuous scale).

In this work, we propose a multivariate, possibly high-dimensional extension of \cite{jia2023latent} where the latent Gaussian series follows a DFM. A recent study by \cite{duker2024high} considered a similar extension but with the latent Gaussian series following a high-dimensional, sparse VAR model. \cite{jia2023latent} considered several estimation methods, including the efficient maximum likelihood estimation based on particle approximations of likelihood function via sequential Monte Carlo. The likelihood approximation becomes computationally infeasible with available tools in higher dimensions. Both here and \cite{duker2024high}, a computationally efficient scheme is employed based on a relationship between second-order properties of the observed and latent Gaussian series. While \cite{duker2024high} consider a sparse VAR for the latent Gaussian series, we assume the latter to follow a DFM. DFMs are models of choice when the cross-sectional dependence across the variables is strong, for example, manifested through correlations as in our application (see Section \ref{se:application} and Figure \ref{fig:Figure_param}). Sparse VARs typically do not have this property and, in fact, there is substantial literature combining the two types of models \cite[e.g.,][]{lin2020regularized,chen2023community}. Our approach is also shown to have theoretical guarantees, in the high-dimensional regime. Here, we rely on the concentration bounds for the covariances proved in \cite{duker2024high} but substantial technical work is still needed to exploit them for the considered DFM and also to show that our factor-based model satisfies certain assumptions in \cite{duker2024high}. Furthermore, we suggest a novel cross-validation scheme related to model selection, namely, the number of factor series. Though we do not employ likelihood estimation based on particle approximations, we use the latter for forecasting once the model is fitted to data. This still carries an expensive computational cost but it is manageable compared to likelihood inference. Kalman filtering techniques alleviate some of the computational cost. Our modeling and forecasting approaches are illustrated in simulations and on real psychometrics data.

We also note that our model is closely related to another strand of the literature, namely, that on copula models and particularly Gaussian copula models. Indeed, as noted in \cite{jia2023latent}, for any finite collection of times, our model can be expressed in terms of a Gaussian copula function, depending on our latent model parameters. Related work includes Gaussian copula regression \cite[e.g.,][]{masarotto2012gaussian}. Factor-type copula models, not necessarily Gaussian or specifically discrete-valued, were considered by \cite{murray2013bayesian,nikoloulopoulos2015factor,kadhem2021factor}, and others. These references are far from being exhaustive. What sets this work apart are the time series setting, a new estimation method, the possibility of high-dimensional regime, and theoretical guarantees.

In summary, our main contributions and highlights of the paper are as follows:
\begin{itemize}\itemsep=0.1em
    \item The introduction of a dynamic factor model for multivariate, possibly high-dimensional discrete-valued time series.
    
    \item The proposal of a relatively simple approach to parameter estimation with theoretical guarantees. Some theoretical developments could be of independent interest, such as the analysis of a certain spectral gap as the dimension $d$ is increasing.

    \item A novel practical approach to model selection including rank and lag order choice.
    
    \item The development of forecasting schemes specific to the discrete and time series nature of the model.
    
    \item A simulation study and application of the proposed model showcasing interpretability and flexibility. 
\end{itemize}

The rest of the paper is organized as follows. Section \ref{se:model} introduces the latent Gaussian dynamic factor model and establishes relationships between the second-order dependence structures of count and underlying Gaussian models. The estimation procedure is described in Section \ref{se:estimation}, followed by Section \ref{se:theory}, providing its theoretical guarantees. Section \ref{se:forecast} concerns forecasting. Numerical experiments can be found in Section \ref{se:simulation}, followed by an illustrative application in Section \ref{se:application}. We close our paper with comments for future work in Section \ref{se:conclusion}. Finally, Appendix \ref{ap:consistency} contains the proofs of our theoretical results and Appendix \ref{ap:kalman} contains some details for our forecasting approach.

\section{Latent Gaussian dynamic factor model}
\label{se:model}

\subsection{Model formulation}
\label{sse:model}

For a $d-$vector time series $\{X_{t}\}_{t\in\mathbb{Z}}=\{(X_{i,t})_{i=1,\ldots,d}\}_{t\in\mathbb{Z}}$, a latent Gaussian dynamic factor model is defined as follows. For $i=1,\ldots,d$, each component series $X_{i,t}$ at time $t$ is given by
\begin{equation}\label{e:generalized_inverse}
    X_{i,t}=F_{i}^{-1}(\Phi(Z_{i,t})) := G_{i}(Z_{i,t}),
\end{equation}
where $F_{i}$ is a cumulative distribution function (CDF), $F_{i}^{-1}(u)=\inf\{v:F_{i}(v)\geq{u}\}$ is its (generalized) inverse, $\Phi(z)$ is the CDF of $\mathcal{N}(0,1)$ distribution, and $Z_{i,t}$ is a zero mean, unit variance, Gaussian stationary series defined below. The CDFs $F_{i}$ are thought to come from parametric families, parameterized by a (possibly vector) parameter $\theta_{i}$. Note that by construction \eqref{e:generalized_inverse}, $\{X_{i,t}\}$ is stationary and its marginal distribution is $F_{i}$. We focus here on discrete distributions $F_{i}$ taking nonnegative integer values $\mathbb{N}_0$. For example, if $F_{i}$ is the CDF of a Bernoulli distribution with parameter $p_{i}$, Bern$(p_{i})$, the model \eqref{e:generalized_inverse} becomes $X_{i,t}=\boldsymbol{1}_{\{\Phi(Z_{i,t})>1-p_{i}\}}=\boldsymbol{1}_{\{Z_{i,t}>\Phi^{-1}(1-p_{i})\}}$, where $\boldsymbol{1}_{\{\cdot\}}$ is the indicator function. More generally, if $F_i$ is a CDF whose support lies in $\mathbb{N}_0$, for example a Poisson distribution with parameter $\theta_{i}$, then $X_{i,t}$ is represented through $Z_{i,t}$ by
\begin{equation}\label{e:ar1example_poisson}
    X_{i,t} = \sum_{n=1}^{\infty} n \boldsymbol{1}_{\{\Phi^{-1}(C_{i,n-1}) < Z_{i,t} \leq \Phi^{-1}(C_{i,n}) \}}, \quad C_{i,n}  = \mathbb{P}(X_{i,t} \leq n)=F_{i}(n),\ n=0,1,\ldots.
\end{equation}
In view of \eqref{e:ar1example_poisson}, the model discretizes the continuous-valued series $\{Z_{i,t}\}$ and is particularly natural to use in the context where $\{X_{i,t}\}$ can be thought as resulting from such discretization. In fact, \eqref{e:ar1example_poisson} defines a count random variable $X_{i,t}$ represented through $Z_{i,t}$ that follows any marginal distribution $F_{i}$. However, we shall use the examples of Bernoulli, Poisson, negative binomial, and categorical marginal distributions for illustration throughout the paper. For example, for categorical marginal counts, the infinite sum in \eqref{e:ar1example_poisson} reduces to a finite sum by the number of categories, say $N$, where $C_{i,N} = 1$. Whereas the Poisson distribution with a large parameter is close to Gaussian and the latent $\{Z_{i,t}\}$'s are effectively observed, note that the Bernoulli case lies at the other extreme and is expected to be most difficult to deal with in our tasks. 

We are interested in the scenario where the underlying Gaussian series $\{Z_{i,t}\}$ obeys a DFM. More specifically, we suppose that the $d-$vector time series $\{Z_{t}\}_{t\in\mathbb{Z}}=\{(Z_{i,t})_{i=1,\ldots,d}\}_{t\in\mathbb{Z}}$ satisfies
\begin{equation}\label{e:dynamic_factor}
    Z_{t} = \Lambda {Y}_{t}  +\varepsilon_{t},
\end{equation}
where $\Lambda$ is a $d\times{r}$ loadings matrix, and $\varepsilon_{t}$ are $\mbox{i.i.d. }\mathcal{N}(0,\Sigma_{\varepsilon})$ random $d-$vectors (independent of $Y_{t}$'s) and $r-$vector factor series $\{Y_{t}\}_{t\in\mathbb{Z}}=\{(Y_{k,t})_{k=1,\ldots,r}\}_{t\in\mathbb{Z}}$  follows a causal (stable) stationary VAR model of order $p$, VAR($p$), given by
\begin{equation}\label{e:VAR}
    Y_{t}=\Psi_{1}Y_{t-1}+\ldots+\Psi_{p}Y_{t-p}+\eta_{t},
\end{equation}
where $\Psi_{1},\ldots,\Psi_{p}$ are $r\times{r}$ matrices and $\eta_{t}$ are $\mbox{i.i.d. }\mathcal{N}(0,\Sigma_{\eta})$ random $r-$vectors. The VAR model \eqref{e:VAR} is flexible to capture temporal dependence from a practical standpoint. Note that the DFM \eqref{e:dynamic_factor} is in the so-called static form. The generalized DFMs where \eqref{e:dynamic_factor} includes lags of $Y_t$ \cite[e.g.][]{forni2000generalized,forni2005generalized} go beyond the scope of this work. Gaussianity is assumed for the various components of \eqref{e:dynamic_factor} and \eqref{e:VAR} given $Z_{i,t}$ is Gaussian in \eqref{e:generalized_inverse}. The factor structure in \eqref{e:dynamic_factor} imposes dependence of $Z_{i,t}$ and hence also $X_{i,t}$ across $i=1,\ldots,d$.

Note that the unit variance of $Z_{i,t}$ is assumed in \eqref{e:generalized_inverse}. For general $Z_{i,t}$, one can standardize it to have unit variance. More generally, the ACVF $\Sigma_{Z}(h) = \E[ Z_{t+h}Z_{t}']$ of $\{Z_t\}$ at lag $h$ can similarly become ACF $R_Z(h)$ as
\begin{equation}\label{e:standarization}
    R_{Z}(h)=\diag(\Sigma_{Z}(0))^{-1/2}\Sigma_{Z}(h)\diag(\Sigma_{Z}(0))^{-1/2}.
\end{equation}
We use $R_{Z}(h)$ for the rest of the analysis so the unit variance assumption for $Z_{i,t}$ is made throughout.

\subsection{Relation between count and Gaussian correlations}
\label{sse:Gaussian_correlation}

Our estimation procedure is based on the following property of the model \eqref{e:generalized_inverse}. It is known \cite[e.g.][]{pipiras2017long} that, for any $i,j=1,\ldots,d$, 
\begin{equation}\label{e:link_function_entrywise}
    R_{X,ij}(h)=L_{ij}(R_{Z,ij}(h))
\end{equation}
or, in short, and entry-wise, 
\begin{equation}\label{e:link_function_matrix}
    R_{X}(h)=L(R_{Z}(h)),
\end{equation}
where $L_{ij}:[-1,1]\mapsto[-1,1]$ are functions to be referred to as link functions (and $L$ as a link function). Furthermore, $L_{ij}$ depends only on the CDFs $F_{i}$ and $F_{j}$ and can be expressed as described below. As $F_i$ depends on parameter $\theta_i$, we shall sometimes write $L_{\theta_i,\theta_j}$ instead of $L_{ij}$ to indicate this dependence.

For $k=0,1,\ldots$, let $H_{k}(z)=(-1)^{k}e^{z^{2}/2}(d^{k}e^{-z^{2}/2}/dz^{k})$ be the Hermite polynomial of order $k$ and
\begin{equation*}
    g_{i,k}=\frac{1}{k!}\int_{-\infty}^{\infty}G_{i}(z)H_{k}(z)\frac{e^{-z^{2}/2}}{\sqrt{2\pi}}dz=\frac{1}{k!}\E[G_{i}(Z_{i,0})H_{k}(Z_{i,0})]
\end{equation*}
be the corresponding Hermite coefficient of the function $G_{i}(z)$ in \eqref{e:generalized_inverse}, so that $G_{i}(z)=\sum_{k=0}^{\infty}g_{i,k}H_{k}(z)$. For $G_{i}(z)$ associated with the CDF $F_i$ on nonnegative integers, \cite{jia2023latent} showed that
\begin{equation}\label{e:hermite_coefficient}
    g_{i,k}=\frac{1}{k!\sqrt{2\pi}}\sum_{n=0}^{\infty}e^{-Q_{i,n}^{2}/2}H_{k-1}(Q_{i,n}),
\end{equation}
where $Q_{i,n}=\Phi^{-1}(C_{i,n})$ and $C_{i,n}=\bbP(X_{i,t}\leq{n})=F_{i}(n)$. When $Q_{i,n}=\pm\infty$ for $C_{i,n}=0$ or $1$, the summand $e^{-Q_{i,n}^{2}/2}H_{k-1}(Q_{i,n})$ is interpreted as zero. For example, for $F_{i}=$ Bern$(p_{i})$, $g_{i,k}=e^{-\Phi^{-1}(1-p_{i})^{2}/2}H_{k-1}(\Phi^{-1}(1-p_{i}))/(k!\sqrt{2\pi})$. Similarly, $g_{i,k}$ can be computed when $F_i = \textrm{Pois}(\lambda_i)$ but \eqref{e:hermite_coefficient} will have infinitely many terms. However, the number of terms will be finite and small in practice. This is because the Poisson distribution is light-tailed so that $Q_{i,n}$ is indistinguishable from 1 numerically even at moderate $n$.

The link functions $L_{ij}$ can now be expressed as
\begin{equation}\label{e:link_function_expansion}
    L_{ij}(u)=\sum_{k=1}^{\infty}\frac{k!g_{i,k}g_{j,k}}{\Sigma_{X,ii}(0)^{1/2}\Sigma_{X,jj}(0)^{1/2}}u^{k} := \sum_{k=1}^{\infty}\ell_{ij,k}u^{k}.
\end{equation}
Under mild assumptions, they can be shown to be monotonically increasing on the interval $(-1,1)$ \citep[see Proposition A.1 in the appendix of][]{jia2023latent} with values in $(L_{ij}(-1),L_{ij}(1))$. Note that $L_{ij}(0)=0$ regardless of the marginal distribution. The quantities $\rho_{+,ij} := L_{ij}(1)$ and $\rho_{-,ij} := L_{ij}(-1)$ are given by
\begin{equation}\label{e:link_function_correlation}
    \rho_{+,ij} = L_{ij}(1)=\corr(G_{i}(Z),G_{j}(Z)),\quad \rho_{-,ij} = L_{ij}(-1)=\corr(G_{i}(Z),G_{j}(-Z))
\end{equation}
for $Z=\mathcal{N}(0,1)$. When $i=j$, $\rho_{+,ij}=1$ but usually $\rho_{-,ij}>-1$. As noted in \cite{jia2023latent}, $\rho_{+,ij}$ and $\rho_{-,ij}$ are the largest and smallest correlations that two dependent count variables with marginals $F_i$ and $F_j$ can achieve; by \eqref{e:link_function_correlation}, they are achieved with construction $G(Z)$ and hence within our considered model. For example, when $F_{i}=$ Bern$(p_{i})$, it can be shown by using \eqref{e:link_function_correlation} that
\begin{eqnarray*}
    \rho_{+,ij}=\left\{\begin{array}{ll}
                \sqrt{\frac{p_{i}(1-p_{j})}{p_{j}(1-p_{i})}}, & \textrm{if }p_{i}\leq{p}_{j}, \\
                \sqrt{\frac{p_{j}(1-p_{i})}{p_{i}(1-p_{j})}}, & \textrm{if }p_{j}<{p}_{i}, \qquad
                \end{array}\right.
    \rho_{-,ij}=\left\{\begin{array}{ll}
                -\sqrt{\frac{(1-p_{i})(1-p_{j})}{p_{i}p_{j}}}, & \textrm{if }p_{i}+{p}_{j}\geq1, \\
                -\sqrt{\frac{p_{j}p_{i}}{(1-p_{i})(1-p_{j})}}, & \textrm{if }p_{i}+{p}_{j}<1.
                \end{array}\right.
\end{eqnarray*}

Since the link functions $L_{ij}$ are monotonically increasing, the inverse link functions can be defined as $L_{ij}^{-1}:[\rho_{-,ij},\rho_{+,ij}]\mapsto[-1,1]$. We will discuss the numerical calculation of the inverse $L_{ij}^{-1}$ in Section \ref{sse:estimation-unknown} below. Thus, \eqref{e:link_function_entrywise} and \eqref{e:link_function_matrix} imply that
\begin{equation*}
    R_{Z,ij}(h)=L_{ij}^{-1}(R_{X,ij}(h))
\end{equation*}
or, in short and entrywise,
\begin{equation}\label{e:inverse_link_matrix}
    R_{Z}(h)=L^{-1}(R_{X}(h)).
\end{equation}
We will exploit \eqref{e:inverse_link_matrix} in estimation in the next section.

\section{Estimation of model parameters}
\label{se:estimation}

The relation \eqref{e:inverse_link_matrix} suggests a natural estimation procedure for the model parameters of the latent series $\{Z_t\}$. Indeed, recall that the component function $L_{ij}$ of $L$ is defined through the marginal CDFs $F_i$ and $F_j$ which depend on the marginal parameters $\theta_i$ and $\theta_j$, respectively. As expanded upon in Section \ref{se:estimation-marginal} of Supplemental Material, the marginal parameters $\theta_i$ can be estimated marginally from the stationary data $X_{i,1},\ldots,X_{i,T}$. This leads to estimates $\widehat{\theta}_i$ and $\widehat{\theta}_j$, and hence $\widehat{L}_{ij}$ and $\widehat{L}$. Computation of the inverse link functions $L_{ij}^{-1}$ (or $\widehat{L}_{ij}^{-1}$) is discussed in Section \ref{sse:estimation-unknown} below. On the other hand, $R_{X}(h)$ in \eqref{e:inverse_link_matrix} can be estimated through sample ACF $\widehat{R}_X(h)$ of the data. Substituting the sample quantities $\widehat{L}$ and $\widehat{R}_X(h)$ into the right-hand side of \eqref{e:inverse_link_matrix} leads to estimate $\widehat{R}_Z(h)$, which characterizes the second-order properties of the latent process $\{Z_t\}$. For the latent VAR-DFM considered in this work, the model parameters can be estimated from the model second-order properties and hence $\widehat{R}_Z(h)$ as described in Section \ref{sse:estimation-known}. Taken together, the approach allows estimating the marginal and latent series parameters of our model.

\subsection{Estimation of parameters of latent series}
\label{sse:estimation-known}

We describe here estimation of the parameters $\Lambda,\Psi_{1},\ldots,\Psi_{p},\Sigma_{\varepsilon}$ and $\Sigma_{\eta}$ of the latent series. We assume here that the marginal CDFs $F_{i}$ and hence the link functions $L_{ij}$ are known. In practice, the link functions can be estimated from marginal stationary data as noted above. We also assume here that $r$ and $p$ are known. Their estimation is descibed in Section \ref{sse:estimation-r} below and Section \ref{se:selection-p} of Supplemental Material. Note that the relation \eqref{e:inverse_link_matrix} allows one to estimate $R_{Z}(h)=\Sigma_{Z}(h)$, $h=0,\ldots,p$, as
\begin{equation}\label{e:matrix_inv_link_function}
    \widehat{R}_{Z}(h)=L^{-1}(\widehat{R}_{X}(h)),
\end{equation}
where $\widehat{R}_{X}(h)$ is the sample matrix ACF of the data $X_{1},\ldots,X_{T}$. $\widehat{R}_{X}(h)$, $h\neq0$, are not necessarily symmetric. Similarly, $\widehat{R}_Z(0)$ is symmetric but not necessarily nonnegative definite. The estimated covariance $\widehat{R}_Z(0)$ of the latent Gaussian process will be used in forecasting described in the next section. If needed, we employ a small positive shift of the eigenvalues to make $\widehat{R}_Z(0)$ nonnegative definite. However, we do not shift eigenvalues through the estimation procedure.

Since $\{Z_t\}$ follows the dynamic factor model \eqref{e:dynamic_factor} and \eqref{e:VAR}, the loadings matrix $\Lambda$, $\Sigma_{Y}(0)$, and $\Sigma_{\varepsilon}$ can be estimated through principal component analysis (PCA). More specifically, since the dynamic factor model \eqref{e:dynamic_factor} implies
\begin{equation}\label{e:cov_Z_0}
    R_{Z}(0)=\Lambda\Sigma_{Y}(0)\Lambda'+\Sigma_{\varepsilon},
\end{equation}
it is natural to estimate $\Lambda\Sigma_{Y}(0)\Lambda'$ as an $r-$rank approximation of $R_{Z}(0)$, and take $\Sigma_{\varepsilon}$ as the approximation error. We thus proceed as follows. Consider the covariance $\widehat{R}_Z(0)$ estimated by \eqref{e:matrix_inv_link_function}. Let $\widehat{R}_{Z}(0)=\widehat{U}\widehat{E}\widehat{U}'$ be the eigendecomposition with $\widehat{E}=\diag(\widehat{e}_{1},\ldots,\widehat{e}_{d})$ consisting of ordered eigenvalues $\widehat{e}_{1}\geq\ldots\geq\widehat{e}_{d}$, and $\widehat{U}=(\widehat{u}_{1},\ldots,\widehat{u}_{d})$ being the orthonormal eigenvector matrix. Setting $\widehat{U}_{r}=(\widehat{u}_{1},\ldots,\widehat{u}_{r})$ and $\widehat{E}_{r}=\diag(\widehat{e}_{1},\ldots,\widehat{e}_{r})$ as $d\times r$ and $r \times r$ matrices, respectively, a rank-$r$ approximation of $\Sigma_{Z}(0)$ can be taken as
\begin{equation}\label{e:matrix_approximation}
    \widehat{U}_{r}\widehat{E}_{r}\widehat{U}_{r}'=(\widehat{U}_{r}\widehat{E}_{r}^{1/2})(\widehat{U}_{r}\widehat{E}_{r}^{1/2})'.
\end{equation}
The relations \eqref{e:matrix_approximation} and \eqref{e:cov_Z_0} suggest estimating the $d \times r$ loadings matrix $\Lambda$ and the $r \times r$ covariance matrix $\Sigma_{Y}(0)$ of the factor series as
\begin{equation} \label{e:est_Lambda_Cov_F}
    \widehat{\Lambda}=\widehat{U}_{r}\widehat{E}_{r}^{1/2}, \quad \widehat{\Sigma}_{Y}(0) = I_r.
\end{equation}
The choice \eqref{e:est_Lambda_Cov_F} identifies $\Lambda,\Sigma_{Y}(0)$ up to a non-singular $r \times r$ transformation, assuming that
\begin{equation} \label{e:identifiability_cond_typical}
    \Lambda'\Lambda= \textrm{diagonal},\quad \Sigma_{Y}(0)=I_r.
\end{equation}
The identifiability condition \eqref{e:identifiability_cond_typical} is common in factor models \cite[e.g.][]{doz2011two,bai2013principal}. Another identifiability condition used in DFMs proposed by \cite{bai2015identification} is to make the first $r \times r$ block of the loadings matrix be identity, that is,  
\begin{equation} \label{e:identifiability_cond}
    \Lambda 
    = \begin{pmatrix} I_r \\ 
                     \Lambda_2
      \end{pmatrix},
\end{equation}
where $\Lambda_{2}$ is $(d-r)\times r$. Note that with the convention \eqref{e:identifiability_cond} above, 
\begin{equation} \label{e:identifiability_est}
    \Lambda \Sigma_{Y}(0) \Lambda' 
    = \begin{pmatrix}
    \Sigma_{Y}(0) & \Sigma_{Y}(0)\Lambda_{2}' \\
    \Lambda_{2}\Sigma_{Y}(0) & \Lambda_{2}\Sigma_{Y}(0)\Lambda_{2}'
    \end{pmatrix}
    =\begin{pmatrix}
    \Sigma_{Y}(0)^{1/2} \\
    \Lambda_{2}\Sigma_{Y}(0)^{1/2}
    \end{pmatrix}
    \begin{pmatrix}
    \Sigma_{Y}(0)^{1/2} & \Sigma_{Y}(0)^{1/2}\Lambda_{2}'
    \end{pmatrix}.
\end{equation}
The relations \eqref{e:identifiability_est} and \eqref{e:matrix_approximation} also suggest setting
\begin{equation} \label{e:est_Lambda_Cov_F_another}
    \begin{pmatrix}
    \Sigma_{Y}(0)^{1/2} \\
    \Lambda_{2}\Sigma_{Y}(0)^{1/2}
    \end{pmatrix} = \widehat{U}_{r}\widehat{E}_{r}^{1/2}
\end{equation}
to define both $\widehat{\Sigma}_{Y}(0)^{1/2}$ and $\widehat{\Lambda}_{2}$. Either \eqref{e:est_Lambda_Cov_F} or \eqref{e:est_Lambda_Cov_F_another} lead to estimators $\widehat{\Lambda}$ and $\widehat{\Sigma}_{Y}(0)$. The estimator $\widehat{\Sigma}_{\varepsilon}$ can now be defined as
\begin{equation}\label{e:estim_cov_eps}
    \widehat{\Sigma}_{\varepsilon} = \widehat{R}_{Z}(0)-\widehat{U}_{r}\widehat{E}_{r}\widehat{U}_{r}'.
\end{equation}
We consider the first identifiability condition \eqref{e:identifiability_cond_typical} for the various simulation settings in Section \ref{se:simulation}, while the second condition \eqref{e:identifiability_cond} is used in Section \ref{se:application} with the application.

Note also that DFM \eqref{e:dynamic_factor} yields
\begin{equation} \label{eq:RZdecomp}
    R_{Z}(h)=\Lambda\Sigma_{Y}(h)\Lambda',\quad h=1,\ldots,p.
\end{equation}
Setting $\widehat{R}_{Z}(h)=L^{-1}(\widehat{R}_{X}(h))$ as in \eqref{e:matrix_inv_link_function} naturally suggests the estimators
\begin{equation}\label{e:optimization_lag1}
    \widehat{\Sigma}_{Y}(h) = (\widehat{\Lambda}'\widehat{\Lambda})^{-1} (\widehat{\Lambda}' \widehat{R}_{Z}(h) \widehat{\Lambda}) (\widehat{\Lambda}'\widehat{\Lambda})^{-1}, \quad h=1,\ldots,p.
\end{equation}
Alternatively, the estimators \eqref{e:optimization_lag1} also solve
\begin{equation*}
    \widehat{\Sigma}_{Y}(h)=\argmin_{\Sigma_{Y}(h)\in\mathbb{R}^{r \times r}}\left\|\widehat{R}_{Z}(h)-\widehat{\Lambda}\Sigma_{Y}(h)\widehat{\Lambda}'\right\|_{F}^{2},\quad h=1,\ldots,p,
\end{equation*}
where $\| A \|_{F}^2 = \sum_{i,j=1}^d |A_{ij}|^2$ denotes the Frobenius norm for a matrix $A =(A_{ij})_{i,j=1,\dots,\Dim} \in \RR^{\Dim \times \Dim}$.
Note that the estimator $\widehat{\Sigma}_{Y}(h)$ can be obtained regardless of the identifiability condition, either \eqref{e:identifiability_cond_typical} or \eqref{e:identifiability_cond}. Having these estimators, the Yule-Walker equations can now be used to obtain the rest of the required estimators $\widehat{\Psi}_{1},\ldots,\widehat{\Psi}_{p}$ and $\widehat{\Sigma}_{\eta}$. That is, $\widehat{\Psi}_{1},\ldots,\widehat{\Psi}_{p}$ solve the system of matrix linear equations
\begin{equation}\label{e:yule_walker}
    \left(\begin{array}{cccc}  
    \widehat{\Sigma}_{Y}(0) & \widehat{\Sigma}_{Y}(1) & \ldots & \widehat{\Sigma}_{Y}(p-1) \\
    \widehat{\Sigma}_{Y}(1)' & \widehat{\Sigma}_{Y}(0) & \ldots & \widehat{\Sigma}_{Y}(p-2) \\
    \vdots & \vdots & \ddots & \vdots \\
    \widehat{\Sigma}_{Y}(p-1)' & \widehat{\Sigma}_{Y}(p-2)' & \ldots & \widehat{\Sigma}_{Y}(0)
    \end{array}\right)
     \left(\begin{array}{c} \widehat{\Psi}_{1}' \\ \widehat{\Psi}_{2}' \\ \vdots \\ \widehat{\Psi}_{p}' \end{array}\right)
    =  \left(\begin{array}{c} \widehat{\Sigma}_{Y}(1)' \\ \widehat{\Sigma}_{Y}(2)' \\ \vdots \\ \widehat{\Sigma}_{Y}(p)' \end{array}\right)
\end{equation}
and $\widehat{\Sigma}_{\eta}$ is 
\begin{equation*}
    \widehat{\Sigma}_{\eta}=\widehat{\Sigma}_{Y}(0)-\sum_{h=1}^{p}\widehat{\Psi}_{h}\widehat{\Sigma}_{Y}(h)'.
\end{equation*}

\begin{remark}\label{rem:latent_factor}
To reiterate the idea of our approach, any estimation of the latent process (principal component analysis, Yule-Walker equations, etc.) that can be carried out on the process in terms of its second-order properties, will have its counterpart for the considered model in terms of the observable process by using relation \eqref{e:inverse_link_matrix}. We shall exploit this idea again in cross-validation below (Section \ref{sse:estimation-r}) when selecting $r$. (See also Supplemental Material for choosing $p$.) For principal component analysis, in particular, the relation \eqref{e:inverse_link_matrix} shows how the data covariance matrix $\Sigma_{X}(0)$ should be transformed entry-wise before applying the analysis.
\end{remark}

\subsection{Calculating inverse of link function}
\label{sse:estimation-unknown}

In estimation, the link functions $L_{ij}$ entering $L$ in \eqref{e:matrix_inv_link_function} are replaced by their sample counterparts $\widehat{L}_{ij}$. We expand here on how $\widehat{L}_{ij}$ and their inverse $\widehat{L}_{ij}^{-1}$ are computed.

Link functions $L_{ij}$ in \eqref{e:link_function_expansion} are defined in terms of the Hermite coefficients $g_{i,k},g_{j,k}$ which are obtained from marginal distributions $F_{i},F_{j}$ through \eqref{e:hermite_coefficient}. If $F_{i},F_{j}$ are characterized by model parameters $\theta_i,\theta_j$, then $g_{i,k}=g_{i,k}(\theta_i),g_{j,k}=g_{j,k}(\theta_j)$ are functions of these parameters. The parameters $\theta_i,\theta_j$ can be estimated through observations for marginal $i,j$ as discussed in Section \ref{se:estimation-marginal}, leading to estimated Hermite coefficients $\widehat{g}_{i,k}=g_{i,k}(\widehat{\theta}_i),\widehat{g}_{j,k}=g_{j,k}(\widehat{\theta}_j)$ and link function coefficients $\widehat{\ell}_{ij,k}$. Link function estimates are then $\widehat{L}_{ij}(u)=\sum_{k=1}^K \widehat{\ell}_{ij,k}u^k$ for large $K$, say $K=100$ or more. To simplify the notation, we write $L_{ij}$ for $\widehat{L}_{ij}$, and consider the computation of $L_{ij}^{-1}$ next.

The idea to calculate $L_{ij}^{-1}$ is as follows. Partition the interval $[-1,1]$ into $u_0<u_1<\dots<u_M$ that satisfies $u_{0}=-1$, $u_{M}=1$, and set
\begin{equation*}
    v_{m} := L_{ij}(u_{m})\qquad \textrm{or}\qquad L_{ij}^{-1}(v_{m})=u_{m},
\end{equation*}
so that one has the points $(v_{m},L_{ij}^{-1}(v_{m}))=(v_{m},u_{m})$ on the curve $L_{ij}^{-1}(v)=u$. The value of $L_{ij}^{-1}(v)$ for other points $v$ can then be obtained through piecewise linear interpolation. In addition, we use finer grids for $u$ near $\pm1$ while wider grids are used around $u=0$. From a numerical standpoint, the $M+1$ points $(v_m,u_m)$ satisfy $v_m = L_{ij}^{-1}(u_m)$, and the interpolation of $L_{ij}^{-1}$ is defined as
\begin{equation*}
    \widetilde{L}_{ij}^{-1}(v) 
    := \sum_{m=1}^{M}\widetilde{L}_{ij,m}^{-1}(v)\boldsymbol{1}_{[v_{m-1},v_{m})}(v)
\end{equation*}
for $v\in(v_{m},v_{m+1})$, $m=0,1,\ldots,M$, with $M$ pieces of the linear spline polynomials
\begin{equation*}
    \widetilde{L}_{ij,m}^{-1}(v) 
    = \widetilde{L}_{ij,m}^{-1}(v_{m-1}) + \frac{\widetilde{L}_{ij,m}^{-1}(v_{m})-\widetilde{L}_{ij,m}^{-1}(v_{m-1})}{v_{m}-v_{m-1}}(v-v_{m-1}).
\end{equation*}

Figure \ref{fig:Figure_link} depicts $L_{ij}(u),L_{ij}^{-1}(v)$ and its interpolation $\widetilde{L}_{ij}^{-1}(v)$ for four representative marginal count distributions: Bernoulli, categorical, Poisson, and negative binomial. For example, the Bernoulli case considers a pair of $F_i=\textrm{Bern}(p_i)$ and $F_j=\textrm{Bern}(p_j)$ with four different choices of combinations $(p_i,p_j)$. Several combinations of parameters for other types of distributions, including categorical, Poisson, and negative binomial distributions, are also considered. As seen from the plots, the inverse link function $L_{ij}^{-1}$ obtained by flipping the axes and numerical inverse $\widetilde{L}_{ij}^{-1}$ obtained through the linear interpolation are nearly indistinguishable. These combinations of marginal distributions are used in the simulation study in Section \ref{se:simulation}.

\subsection{Selecting the number of factors}
\label{sse:estimation-r}

In practice, the number of factor series $r$ is unknown and needs to be chosen for the model estimation. We consider here several methods for this task based on available approaches and also introduce cross-validation schemes tailored for our model.

One practical approach is to examine a scree plot of the eigenvalues of $\widehat{\Sigma}_{Z}(0) = \widehat{L}^{-1}(\widehat{\Sigma}_{X}(0))$ where the presence of a ``knee'' suggests the value of $r$. More formally, one can design an algorithm that determines the ``knee.'' For example, \cite{onatski2010determing} suggested the Edge Distribution (ED) estimator, whereby
\begin{equation}\label{e:estim_ED}
    \widehat{r}(\delta) 
    := \max\{k \leq r_{\max} : \widehat{e}_k - \widehat{e}_{k+1} \geq \delta\}, 
\end{equation}
where $\widehat{e}_1 \geq \widehat{e}_2 \geq \ldots \geq \widehat{e}_{d}$ are the ordered eigenvalues of $\widehat{R}_Z(0)$ and $\delta$ is calibrated through the algorithm described in Section 4 of that paper. 

Alternatively, one could rely on information criteria (IC). That is, from $q=1,\ldots,r_{\max}$, we choose $q$ as the estimate $\widehat{r}$ that minimizes
\begin{equation}\label{e:information_criteria}
    \textrm{IC}(q) = \ln\left( \frac{1}{dT}\|\widehat{\Sigma}_{\varepsilon}(q)\|_F^2\right) + qg_i(d,T),
\end{equation}
where $\widehat{\Sigma}_{\varepsilon}(q)$ is the estimator in \eqref{e:estim_cov_eps} from the rank-$q$ approximation of $\Sigma_{Z}(0)$ in \eqref{e:matrix_approximation}, and $g_{i}(d,T)\to 0$ and $\min(d,T)g_i(d,T)\to \infty$ as $d,T \to \infty$. The recommended choices of the penalty functions $g_{i}(d,T)$ are
\begin{equation} \label{e:penalties_rank}
    \begin{gathered}
    g_{1}(d,T) = \frac{d+T}{dT}\ln\left(\frac{dT}{d+T}\right), \quad g_{2}(d,T) = \frac{d+T}{dT}\ln\left(C_{dT}^2\right), \\
    g_{3}(d,T) = \frac{\ln(C_{dT}^2)}{C_{dT}^2}, \quad C_{dT}=\min(\sqrt{d},\sqrt{T}).
    \end{gathered}
\end{equation}
which corresponds to $\textrm{IC}_{p1}(r)$--$\textrm{IC}_{p3}(r)$ studied by \cite{bai2002determining}. 

We now propose a block cross-validation (BCV)-based rank selection. The idea goes back at least to \cite{browne1989single}, and continues being utilized in psychometrics \cite[e.g.][]{haslbeck2022estimating}. Differences from the setting considered here are that our observations are serially correlated and that our factor model is latent. To account for temporal dependence, we do not partition observations randomly but rather into equally sized consecutive blocks. The latent nature of the factor model will be dealt with by exploiting the idea in Remark \ref{rem:latent_factor}. We thus fold $\{X_t\}$ along the time into $B$ blocks. The superscript $(b)$ will refer to the $b$th block, to be used for test data. The superscript $(-b)$ will refer to the $b$th block being excluded, to be used for training data. Let $\widehat{R}_{Z}^{(b)}(0) = \widehat{L}^{-1}(\widehat{R}_{X}^{(b)}(0))$ be the sample matrix ACF of the latent Gaussian series at lag 0, computed from the sample matrix ACF of the observations from the $b$th block, substituted into the inverse link function. Similarly, one can compute $\widehat{R}_{Z}^{(-b)}(0) = \widehat{L}^{-1}(\widehat{R}_{X}^{(-b)}(0))$ that excludes the $b$th block. Then, for each candidate rank $q$ on a grid $q=1,\ldots,r_{\max}$, the mean square error (MSE) of BCV is defined as 
\begin{equation*}
    \textrm{MSE}(q) 
    = \frac{1}{B}\sum_{b=1}^B \left\|\widehat{R}_{Z}^{(b)}(0) - \widehat{R}_{Z}^{(-b,q)}(0) \right\|_F^2,
\end{equation*}
where $\widehat{R}_{Z}^{(-b,q)}(0)$ is the rank-$q$ approximation of $\widehat{R}_{Z}^{(-b)}(0)$ plus a diagonal matrix from the estimated covariance matrix of the innovations of the factor model. The PCA-based estimation procedure described in Section \ref{sse:estimation-known} can be used. The minimizer $q$ of the MSE is chosen as the estimate of the number of factor series $r$.

Instead of our PCA-based estimation, other estimation approaches can and have been used as well. For example, in the minimum residual factor analysis \cite[MINRES,][]{harman1966factor}, $\widehat{R}_{Z}^{(-b,q)}(0)$ is the rank-$q$ approximation of $\widehat{R}_{Z}^{(-b)}(0)$ obtained by minimizing the difference between $\widehat{R}_{Z}^{(-b)}(0)$ and the sum of $\widehat{R}_{Z}^{(-b,q)}(0)$ and a diagonal matrix, the latter accounting for the variances of the error terms. This estimation approach is quite popular in factor analysis, with the estimation procedure implemented through R package \texttt{psych} by \cite{revelle2023psych}. Other estimation approaches for factor analysis can be found in \cite{bertsimas2017certifiably}.

The selection of lag order $p$ in \eqref{e:VAR} can be similarly conducted by using cross-validation, which is explained in Supplemental Material. Note that determining the number of factors can be performed regardless of the lag order of the factor series. We thus recommend choosing $r$ first, followed by selecting $p$.

\section{Theoretical properties} 
\label{se:theory}

In this section, we investigate theoretical properties of our estimators for the latent factor model \eqref{e:dynamic_factor}. Our results are derived for the PCA estimators of the latent factor model in Section \ref{se:estimation}. The proofs are based on \cite{duker2024high} who provide a theoretical foundation to investigate the model \eqref{e:generalized_inverse} for general latent Gaussian processes. We base our investigations on results in \cite{doz2011two} who prove that the PCA estimators are consistent when the series following a factor model is observed and also on concentration results of bounded functions of Markov chains in \cite{fan2021hoeffding}. We start with briefly recalling our estimation procedure and introduce some additional notation in Section \ref{sec:TP_estimation}. We collect assumptions in Section \ref{se:assumptions} and then state our main results in Section \ref{se:main_results}. Section \ref{se:example} concludes with a particular case satisfying the assumptions made in Section \ref{se:assumptions}.

\subsection{Estimation procedure with additional notation}
\label{sec:TP_estimation}

We first recall and supplement some of the quantities and their estimated counterparts given in Section \ref{sse:estimation-known}. Recall the function $L$ from \eqref{e:link_function_expansion} and introduce its non-standardized counterpart 
\begin{align} \label{eq:function_ell}
\ell(u) = (\ell_{ij}(u))_{i,j = 1, \dots, \Dim}
\hspace{0.2cm} \text{ with } \hspace{0.2cm} 
\ell_{ij}(u) = \sum_{k=1}^{\infty} k! g_{i,k} g_{j,k} u^k,
\end{align}
where $g_{i,k}$ are the Hermite coefficients defined in \eqref{e:hermite_coefficient}. Then, using \eqref{eq:function_ell}, \eqref{e:link_function_matrix} can also be written as
\begin{equation} \label{eq:pupulation_Rsigmaell}
    \Sigma_{X}(h)=\ell(R_{Z}(h))
\end{equation}
with $R_{Z}(h)$ as in \eqref{e:standarization}. We write $C_{n}(\theta_{i}):=C_{i,n}$ and $Q_{n}(\theta_{i}):=Q_{i,n}$ to emphasize dependence on $\theta_{i}$. An estimator of $\ell$ is written as $\widehat{\ell}$ and computed by replacing $C_{i,n}$ with $\widehat{C}_{i,n} = C_{n}( \widehat{\theta}_{i})$ in $Q_{i,n} = \Phi^{-1}(C_{i,n})$ in \eqref{e:hermite_coefficient}.
Estimation can then be conducted using \eqref{eq:pupulation_Rsigmaell} and a regular autocovariance estimator $\widehat{\Sigma}_{X}(h)$ for $\Sigma_{X}(h)$ based on our observed count series $\{X_t\}$ such that
\begin{equation*}
    \widehat{R}_{Z}(h)=\widehat{\ell}^{-1}(\widehat{\Sigma}_{X}(h)).
\end{equation*}
We also introduce $\bmR = (R_Z(r-s))_{r,s=1,\dots,\VO}$ and $\bmS = (\Sigma_X(r-s))_{r,s=1,\dots,\VO}$ with $\VO$ denoting the lag order of the latent factors. 
With a slight abuse of notation, we write
\begin{equation} \label{eq:boldRandR}
    \bmR = \ell^{-1}(\bmS)
    \hspace{0.2cm}
    \text{ and }
    \hspace{0.2cm}
    \bmhR = \widehat{\ell}^{-1} ( \bmhS ).
\end{equation}

We now move on to estimation of the latent model parameters in \eqref{e:dynamic_factor} and \eqref{e:VAR}. Based on the decomposition \eqref{e:matrix_approximation}, we get
\begin{equation} \label{eq:Lambda_hat}
    \widehat{\Lambda} = \widehat{U}_{r}\widehat{E}_{r}^{1/2}.
\end{equation}
On the population level, we write $E_r$ for the diagonal matrix whose diagonal entries are the eigenvalues of $\Lambda' \Lambda$ and $Q_r$ for the matrix of a set of unitary eigenvectors associated with $E_r$. Set further 
\begin{equation} \label{eq:other_relations}
    U_r = \Lambda Q_r E_r^{-1/2}
\end{equation}
and recall from \eqref{eq:RZdecomp} the relations
\begin{equation*}
    R_{Z}(h)
    = \Lambda\Sigma_{Y}(h)\Lambda'
    = \Lambda Q_r Q'_r \Sigma_{Y}(h) Q_r Q'_r \Lambda' 
    ,\quad h=1,\ldots,p.
\end{equation*}

In order to estimate the transition matrices in the VAR($\VO$) model \eqref{e:VAR}, 
we write \eqref{e:VAR} as a $\VO r$-dimensional VAR($1$) model, that is, 
\begin{equation} 
\label{eq:writtenasVAR1}
\begin{pmatrix}
Y_{t} \\
Y_{t-1} \\
\vdots \\
Y_{t-p+1}
\end{pmatrix}
=
\begin{pmatrix}
\TM_{1}	&	\cdots	& \TM_{\VO-1}	& \TM_{\VO} \\
I_{r}		&	\cdots	& 0			&	0 \\
\vdots	&	\ddots	&	\vdots	&	\vdots \\
0		&	\cdots	&	I_{r}		&	0\\
\end{pmatrix}
\begin{pmatrix}
Y_{t-1} \\
\vdots \\
Y_{t-p}
\end{pmatrix}
+
\begin{pmatrix}
\eta_{t} \\
0 \\
\vdots \\
0
\end{pmatrix}
\hspace{0.2cm}
\text{ or }
\hspace{0.2cm}
\mathcal{Y}_{t} = \bm{\TM} \mathcal{Y}_{t-1} + \widetilde{\eta}_{t}.
\end{equation}
On the population level, we introduce
$\Sigma_{Y}(h) = \E[ Y_{t+h} Y_{t}']$, $\bm{\Sigma}_Y^{(p)} := (\Sigma_Y(r-s))_{r,s=1,\dots,\VO}$ such that $\bm{\Sigma}_{QY}^{(p)} := (Q\Sigma_Y(r-s)Q')_{r,s=1,\dots,\VO} = \bm{Q} \bm{\Sigma}_{Y}^{(p)} \bm{Q}'$ with $\bm{Q} = I_p \otimes Q_r$.
Consistency results are derived below up to transformation with the orthogonal matrix $Q_r$.
Since the factors are defined up to a pre-multiplication by an invertible matrix, we choose $\Sigma_Y(0) = I_r$ and maintain this assumption throughout the paper.

Using the introduced notation, the equation in \eqref{e:yule_walker} can be written as
\begin{equation} \label{e:yule_walker_rewirtten} 
    \bm{\widehat{\Sigma}}_Y^{(p)}
    \widehat{\bm{\TM}}'
    =  (S_1 \bm{\widehat{\Sigma}}_Y^{(p+1)} S'_2)'
\end{equation}
with $S_1 =
\begin{pmatrix}
0_{p} & I_{p}
\end{pmatrix}$, $S_2 =
\begin{pmatrix}
I_{p} & 0_{p} 
\end{pmatrix}$, $0_{p}$ being a $p-$dimensional column vector with all entries set to zero. Furthermore, we can infer from \eqref{e:yule_walker_rewirtten} that 
\begin{equation} \label{eq:widehatbmTM}
    \widehat{\bm{\TM}} = S_2 \bm{\widehat{\Sigma}}_Y^{(p+1)} S'_1 (\bm{\widehat{\Sigma}}_Y^{(p)})^{-1}
\end{equation}
is the solution of \eqref{e:yule_walker} and we get $\bm{\TM} = S_2 \bm{\Sigma}_{QY}^{(p+1)} S'_1 (\bm{\Sigma}_{QY}^{(p)})^{-1}$ on the population level.
We further have $
    \widehat{\Sigma}_{Y}(h) = 
    \widehat{E}_r^{-1/2} \widehat{U}'_r \widehat{R}_Z(h) \widehat{U}_r \widehat{E}_r^{-1/2}
$
and
\begin{equation} \label{eq:estimatorSigmaY}
    \bm{\widehat{\Sigma}}_{Y} = 
    (I_p \otimes \widehat{E}_r^{-1/2} \widehat{U}'_r ) \bm{\widehat{R}}_Z 
    (I_p \otimes \widehat{E}_r^{-1/2} \widehat{U}'_r )'.
\end{equation}
%

\subsection{Assumptions} 
\label{se:assumptions}

We work with two sets of assumptions. The first assumption set applies to Proposition 3.1 in \cite{duker2024high}, which relates the probability of how much $\bm{\widehat{R}}_Z$ deviates from $\bm{R}_{Z}$ to the analogous probabilities for the respective quantities of the observed series $\{ X_t \}$.

\begin{Lassump} \label{ass:L1}
There is a constant $\conZ \in (0,1)$ such that $| R_{Z,ij}(h) | < \conZ$ for $h \neq 0$, $i, j=1, \dots, d$ and $| R_{Z,ij}(0) | < \conZ$ for all $i \neq j$.
\end{Lassump}

\begin{Lassump} \label{ass:L2}
For each $\theta_{i} = (\theta_{i1}, \dots, \theta_{iK_{i}})'$, there exists an open neighborhood $S$ of $\theta_{i}$ such that  
$\sup_{\theta_{i} \in S}\E[ |X_{i,t}|^p] = \sup_{\theta_{i} \in S}\E_{\theta_{i}}[ |X_{i,t}|^p] <\infty$ for some $ p > 2$.
\end{Lassump}

\begin{Lassump} \label{ass:L3}
For each $\theta_{i} = (\theta_{i1}, \dots, \theta_{iK_{i}})'$, there exists an open neighborhood $S$ of $\theta_{i}$ such that 
\begin{align*}
&\sup_{\theta_{i} \in S} \sum_{n = 0}^{\infty} (1-C_{n}(\theta_{i}))^{-1/2} 
\sum_{j=1}^{K_{i}}
\left| \frac{\partial}{\partial \theta_{ij}} C_{n}(\theta_{i}) \right| 
< \infty.
\end{align*}
\end{Lassump}

\begin{Lassump} \label{ass:L4}
For each $\theta_{i} = (\theta_{i1}, \dots, \theta_{iK_{i}})'$, there exist an open neighborhood $S$ of $\theta_{i}$ and at least one $n$ such that $\inf_{\theta_{i} \in S} C_{n}(\theta_{i}) > 0$.
\end{Lassump}

The constant $\conZ$ in Assumption \ref{ass:L1} controls the dependence of the series $\{Z_t\}$ and appears in the non-asymptotic bound derived in \cite{duker2024high} which are used in our work. The assumption is expected to hold for some $\conZ$ for the processes considered both here and in \cite{duker2024high}. Assumption \ref{ass:L3} is shown to hold for several common count distributions under Assumption \ref{ass:L2}; see Appendix E in \cite{duker2024high}. Note that we require our moment conditions to hold uniformly in a neighborhood around $\theta_{i}$. This allows us to infer finiteness on a compact subset of the parameter space of $\theta_{i}$. The second set of assumptions is needed to derive consistency results for estimators of the latent factor model. The asymptotics described below are understood as $d,T\to\infty$.

\begin{Cassump} \label{Ass:C1}
Suppose
    \begin{equation*}
    \| \widehat{\theta} - \theta \|_{\maxF}
    =
    \mathcal{O}_{\prob}\left( \sqrt{ \frac{\log(dT)}{T}} \right).
\end{equation*}
\end{Cassump}

\begin{Cassump} \label{Ass:C2}
Set $\bmS^{(p)} := (\Sigma_X(r-s))_{r,s=1,\dots,\VO}$ and similarly for its estimated counterpart $\bmhS^{(p)} := (\widehat{\Sigma}_X(r-s))_{r,s=1,\dots,\VO}$. Suppose
    \begin{equation*}
    \| \bmhS^{(p+1)} - \bmS^{(p+1)} \|_{\maxF}
    =
    \mathcal{O}_{\prob}\left( \sqrt{ \frac{\log(dT)}{T}} \right).
\end{equation*}
\end{Cassump}

Assumptions \ref{Ass:C1} and \ref{Ass:C2} are crucial to ensure consistent estimation of the transition matrices of the latent factors. We point out that the rate $\sqrt{ \frac{\log(dT)}{T}}$ might not be optimal. Results in \cite{doz2011two}, where estimation for the latent factors is done assuming that the series following a factor model is observed, suggest that a rate of $\frac{1}{\sqrt{T}}$ might be possible. See also Remark \ref{re} below. We prove that Assumptions \ref{Ass:C1} and \ref{Ass:C2} are satisfied under mild assumptions on the marginal distributions in Section \ref{se:example}. 

We remark also that Assumptions \ref{Ass:C1} and \ref{Ass:C2} are the analogues of Assumptions C.1 and C.2 in \cite{duker2024high}. Assumptions C.1 and C.2 in \cite{duker2024high} ensure consistent estimation of $R_Z$ when $\{Z_t\}_{t\in\mathbb{Z}}$ follows a $d-$dimensional VAR($p$) model.
Those assumptions are concentration inequalities, suggesting non-asymptotic bounds for the deviation between estimated and true quantities. We use here a representation in terms of stochastic boundedness which is a weaker assumption but sufficient for our purposes since we aim to recover the results in \cite{doz2011two} who also state their results in terms of stochastic boundedness. 

The third set of assumptions is on the latent factor model. The assumptions below and their interpretation are essentially the same as in \cite{doz2011two}. The notation $\| \cdot \|$ stands for the spectral norm of a matrix, and $\| \cdot \|_{\max}$ for the maximum of its elements (in absolute value). For a symmetric matrix $M$, $\lambda_{\min}(M)$ and $\lambda_{\max}(M)$ will refer to its minimum and maximum eigenvalues.

\begin{Fassump} \label{ass:2}
$\{Y_t\}_{t\in\mathbb{Z}}$ and $\{\varepsilon_t\}_{t\in\mathbb{Z}}$ are independent and:
\begin{itemize}
\item
$r-$vector factor series $\{Y_{t}\}_{t\in\mathbb{Z}}=\{(Y_{i,t})_{i=1,\ldots,r}\}_{t\in\mathbb{Z}}$ follows a stationary VAR($p$) model given by
\begin{equation}\label{e:VAR2}
    Y_{t}=\Psi_{1}Y_{t-1}+\ldots+\Psi_{p}Y_{t-p}+\eta_{t},
\end{equation}
where $\Psi_{1},\ldots,\Psi_{p}$ are $r\times{r}$ matrices and $\eta_{t}$ are i.i.d. $\mathcal{N}(0,\Sigma_{\eta})$ random $r-$vectors.
\item
$\varepsilon_{t}$ are i.i.d. $\mathcal{N}(0,\Sigma_{\varepsilon})$ $d-$vectors with $\Sigma_{\varepsilon} = \diag(\sigma_1^2, \dots, \sigma_d^2)$ and $\|\Sigma_{\varepsilon} \|<\infty$.
\end{itemize}
\end{Fassump}

\begin{Fassump} \label{ass:8}
The eigenvalues of $\Lambda' \Lambda$ are distinct.
\end{Fassump}

\begin{Fassump} \label{ass:4}
 $\liminf_{d\to\infty} \frac{\lambda_{\min}(\Lambda'\Lambda) }{d} > 0$.
\end{Fassump}

\begin{Fassump} \label{ass:7}
 $\limsup_{d\to\infty}  \frac{\lambda_{\max}(\Lambda'\Lambda)}{d}   = \limsup_{d\to\infty} \frac{\|\Lambda\|^2}{d} <\infty$ and $\| \Lambda \|_{\max} \leq \widebar{\lambda} < \infty$.
\end{Fassump}

\subsection{Main results} 
\label{se:main_results}

Our main results state that certain quantities of the latent model can be consistently estimated using the relationship \eqref{e:matrix_inv_link_function}. 
This will yield consistency of the loadings matrix and the transition matrices in the VAR$(p)$ representation of the latent factors. The proofs of the stated results are in Appendix \ref{ap:consistency}.

The following results are analogous to Lemma 2(i), and Proposition 3 in \cite{doz2011two}. While \cite{doz2011two} assume that the observed series follows a factor model, we recover their results even though we observe a count series with latent factor model structure. Note that our rate differs from that derived by \cite{doz2011two} by a $\sqrt{\log(dT)}$ term; see also Remark \ref{re} below.

\begin{proposition}\label{prop1}
Recall $\bmhR$ from \eqref{eq:boldRandR}. Suppose Assumptions \ref{ass:L1}--\ref{ass:L4}, \ref{Ass:C1}--\ref{Ass:C2}, \ref{ass:2}--\ref{ass:7}. Then, 
\begin{equation*}
\frac{1}{d} \| \bmhR - \bmL \bm{\Sigma}_Y \bmL' \| = 
\mathcal{O}_{\prob}\left( \sqrt{\frac{\log(dT)}{T}} \right) + \mathcal{O}\left(\frac{1}{d}\right)
\end{equation*}
with $\bm{\Lambda} = I_p \otimes \Lambda$.
\end{proposition}

\begin{proposition} \label{prop2}
Recall $\widehat{E}_r$, $\widehat{U}_r$ and $Q_r$ from \eqref{eq:Lambda_hat} and \eqref{eq:other_relations}. Suppose Assumptions \ref{ass:L1}--\ref{ass:L4}, \ref{Ass:C1}--\ref{Ass:C2}, \ref{ass:2}--\ref{ass:7}. Then, 
\begin{equation*}
\| \bm{\widehat{\Sigma}}_{Y} - \bm{Q}' \bm{\Sigma}_{Y} \bm{Q} \| = 
\mathcal{O}_{\prob}\left( \sqrt{\frac{\log(dT)}{T}} \right) + \mathcal{O}_{\prob}\left(\frac{1}{d}\right)
\end{equation*}
with $\bm{\widehat{\Sigma}}_{Y}$ as in \eqref{eq:estimatorSigmaY} and $\bm{Q} = I_p \otimes Q_r$.
\end{proposition}

\begin{corollary} \label{cor1}
Recall $\widehat{\bm{\TM}}$ from \eqref{eq:widehatbmTM}. Suppose Assumptions \ref{ass:L1}--\ref{ass:L4}, \ref{Ass:C1}--\ref{Ass:C2}, \ref{ass:2}--\ref{ass:7}. Then, 
\begin{equation*}
\| \widehat{\bm{\TM}} - \bm{\TM} \|
= \mathcal{O}_{\prob}\left( \sqrt{\frac{\log(dT)}{T}} \right) + \mathcal{O}_{\prob}\left(\frac{1}{d}\right)
\end{equation*}
with $\bm{\TM} = S_2 \bm{\Sigma}_{QY}^{(p+1)} S'_1 (\bm{\Sigma}_{QY}^{(p)})^{-1}$.
\end{corollary}

\begin{lemma} \label{lem1}
Recall $\widehat{\Lambda}$ from \eqref{eq:Lambda_hat}. Under Assumptions \ref{ass:L1}--\ref{ass:L4}, \ref{Ass:C1}--\ref{Ass:C2}, \ref{ass:2}--\ref{ass:7}, for fixed $i=1,\dots,d$, 
\begin{equation*}
    (\widehat{\Lambda} - \Lambda)_i =
    \mathcal{O}_{\prob}\left( \sqrt{\frac{\log(dT)}{T}} \right) +
    \mathcal{O}_{\prob}\left(\frac{1}{d}\right),
\end{equation*}
where $(\widehat{\Lambda} - \Lambda)_i$ denotes the $i$th row of $\widehat{\Lambda} - \Lambda$. 
\end{lemma}

\subsection{Assumptions \ref{Ass:C1} and \ref{Ass:C2}} 
\label{se:example}

Under some additional assumptions on the marginal distribution of the observed series in \eqref{e:generalized_inverse}, one can verify Assumptions \ref{Ass:C1} and \ref{Ass:C2}.

\begin{Eassump} \label{Ass:E1}
The parameter $\theta_{i}$ is such that $\theta_{i} = \E [X_{i,t}]$ which allows  estimating it via $\widehat{\theta}_{i} = \frac{1}{T} \sum_{t = 1}^{T} X_{i,t}$.
\end{Eassump}

Define the function $G: \RR^d \to \RR^d$ with $z \mapsto G(z)$ such that our multivariate count model can be written as
\begin{equation} 
\label{eq:subo}
X_{t} 
= (X_{1,t}, \dots, X_{\Dim,t})'
= (G_{1}(Z_{1,t}), \dots, G_{\Dim}(Z_{\Dim,t}))'
= G(Z_{t}).
\end{equation}

\begin{Eassump} \label{Ass:E2}
The function $G$ in \eqref{eq:subo} satisfies $G: \RR^{\Dim} \to [a,b]^{\Dim}$.
\end{Eassump}

Assumptions \ref{Ass:E1} and \ref{Ass:E2} cover important cases in the modeling of count time series including Bernoulli marginals. We expect that the boundedness of $G$ in Assumption \ref{Ass:E2} can be relaxed but are constrained by concentration results currently available in the literature.

Our proposed verification of Assumptions \ref{Ass:C1} and \ref{Ass:C2} employs concentration results for Markov chains. In Appendix \ref{se:proofs_verif_ass}, we define a reversible Markov chain $\{ \mathcal{V}_t \}$ with Gaussian transition kernel $K(x,\cdot)$ with mean $\bm{\TM}_{\mathcal{V}} x$ and covariance matrix $\Sigma_{\xi}$, where, with $\aaa = d(p-1)$, 
\begin{equation}
    \bm{\TM}_{\mathcal{V}}
    =
    \begin{pmatrix}
        \bm{\Phi} & \widetilde{\bm{\Phi}} \\
        0_{dp \times dp} & 
        \begin{matrix}
            0_{d \times \aaa} & 0_{d \times d} \\
            I_{\aaa} & 0_{\aaa \times d}
        \end{matrix}
    \end{pmatrix},
    \hspace{0.2cm}
    \Sigma_{\xi} =
    \begin{pmatrix}
        \widetilde{\Sigma} & 0_{d \times \aaa} & \Sigma_{\varepsilon} & 0_{d\times \aaa}  \\
        0_{\aaa \times d} & 0_{\aaa \times \aaa} & 0_{\aaa \times d} & 0_{\aaa \times \aaa} \\
        \Sigma_{\varepsilon} & 0_{d \times \aaa} & \Sigma_{\varepsilon} & 0_{d \times \aaa} \\
        0_{\aaa \times d} & 0_{\aaa \times \aaa} & 0_{\aaa \times d} & 0_{\aaa \times \aaa} 
    \end{pmatrix}
\end{equation}
with
\begin{equation*}
    \bm{\Phi} =
    \begin{pmatrix}
    \Phi_{1}	&	\cdots	& \Phi_{\VO-1}	& \Phi_{\VO} \\
    I_{d}		&	\cdots	& 0			&	0 \\
    \vdots	&	\ddots	&	\vdots	&	\vdots \\
    0		&	\cdots	&	I_{d}		&	0\\
    \end{pmatrix},
    \hspace{0.2cm}
    \widetilde{\bm{\Phi}} =
    \begin{pmatrix}
    -\Phi_{1}	&	\cdots	& -\Phi_{\VO} \\
    0_{\aaa \times d}&	\cdots	&	0_{\aaa \times d} \\
    \end{pmatrix},
\end{equation*}
\begin{equation*}
    \Phi_i = \frac{1}{d} \Lambda \TM_i \Lambda'
    \hspace{0.2cm}
    \text{ and }
    \hspace{0.2cm}
    \widetilde{\Sigma} = \E[ (\Lambda \eta_t + \varepsilon_t)(\Lambda \eta_t + \varepsilon_t)']
= \Lambda \Sigma_{\eta} \Lambda' + \Sigma_{\varepsilon}.
\end{equation*}
We assign more meaning to the above quantities in the proofs; see Appendix \ref{se:proofs_verif_ass}. 

The used concentration results for Markov chains are expressed in terms of $\lambda_{\rig}:=\lambda_{\rig}(P)$, the rightmost value of the spectrum $[-\lambda, \lambda]$ of the Markov operator $P$ induced by the transition kernel $K$. We refer to $1-\lambda_{\rig}$ as the right spectral gap of the Markov chain.
For the reversible Markov chain $\{ \mathcal{V}_t \}$ with Markov operator $P$ and stationary distribution $\pi$, $\lambda_{\rig}$ is defined as 
\begin{equation} \label{def:spectral_gap}
    \lambda_{\rig}:=\lambda_{\rig}(P) := \sup\{ \langle Ph,h \rangle_{\pi} ~|~ \| h \|_{\pi} = 1, \pi(h) = 0 \},
\end{equation}
where $\langle h_1,h_2 \rangle_{\pi} = \pi(h_1 h_2)$ and $\| h \|_{\pi}$ is its induced norm.

\begin{Eassump} \label{Ass:E3}
The Markov chain $\{ \mathcal{V}_t \}$ with transition kernel $K(x,\cdot)$ admits a spectral gap $1-\lambda_{\rig}>0$ that satisfies $\limsup_{d \to \infty} \lambda_{\rig} < 1 $.
\end{Eassump}
Assumption \ref{Ass:E3} is expected to be satisfied under quite general assumptions. We refer to Section \ref{se:discussion_Ass_E3} for a discussion on Assumption \ref{Ass:E3}.

The following lemma formalizes that the described setting is sufficient for Assumptions \ref{Ass:C1} and \ref{Ass:C2} to be satisfied.
\begin{lemma} \label{le:justifyAss}
Suppose Assumptions \ref{ass:2}, \ref{ass:4}, \ref{ass:7} and \ref{Ass:E1}--\ref{Ass:E3} are satisfied. Then, Assumptions \ref{Ass:C1} and \ref{Ass:C2} hold.
\end{lemma}

\section{Forecasting}
\label{se:forecast}

\subsection{Particle-based sampling procedure}
\label{sse:particle}

A fitted latent Gaussian dynamic factor model in \eqref{e:generalized_inverse}--\eqref{e:VAR} can naturally be used to forecast the series $X_{t}$. We will show how this can be carried out for fixed model parameters. For example, the model parameters can be obtained through estimation (in which case our forecast will not reflect any uncertainty from estimation error). More specifically, for a given $t$ (typically, $t=T$, the sample length), we are interested in the distribution of
\begin{equation}\label{e:forecasting}
    \widehat{X}_{t+h|t}=(X_{t+h}|X_{1}=x_{1},\ldots,X_{t}=x_{t}),
\end{equation}
where $h=1,2,\ldots$, the vertical bar indicates the conditioning and $x_{1},\ldots,x_{t}$ are the observed values of $X_{1},\ldots,X_{t}$. This is equivalent to finding
\begin{equation}\label{e:filtering}
    \mathbb{E}_{x_{1:t}}\left[V(\widehat{X}_{t+h|t})\right]
\end{equation}
for arbitrary function $V$, where the subscript $x_{1:t}$ in $\mathbb{E}_{x_{1:t}}$ refers to the conditioning on $\{x_{1},\ldots,x_{t}\}$ as in \eqref{e:forecasting}. For example, suppose we have a $d-$dimensional vector $x$. With $V(x) = V((x_{i})_{i=1,\ldots,d})=\boldsymbol{1}_{\{x_{i}=n_{i},\ i=1,\ldots,d\}}$ and $d-$dimensional integer $n$, the quantity \eqref{e:filtering} becomes $\mathbb{P}_{x_{1:t}}(\widehat{X}_{t+h|t}=n)$, which is of primary interest in forecasting $X_{t+h}$, when the components of $X_t$ are integer-valued. 

Let $\widehat{Z}_{t+h|t}=\widehat{Z}_{t+h}(Z_{1:t})=H_{t1}^{(h)}Z_{t}+\ldots+H_{tt}^{(h)}Z_{1}$ be the $h$-step-ahead linear prediction of $Z_{t+h}$ from $Z_{1:t}$. Define $\widehat{R}_{t+h|t} = \mathbb{E}[(Z_{t+h}-\widehat{Z}_{t+h|t})(Z_{t+h}-\widehat{Z}_{t+h|t})']$ as the corresponding covariance matrix of prediction error of $Z_{t+h}$. One can compute the latent prediction $\widehat{Z}_{t+h|t}$ by using Kalman recursions as recalled in Appendix \ref{ap:kalman}. This exploits the state-space structure of the model and is more efficient computationally than a direct application of e.g. Durbin-Levinson algorithm. Then, the quantity \eqref{e:filtering} can be expressed as
\begin{equation} \label{e:SMC_approx}
    \mathbb{E}_{x_{1:t}}\left[V(\widehat{X}_{t+h|t})\right]=\mathbb{E}_{x_{1:t}}\left[D_{V,t+h}(\widehat{Z}_{t+h|t})\right],
\end{equation}
where
\begin{equation}\label{e:integral_D}
    D_{V,t+h}(z)
    =\int_{\mathbb{R}^d} {V}(G(z_{t+h}))\frac{ \exp\left( -1/2(z-z_{t+h})'\widehat{R}_{t+h|t}^{-1}(z-z_{t+h}) \right)} {(2\pi)^{d/2}|\widehat{R}_{t+h|t}|^{1/2}}dz_{t+h}
\end{equation}
\cite[see equations (23)--(24) in ][]{jia2023latent}. The right-hand side of \eqref{e:SMC_approx} will be approximated through a Monte Carlo scheme below; direct numerical calculation of underlying integrals is too cumbersome. 

Note that conditioning on $x_{1:t}$ does not determine the exact path of $z_{1:t}$. Indeed, recall from \eqref{e:ar1example_poisson} that for $i=1,\ldots,d$,
\begin{equation*}
    \{X_{i,t}=x_{i,t}\} 
    = \left\{ \Phi^{-1}(C_{i,x_{i,t}-1}) < Z_{i,t} \leq \Phi^{-1}(C_{i,x_{i,t}}) \right\}=\{ Z_{i,t}\in A_{i,x_{i,t}} \},
\end{equation*}
where $C_{i,n}=\mathbb{P}(X_{i,t}\leq n)$ and $A_{i,x_{i}}=\left(\Phi^{-1}(C_{i,x_{i}-1}),\Phi^{-1}(C_{i,x_{i}}) \right]$. That is, each entry of the realization of $X_{t}$ is determined by the range of the corresponding entry of $Z_t$ at each time $t$. For $d-$dimensional observations, one has
\begin{equation*}
    \{X_t = x_t\} 
    = \bigcap_{i=1}^d\left\{ Z_{i,t}\in A_{i,x_{i,t}}\right\} = \{Z_t \in A_{x_{t}}\},
\end{equation*}
where $A_{x_t}=A_{1,x_{1,t}}\times \ldots \times A_{d,x_{d,t}}$. The notation $A_{i,x_{i}}$ and $A_x$ will be used below. In the Bernoulli case, for example,
\begin{equation*}
    A_{i,x_{i,t}}= \left( \Phi^{-1}(C_{i,x_{i,t}-1}),\Phi^{-1}(C_{i,x_{i,t}}) \right] 
    = \left\{\begin{array}{ll}
    \left(\Phi^{-1}(1-p_{i}),\infty \right) , & \textrm{if}\ x_{i,t}=1, \\
    \left(-\infty,\Phi^{-1}(1-p_{i}) \right], & \textrm{if}\  x_{i,t}=0.
    \end{array}\right.
\end{equation*}
The Bernoulli marginal distribution thus has the largest ranges for $Z_t$. In a Monte Carlo approximation of \eqref{e:SMC_approx}, one will be generating $Z_{i,t}\in A_{i,x_{i,t}}$ and producing their forecast $\widehat{Z}_{i,t+h}$.

The following presentation extends that of \cite{jia2023latent} to the multivariate setting relevant to the Monte Carlo approximation problem. The quantity \eqref{e:SMC_approx} is known to be well approximated through Sequential Monte Carlo (SMC) by generating ``particles'' over time $t$ \cite[e.g.][]{doucet2001introduction,doucet2009tutorial}. The main difference from \cite{jia2023latent} is that the model here has two latent processes $\{Z_t\}_{t\in\mathbb{Z}}$ and $\{Y_t\}_{t\in\mathbb{Z}}$. To deal with the two latent processes, we use Kalman recursions to forecast and update the latent process $\{Y_t\}$ and approximate the distribution of $\{Z_t\}$ conditioning on $\{X_t\}$. This approach is called Rao-Blackwellization and the method is adapted from the partially observed Gaussian state-space models \cite[e.g.][]{andrieu2002particle,briers2010smoothing}.

The following is the SMC algorithm for particle filtering to generate particles $\{\widetilde{Z}_t^{(k)}\}_{t=1,\ldots,T}$, $k=1,\ldots,N$, over time, whose weighted average then approximates \eqref{e:SMC_approx}. The particle $\{\widetilde{Z}_{t}^{(k)}\}_{t=1,\ldots,T}$ can be regarded as the realization of the underlying latent process. Additionally, we add a resampling step which is often used in sequential Monte Carlo algorithms. We will explain the necessity of resampling below.

\textbf{Sequential Importance Sampling and Resampling (SIS/R)}: Set the initial importance weights $w_{0}^{(k)}=1$ for all $k$, initialize $\widetilde{Y}_{0|0}^{(k)} \sim \mathcal{N}(0,\widetilde{Q}_{0|0})$, where $\widetilde{Q}_{0|0}=\textrm{Var}(Y_0)$. For $p=1$, $\widetilde{Q}_{0|0}$ is approximated by $\sum_{m=0}^M \Psi^m \Sigma_{\eta} (\Psi')^m$ for large $M$. Then, recursively over $t=1,\ldots,T$, do the following steps: For each $k=1,\ldots,N$:
\begin{enumerate}
    \item Forecasting step: Compute $\widehat{Y}_{t|t-1}^{(k)}$, $\widehat{Q}_{t|t-1}$, $\widehat{Z}_{t|t-1}^{(k)}$ and $\widehat{R}_{t|t-1}$ via Kalman recursions (see Appendix \ref{ap:kalman}).
 
    \item Importance sampling step: Sample residual $\xi_{t}^{(k)}$ satisfying
    \begin{equation}\label{e:sampling_error}
        \xi_{t}^{(k)}\stackrel{d}{=}
        \mathcal{N}_d\left(0_{d},I_{d} \Big|\Phi^{-1}(C_{x_t-1}) 
        < \widehat{Z}_{t|t-1}^{(k)} + \widehat{R}_{t|t-1}^{1/2}\xi_{t}^{(k)} \leq \Phi^{-1}(C_{x_t}) \right),
    \end{equation}
    where $\Phi^{-1}(C_{n})=( \Phi^{-1}(C_{1,n_{1}}), \ldots, \Phi^{-1}(C_{d,n_{d}}))'$ and $\mathcal{N}_d(\mu,\Sigma|A)$ indicates a $d-$dimensional multivariate normal distribution with mean $\mu$ and covariance $\Sigma$ restricted to the set $A$. Then, update the particle as
    \begin{equation*} 
        \widehat{Z}_{t}^{(k)} = \widehat{Z}_{t|t-1}^{(k)} + \widehat{R}_{t|t-1}^{1/2}  \xi_t^{(k)}
    \end{equation*}
    and update the importance weight as $w_{t}^{(k)}=w_{t-1}^{(k)} w_{t}(\widehat{Z}_{t|t-1}^{(k)})$, where
    \begin{equation}\label{e:importance_weight}
        w_{t}(\widehat{Z}_{t|t-1}^{(k)}) = \mathbb{P}\left( \mathcal{N}(\widehat{Z}_{t|t-1}^{(k)},\widehat{R}_{t|t-1}) \in A_{x_t} \right).
    \end{equation}
    
    \item Resampling step: Set $\Omega_{t,N} = \sum_{k=1}^N w_{t}^{(k)}$ and normalize $w_{t}^{(k)}$ by $\widetilde{w}_{t}^{(k)}=w_{t}^{(k)}/\Omega_{t,N}$. Take a quartet $(\widetilde{w}_{t}^{(k)},\widetilde{Y}_{t|t-1}^{(k)},\widetilde{Z}_{t|t-1}^{(k)},\widetilde{Z}_{t}^{(k)})$ as follows.
    \begin{enumerate}
        \item[-] If a resampling criterion described around \eqref{e:ESS} is satisfied, then take $(\frac{1}{N},\widehat{Y}_{t|t-1}^{(I_k)},\widehat{Z}_{t|t-1}^{(I_k)},\widehat{Z}_{t}^{(I_k)})$, where $\{I_k\}$ are chosen indices after resampling.
        
        \item[-] If the criterion is not satisfied, then take $(\widetilde{w}_{t}^{(k)},\widehat{Y}_{t|t-1}^{(k)},\widehat{Z}_{t|t-1}^{(k)},\widehat{Z}_{t}^{(k)})$.
    \end{enumerate}
    
    \item Updating step: Use $\widetilde{Y}_{t|t-1}^{(k)},\widetilde{Z}_{t|t-1}^{(k)},\widetilde{Z}_{t}^{(k)}$ and $\widehat{Q}_{t|t-1}$ to compute $\widetilde{Y}_{t|t}^{(k)}$ and $\widetilde{Q}_{t|t}$ via Kalman recursions (see Appendix \ref{ap:kalman}).
\end{enumerate}

Finally, the SMC approximation of \eqref{e:SMC_approx} becomes
\begin{equation}\label{e:weighted_average}
    \mathbb{E}_{x_{1:t}}[V(\widehat{X}_{t+h|t})] \approx \sum_{k=1}^{N}\widetilde{w}_t^{(k)}D_{V,t+h}\left(\widehat{Z}_{t+h|t}^{(k)}\right),
\end{equation}
where $\widehat{Z}_{t+h|t}$, $h\geq1$, are computed through forecasting step in the Kalman recursions. See equation (25) in \cite{jia2023latent} for the justification of an analogous approximation.

The SMC is known to suffer from the so-called weight degeneracy of particles \cite[e.g.][]{snyder2008obstracles} which occurs when the variance of normalized weights becomes inflated. The latter happens and becomes worse as the sample size increases. To overcome this, it is suggested to remove the particles with small weights. By following \cite{doucet2009tutorial}, we resample only when the effective sample size (ESS) exceeds $N/2$, as a rule of thumb, for the criteria of resampling, where the ESS is defined as
\begin{equation}\label{e:ESS}
    \textrm{ESS}_t = \left(\sum_{k=1}^N \left(\widetilde{w}_t^{(k)}\right)^2 \right)^{-1}.    
\end{equation}
More specifically, we resample particles by following systematic resampling. That is, sample $U_1 \sim \mathcal{U}(0,1/N)$ and set $U_k = U_1 + \frac{k-1}{N}$, $k=2,\ldots,N$. Then, compute 
\begin{equation*}
    I_{k} = \left| \left\{ U_i:\sum_{j=1}^{k-1}\widetilde{w}_{t}^{(j)} 
    \leq U_{i} \leq 
    \sum_{j=1}^{k}\widetilde{w}_{t}^{(j)} \right\} \right|
\end{equation*}
with $\sum_{j=1}^0=0$ as a convention. This is used in Step 3 of the SIS/R algorithm above. Alternatively, one can use categorical resampling, which is resampling particles by regarding $\{\widetilde{w}_t^{(k)}\}$ as a categorical probability distribution. Many other resampling methods exist (see \cite{douc2005comparison} for more information).

\begin{remark}\label{rem:marginal_dist}
The forecasting distribution \eqref{e:filtering} is characterized by the function 
\begin{equation*}
    V(x)=V((x_i)_{i=1,\ldots,d}) = \boldsymbol{1}_{\{x_i = n_i,\ i=1,\ldots,d\}}
\end{equation*} 
for fixed $n=(n_i)_{i=1,\ldots,d}$. For a single $d-$dimensional forecast value, one could take $n=(n_i)_{i=1,\ldots,d}$ for which \eqref{e:filtering} is largest with the corresponding functions $V(x)$. Note, however, that this is a daunting task computationally. For example, even with the Bernoulli marginals where $n_i=0$ or 1, the number of functions $V$ to consider is $2^d$, which grows exponentially in $d$. To sidestep this issue, we only consider $V(x)=\boldsymbol{1}_{\{x_i=n_i\}}$ in practice and take $n_i$ as the forecast value in the $i$th coordinate for which \eqref{e:filtering} is largest. For example in the Bernoulli case, this task computationally is of the order $d$.
\end{remark}

\subsection{Speed-up in forecasting computation}
\label{sse:speed_up}

The computation burden for sequential Monte Carlo sampling is substantial. The majority of the cost is due to sampling (doubly) truncated multivariate Gaussian random variables $\{\xi_{t}^{(k)}\}$ in \eqref{e:sampling_error}, and the fact that the algorithm runs for $t=1,\ldots,T$. Currently, we implement sampling through the R package \texttt{TruncatedNormal} developed by \cite{botev2017normal}. But a significant improvement in the computation speed for generating truncated multivariate normal random variables is not expected.

To reduce the computational cost, we note that the covariance matrices $\widehat{R}_{t|t-1}$ of prediction error typically converge within a few steps. This is due to a similarly quick convergence of the covariance matrix $\widehat{Q}_{t|t-1}$ of prediction error of the factor series in \eqref{e:prediction_Q}, and the covariance matrix $\widetilde{Q}_{t|t}$ and Kalman gain $K_t$ described in \eqref{e:kalman_update_cov}.

From the pair of covariance matrices in \eqref{e:prediction_Q} and \eqref{e:kalman_update_cov}, one has the recursive equation
\begin{equation*}
    \widehat{Q}_{t+1|t} = \Psi\left(\widehat{Q}_{t|t-1} - \widehat{Q}_{t|t-1} \Lambda'(\Lambda\widehat{Q}_{t|t-1}\Lambda'+\Sigma_{\varepsilon})^{-1}\Lambda\widehat{Q}_{t|t-1}\right)\Psi' + \Sigma_{\eta},
\end{equation*}
with the given initial condition $\widehat{Q}_{1|0}=\Psi\widetilde{Q}_{0|0}$. The covariance matrices of the prediction error converge to a positive definite matrix $Q$ satisfying the discrete algebraic Riccati equation (DARE),
\begin{equation*}
    Q = \Psi\left(Q - Q \Lambda'(\Lambda Q \Lambda'+\Sigma_{\varepsilon})^{-1}\Lambda Q\right)\Psi' + \Sigma_{\eta}.
\end{equation*}
One has a similar equation for the covariance matrices $\widetilde{Q}_{t|t}$. It is the convergence to these equations that happens within a few time steps substantially shorter than the length of observations $T$. Since the purpose of the SIS/R algorithm is to obtain the importance weights $\{\widetilde{w}_{T}^{(k)}\}$ along the particles $\{Z_{1:T}^{(k)}\}$ and we presume the stable factor series, it is reasonable to run the forecasting algorithm with only a few last observations. In simulation and application below, we employ the forecasting algorithm with the last 5 observations.

\subsection{On forecasting for longer horizons}
\label{sse:convergence}

In this section, we briefly discuss what to expect from the forecasting method when the forecasting horizon $h$ becomes longer. Recall that the latent factor series is stationary and follows a stable VAR model. The latent process $\{Z_t\}_{t\in\mathbb{Z}}$ is also stationary and its long-term prediction converges to its mean, which is a zero vector. From \eqref{e:forecast_Y_after_T} and \eqref{e:forecast_Z_after_T}, the predicted particles are therefore expected to converge eventually to zero vectors as well. Thus, for longer horizon $h$, we expect
\begin{equation}\label{e:approximation_longer_horizon}
    \mathbb{E}_{x_{1:t}}[V(\widehat{X}_{t+h|t})] \approx D_{V,t+h}(0).
\end{equation}
As in Remark \ref{rem:marginal_dist}, consider $V(x) = V(x_i) = \boldsymbol{1}_{\{x_i=n_i\}}$, $i=1,\ldots,d$. For $\widehat{Z}_{t+h|t}=z=0$, \eqref{e:integral_D} becomes
\begin{eqnarray}
    D_{V,t+h}(0) 
    &=& \int_{\mathbb{R}^d} \boldsymbol{1}_{\{(G(z_{t+h}))_{i}=n_i\}} \frac{ \exp\left( -\frac{1}{2}(z_{t+h})'\widehat{R}_{t+h|t}^{-1}(z_{t+h}) \right)} {(2\pi)^{d/2}|\widehat{R}_{t+h|t}|^{1/2}}dz_{t+h} \nonumber\\
    &=& \int_{\left\{G_i(z_{i,t+h})=n_i\right\}} \frac{ \exp\left( -\frac{1}{2}z_{i,t+h}^2/(\widehat{R}_{t+h|t})_{ii} \right)} {\sqrt{2\pi}|(\widehat{R}_{t+h|t})_{ii}|^{1/2}}dz_{i,t+h}, \label{e:integral_longer_horizon}
\end{eqnarray}
where $(\widehat{R}_{t+h|t})_{ii}$ is the $i$th diagonal entries of covariance matrix $\widehat{R}_{t+h|t}$. For longer horizon $h$, $(R_{t+h|t})_{ii}\approx \textrm{Var}(Z_{i,t})=1$. We thus expect from \eqref{e:approximation_longer_horizon} and \eqref{e:integral_longer_horizon} that for longer horizon $h$,
\begin{equation}\label{e:long_run_prob}
    \mathbb{E}_{x_{1:t}}[\boldsymbol{1}_{\{(\widehat{X}_{t+h|t})_{i}=n_i\}}] 
    \approx D_{V,t+h}(0) \approx \mathbb{P}( Z \in A_{i,n_i} ) = \mathbb{P}( X_i = n_i ),
\end{equation}
where $Z \sim \mathcal{N}(0,1)$ and $X_i$ has a CDF $F_i$ with the parameter $\theta_i$. Hence, for longer horizon $h$, our forecasting method can be thought as choosing $n_i$ among possible values that maximizes the most likely count value according to the distribution $F_i$ with the parameter $\theta_i$. Two observations are worth making in this regard. 

First, in some instances, e.g. Bernoulli and categorical distributions, $\theta_i$ represents proportion of count values and is estimated as the corresponding sample proportions. In these cases, for long horizon $h$,  we therefore expect our forecast to yield the most likely observed count (modulo the issue of ties). This is the case with the application considered in Section \ref{se:application}. On the other hand, for many other distributions, this observation may not necessarily hold. For example, for the Poisson distribution, the parameter is taken as the sample mean and the most likely count according to this Poisson distribution does not need to be the most likely observed value, let alone be in the sample.

Second, the relation \eqref{e:long_run_prob} might be confusing from the following point of view. As argued above, for longer horizon $h$, we expect $\widehat{Z}_{t+h|t}\approx0$. In fact, we see this clearly in the application of Section \ref{se:application}. The relation \eqref{e:long_run_prob} might be read as saying that 0 belongs to the bin $A_{i,n_i}$ with the highest standard normal probability. This is, however, not necessarily the case. It will be the case when $n_i$ is the median of the distribution $F_i$ (i.e. $F_{i}(n_i-1)<1/2$ and $F_{i}(n_i)\geq1/2$), a quite likely scenario in practice especially for ``bell-shaped'' distribution with the most likely value $n_i$ being at the center, but the statement will not hold in general.

\subsection{Model diagnostics}
\label{sse:diagnostics}

The goodness of fit of a time series model for univariate count data can be assessed through a probability integral transformation (PIT) histogram. See \cite{czado2009predictive,kolassa2016evaluating,jia2023latent}. PIT involves computing predictive distributions (see also below). We are not aware of PIT extensions to the multivariate setting, though natural possibilities can certainly be proposed. As with forecasting discussed in Sections \ref{sse:particle}--\ref{sse:convergence}, the bigger issue for our model and potentially high-dimensional setting is that obtaining a multivariate predictive distribution would be computationally infeasible. As in those sections, we shall compromise by computing marginal predictive distributions but conditioning on the past of all variables. More specifically, we proceed as described next.

Suppose that our model parameters $\theta,\Lambda,\Psi_{1},\ldots,\Psi_{p},\Sigma_{\varepsilon}$ and $\Sigma_{\eta}$ are estimated as in Sections \ref{sse:estimation-known}, \ref{sse:estimation-unknown}, and \ref{se:estimation-marginal}. Then, the marginal predictive distribution of $i$th variable is defined as
\begin{displaymath}
    P_{i,t}(x_i) = \mathbb{P}(X_{i,t} \leq x_i | X_{1} = x_1,\ldots,X_{t-1}=x_{t-1}) = \mathbb{P}(X_{i,t}\leq x_{i}|x_{0:t-1}).
\end{displaymath}
It can be approximated similarly to the SMC approximation \eqref{e:weighted_average} by
\begin{displaymath}
    \widehat{P}_{i,t}(x_{i}) 
    = \sum_{\ell=0}^{x_i}\widehat{\mathbb{E}}_{x_{1:t}}[1_{\{x_i=\ell\}}(X_t)] = \sum_{\ell=0}^{x_i}\widehat{\mathbb{E}}_{x_{1:t}}[D_{1_{\{x_i = \ell\},t}}(\widehat{Z}_{t|t-1})] 
    = \sum_{\ell=0}^{x_i}\sum_{k=1}^N \frac{w_{t}^{(k)}}{\Omega_{N,t}}D_{1_{\{x_i = \ell\},t}}(\widehat{Z}_{t|t-1}^{(k)}),
\end{displaymath}
where $w_{t}^{(k)}=w_{t-1}^{(k)}w_{t}(\widehat{Z}_{t|t-1}^{(k)})$ as in \eqref{e:importance_weight} and $\Omega_{N,t}=\sum_{k=1}^Nw_{t}^{(k)}$, and by using \eqref{e:integral_D},
\begin{align*}
    D_{1_{\{x_i = \ell\},t}}(z) 
    &= \int_{\{G_i(z_{i,t})=\ell\}} \frac{\exp(-(z_i - z_{i,t})^2/(2(\widehat{R}_{t|t-1}^{1/2})_{ii}))}{\sqrt{2\pi}|(\widehat{R}_{t|t-1})_{ii}|^{1/2}}dz_{i,t} \\
    &= \Phi\left((\Phi^{-1}(C_{i,\ell})-z)/(\widehat{R}_{t|t-1}^{1/2})_{ii}\right) - \Phi\left((\Phi^{-1}(C_{i,\ell-1})-z)/(\widehat{R}_{t|t-1}^{1/2})_{ii}\right),
\end{align*}
where $\widehat{Z}_{t|t-1}^{(k)}$ is computed as in Section \ref{sse:particle} using $x_1,\ldots,x_{t-1}$, and $\widehat{R}_{t|t-1}$ is derived by Kalman recursions in Appendix \ref{ap:kalman}.

The (nonrandom) sample PIT for the $i$th component is defined as
\begin{displaymath}
    \widebar{F}_{i}(u) = \frac{1}{T+1}\sum_{t=0}^T F_{i,t}(u|x_{t,t}), \quad u\in[0,1],
\end{displaymath}
where
\begin{displaymath}
    F_{i,t}(u|x_{i,t}) 
    = \left\{\begin{array}{ll}
    0, & \textrm{if } u \leq \widehat{P}_{i,t}(x_{i,t}-1), \\
    \frac{u - \widehat{P}_{i,t}(x_{i,t}-1)}{\widehat{P}_{i,t}(x_{i,t}) - \widehat{P}_{i,t}(x_{i,t}-1)}, & \textrm{if } \widehat{P}_{i,t}(x_{i,t}-1) < u < \widehat{P}_{i,t}(x_{i,t}), \\
    1, & \textrm{if } u \geq \widehat{P}_{i,t}(x_{i,t}).
    \end{array}\right.
\end{displaymath}
In practice, the sample PIT of the $i$th component is a histogram with the height $\bar{F}_{i}(m/M) - \bar{F}_{i}((m-1)/M)$, $m=1,\ldots,M$, where $M$ is the number of bins (often $M=10$). If the model fits the data well, the sample PIT histogram should be close to a uniform distribution.

\section{Simulation study}
\label{se:simulation}

In this section, we assess the performance of estimation and forecasting procedures introduced in Sections \ref{se:estimation} and \ref{se:forecast}. We focus on Bernoulli, categorical, Poisson, and negative binomial marginal distributions with several parameter values. Our goal is to check how the performance of our proposed methods compares to that of the traditional PCA methods in the factor modeling literature \cite[e.g.,][]{lam2011estimation,doz2011two,doz2012quasi}.

\subsection{Estimation}
\label{sse:estimation}

\subsubsection{Estimation for known numbers of factors and lag order}
\label{ssse:estimation_main}

The simulation settings are as follows. First, we fix the order $p=1$ for factor series $\{Y_t\}_{t\in\mathbb{Z}}$ in \eqref{e:VAR} and take the number of factor series as $r=2$ or $r=5$. These values are assumed to be given in this section. Then, by following the standard DFM literature \cite[e.g.,][]{doz2011two}, we generate the latent Gaussian DFM \eqref{e:dynamic_factor}--\eqref{e:VAR} with parameters:

\begin{itemize}\itemsep=0.1em
    \item $\Psi_1=(\Psi_{1,ij})$ is diagonal with $\Psi_{1,ii} = 0.9$ and $\Sigma_{\eta} = (1-0.9^2)I_{r}$.
    \item $\Lambda = (\lambda_{ij})$ with $\lambda_{ij} \stackrel{i.i.d.}{\sim} \mathcal{N}(0,1)$.
    \item $\Sigma_{\varepsilon}=(\Sigma_{\varepsilon,ij})$ is diagonal with entries $\Sigma_{\varepsilon,ii}=\frac{c_{ii}}{1-c_{ii}}\sum_{j=1}^r \lambda_{ij}^2$ with $c_{ii}\stackrel{i.i.d.}{\sim}\mbox{Unif}(0.3,0.7)$.
\end{itemize}
These model parameters do not ensure that $\mathbb{E}Z_{i,t}^2=1$ as needed for \eqref{e:generalized_inverse}, so further standardization is applied. We take $d=15,30,60,90$ and $T=100,200$. Finally, to define the marginal count distributions, let $I_1,I_2,I_3$ be the sets of indices that partition $\{1,2,\ldots,d\}$ such that $I_1=\{1,\ldots,d/3\}$, $I_{2}=\{d/3+1,\ldots,2d/3\}$, and $I_3=\{2d/3+1,\ldots,d\}$. Then,
\begin{itemize}\itemsep=0.1em
    \item For Bernoulli marginal distributions, $p_{i}=0.2$, 0.4, and 0.7 for $i \in I_1$, $I_2$, and $I_3$, respectively.
    \item For categorical marginal distributions with $\theta_{i}=(p_{1,i},\ldots,p_{5,i})$ where the number of the categories is given as 5, $\theta_{i}=(0.2,\ldots,0.2)$ (uniform), $(0,0.25,0.5,0.25,0)$ (unimodal), and $(0.45,0,0.1,0,0.45)$ (trimodal) for $i \in I_1$, $I_2$, and $I_3$, respectively.
    \item For Poisson marginal distributions, $\theta_i = 0.1$, 1, and 10 for $i \in I_1$, $I_2$, and $I_3$, respectively.
    \item For negative binomial marginal distributions, $p_{i}=0.2$, 0.4, and 0.7  for $i \in I_1$, $I_2$, and $I_3$, where the number of successes is 3.
\end{itemize}
Monte Carlo simulations are based on 100 replications for each setting. For the 100 replications, we report the mean and the standard deviations of $\ell_2$ losses $\|\widehat{a}-a\|_2/\sqrt{b}$ of the estimators, where $\widehat{a}$ is the (vectorized) estimator of the parameter $a$ and $b$ is a scalar; $b=d$ for $\theta$ in \eqref{e:generalized_inverse} and $\Lambda,\Sigma_{\varepsilon}$ in \eqref{e:dynamic_factor}, and $b=r$ for $\Psi_1,\Sigma_{\eta}$ in \eqref{e:VAR}.

Table \ref{tab:table_estimation} summarizes the estimation results for several $r,d,T$ values, and marginal distributions. The results align with what is expected from the standard factor models. For example, the means of the losses decrease with increasing dimension $d$ and sample size $T$. When the number of factors $r$ increases, both the averages of the losses of the estimators $(\widehat{\Lambda},\widehat{\Sigma}_{\varepsilon})$ for the factor model \eqref{e:dynamic_factor} and those of $(\widehat{\Psi},\widehat{\Sigma}_{\eta})$ for the factor series \eqref{e:VAR} increase. Another interesting point is that the means of the losses for the factor series are larger than those for the factor model. On the observation level, the means of the losses of estimators $\widehat{\theta}$ of marginal distributions are relatively small, regardless of $d,T$, and even $r$. Furthermore, the magnitudes of the means of the losses are not substantially different across different marginal count distributions.

\subsubsection{Selection of the number of factors}
\label{ssse:r}

We investigate the performance of the selection methods of the number of factors $r$, suggested in Section \ref{sse:estimation-r}. The same model parameters as in Section \ref{ssse:estimation_main} with $d=30,50$, and 100 are used with fixed $p=1$. We denote the scree plot method of finding the ``knee'' described in \eqref{e:estim_ED} by ED. The IC methods as combinations of \eqref{e:information_criteria} with the three different penalty functions \eqref{e:penalties_rank} are denoted by IC1--IC3, respectively. For the BCV-based approach, we employ two different estimation procedures. Our principal component-based estimation in Section \ref{sse:estimation-known} is denoted by PC. On the other hand, Fac refers to MINRES estimation by \cite{harman1966factor}, which is briefly described in Section \ref{sse:estimation-r}. For each simulation setting, 100 replications are performed.

Figure \ref{fig:Figure_rank} depicts the frequencies of estimated $r$ in 100 replications. From the figure, PC method outperforms all baselines (IC1--3, ED) in selecting true $r$, and performs similarly to Fac. Interestingly, the traditional information criterion-based or scree plot-based approaches fail for this model. The quality of estimation by the BCV-based approaches follows the pattern in the estimation of model parameters. That is, as the dimension $d$ and sample length $T$ increase, so do the percentages of correctly estimated $r$. Another observation is that the larger number of factor series deteriorates the performance of cross-validations. This can be seen for both cross-validation schemes, which is also an expected phenomenon with the standard factor models. In terms of marginal distributions, cross-validation schemes work best for negative binomial, followed by Poisson, categorical, and Bernoulli marginal distributions. This indicates that cross-validation schemes improve when marginal distributions tend to take a larger number of values so that they are more akin to continuous distributions.

\subsection{Forecasting}
\label{sse:forecasting}

In this section, we assess forecasting performance in the simulation settings considered in Section \ref{ssse:estimation_main}, with sample lengths for the data generation extended by 12 observations. We then hold out the last 12 observations so that the targeted forecasting horizons are $H=1,2,3,6$, and $12$. We assume that the model parameters are given, allowing for the estimation of covariances $\widehat{Q}_{t|t-1}$ and $\widehat{Q}_{t|t}$, the Kalman gain $K_t$, and the importance weights ${\widetilde{w}_{T}^{(k)}}$ to be conducted within the last 5 observations from the sample, as discussed in Section \ref{sse:speed_up}. With these estimators, we generate $h$-step-ahead predictions $V(\widehat{X}_{T+h|T})$ through \eqref{e:weighted_average}. The remainder of the prediction follows the Kalman recursions as discussed in Appendix \ref{ap:kalman}.

Assuming the data is generated from our model, we would like to examine our forecasting scheme, especially compared to alternative approaches. We refrain from comparing forecasting based on other multivariate discrete-valued time series models mentioned in Section \ref{se:introduction}. The MATLAB codes for Bayesian DFMs by \cite{cui2014generalized} and nonstationary DFMs by \cite{wang2018modeling} are publicly available. However, those models were not considered for forecasting, necessitating additional methodological and implementation considerations. Furthermore, the assumed dimensions in those models are relatively low, typically five or fewer, compared to 15 or more in our study. As a result, we consider three naive baselines instead. First, as the most naive forecasting method, we use the last observation. We refer to this method as Last. Second, we consider each dimension $i=1,\ldots,d$ separately and define the predictions as the likeliest previous value for that dimension. For example, if the most frequent value $X_{1,t}$ for the first dimension $i=1$ within the observation window is 3 for the Poisson case, we take 3 as predictions for all 5 steps ahead. We refer to this approach as Marginal; see also Section \ref{sse:convergence}. 
Third, we also consider the case discussed at the end of Section \ref{sse:convergence} where the forecast is taken as the discrete value associated with the bin containing the value $Z=0$.
For example, if the $i$th variable follows Bernoulli distribution with $p_i=0.4$, the forecast is 0 since $0<\Phi^{-1}(1-0.4)$. On the other hand, if $p_i=0.7$, the forecast becomes 1 since $0>\Phi^{-1}(1-0.7)$. Forecasts for other distributions can be defined similarly by using \eqref{e:ar1example_poisson}. We refer to this approach as Null.

The Monte Carlo simulations are based on 100 replications. We consider two types of forecasting measures: one for the latent series and one for the observed series. For the latent Gaussian series, we compute means and standard deviations of the root mean square error (RMSE) of $H$-step ahead forecast error averaged over $N$ particles defined by 
\begin{equation}\label{e:forecasting_error}
    \sqrt{\frac{1}{Nr}\sum_{k=1}^N\left\|\widehat{Y}_{T+H|T}^{(k)}-Y_{T+H}\right\|_2^2}, \quad
    \sqrt{\frac{1}{Nd}\sum_{k=1}^N\left\|\widehat{Z}_{T+H|T}^{(k)}-Z_{T+H}\right\|_2^2}.
\end{equation}
For the observed series, we consider an out-of-sample loss, similar to \eqref{e:forecasting_error} by replacing $N=100$ and $\widehat{Y}_{T+H|T}^{(k)},Y_{T+H}$ or $\widehat{Z}_{T+H|T}^{(k)},Z_{T+H}$ with $N=1$ and $\widehat{X}_{T+H|T},X_{T+H}$, respectively. In addition, applying this performance measure to $X_t$ may be less adequate due to the discrete nature of values. So, forecasting accuracy for the observations using the proposed methods, including the two baselines, is also measured by the accuracy of $H$-step ahead forecasting,
\begin{equation*}
    \mbox{ACC}(H) 
    = \frac{1}{d}\sum_{i=1}^{d}\boldsymbol{1}_{\{\widehat{X}_{i,T+H|T} = X_{i,T+H}\}}.
\end{equation*}

Table \ref{tab:table_forecast1} reports the results of means and standard deviations of the RMSE of $H$-step ahead forecast error from the two latent processes $Z_t$ and $Y_t$, defined in \eqref{e:forecasting_error}. The increasing trends for the RMSEs of those two forecasting errors along the forecasting horizon $H$ are what one would expect from forecasting results for typical time series models. The RMSEs of the forecasting errors of the factor series \eqref{e:VAR} are comparably larger than those of the factor models \eqref{e:dynamic_factor}. Furthermore, note that the RMSEs of the forecasting errors depend on both the dimension $d$ and the number of factors $r$ but in different ways: while the increase of $d$ leads to smaller errors of both the factor series and the factor models, the increase of $r$ causes the increase and decrease of errors in the factor models and factor series, respectively.

Table \ref{tab:table_forecast2} reports the results from the observations $X_t$, in terms of the mean of RMSEs of forecasting errors and accuracy of forecasting including three benchmarks. Similarly to the two latent processes, the RMSEs of the forecasting errors increase as the forecasting horizon increases. Furthermore, the RMSEs of the forecasting errors for the observed series tend to be larger than those for the latent processes. It is interesting to note the larger forecasting errors for marginal distributions that take more values, such as negative binomial, followed by Poisson, categorical, and Bernoulli distributions. This was not the case for the RMSEs of the forecasting errors of latent processes. Similar tendencies as for the RMSEs of the forecasting errors are observed with the accuracy measure. Finally, compared to Last, Marginal, and Null approaches, consistently better forecasting performance is noted for the proposed method (the larger accuracy is associated with better predictors).

\section{Application}
\label{se:application}

To demonstrate the utility of the proposed model, we consider individual-level time series consisting of daily self-reported measures of personality, collected by \cite{borkenau1998big}. That study was designed to explore items describing personal emotions that represent the ``Big Five'' personality factors. The structure of the data is as follows. Time series for 30 emotion-related items were collected, where groups of 6 items are known to correspond to one of the five personality factors. These data were collected for 22 students over 90 days.
 
All 30 items are believed to correspond to at least one of the ``Big Five'' personality factors. These factors are known to follow a structured categorization, denoted as categories 1 through 5 below. Each evening, participants in the study were instructed to evaluate their daily behavior for each item on a scale from 0 to 6, with 6 indicating the strongest endorsement of the emotion that day. For illustrative purposes, we focus on the data from one student out of the 22 available. This choice of the student was largely motivated by the following consideration: for many students, responses to some items exhibited very little variability (i.e., remained mostly constant), making them less interesting from a time series modeling perspective. To mitigate the influence of extreme observations 0 and 6, we merged them with 1 and 5, respectively, resulting in a new scale ranging from 1 to 5. Practitioners intending to use our model (or any other) should first conduct similarly basic exploratory data analysis.

The data set and its time series plots are presented in Figure \ref{fig:Figure_observation}. Among 90 consecutive observations, we use the first 85 observations for estimation and reserve the last 5 as a holdout sample to evaluate our forecasts. The corresponding 30 items are grouped by the identified categories C1 through C5, each containing 6 items. The time series of individual items exhibit substantial variability. In addition, the time plots reveal that the dynamics of the 6 items within each category share common patterns. This suggests the plausible existence of a latent factor structure underlying these dynamics.

The following remarks provide further evidence for the latent factor structure. Several estimated correlation matrices are depicted in Figure \ref{fig:Figure_covariance}. The top panel presents the sample autocorrelation matrices of the observed series $\{X_t\}_{t=1,\ldots,85}$, while the bottom panel shows the estimated autocorrelation matrices of the latent series $\{Z_t\}_{t=1,\ldots,85}$. In both panels, the left plot corresponds to lag 0, and the right plot corresponds to lag 1. One can see that both autocorrelation matrices for the same lag order are nearly indistinguishable. At lag 0, the plots in both panels show clear block patterns characteristic of the factor structure. Furthermore, the factor structure appears to be preserved through temporal dependence, as suggested by the plots of the sample ACFs at lag 1, though it is less discernible.

We work with the model assuming $r=5$ and $p=1$ for the sake of illustration. We also assume categorical distributions on $\{1,2,3,4,5\}$ as marginal distributions. For the loadings matrices, we use the second identifiability condition \eqref{e:identifiability_cond}. The left panel of Figure \ref{fig:Figure_param} presents the estimated loadings matrix, while the right panel depicts the transition matrix of the factor series, estimated using the method described in Section \ref{se:estimation}. This pattern is consistent with what we observe in Figure \ref{fig:Figure_covariance}. Note that the transition matrix has relatively large values not only on the diagonal but also for some off-diagonal entries. The large off-diagonal values indicate that there are cross-correlations across the factor series.

For the model diagnostics, Figure \ref{fig:Figure_pit} shows the PIT histograms for 30 individual items, following the procedure in Section \ref{sse:diagnostics}. With the exception of a few items, the heights of the PIT histograms across 10 bins tend to be uniformly equal, aligning with the 0.1 relative frequencies indicated by the black dashed lines.

With the estimated model, we forecast the next 5 steps as discussed in Remark \ref{rem:marginal_dist} and compare the values with the true ones held out of the sample. For comparison, we consider two simple forecasting approaches, Last and Marginal, as described in Section \ref{sse:forecasting}.

Figure \ref{fig:Figure_particle} presents the generated particles $\{\widehat{Z}_t^{(k)}\}_{t=1,\ldots,T}$ for 30 items within the last 5 observations $t=81,\ldots,85$ and the predicted values of latent process $\{\widehat{Z}_{T+h|T}^{(k)}\}_{h=1,\ldots,5}$ for the next 5 forecasting steps, respectively. The items are grouped into five underlying categories. The three or four parallel lines in Figure \ref{fig:Figure_particle} represent the thresholds, with each pair of lines forming a bin. To distinguish the values, we use different line types for each threshold. As explained above, the particles are generated at each time point to belong to a specific bin that corresponds to the discrete observation for that dimension. This explains why the particles stay within bins at all time points in the left panel. For some of the items, a thresholding line extends beyond the given vertical scale due to the way the bins are defined. Since each bin is estimated through the observations, some values do not appear if they are not realized in the observation period. These values are also excluded from the candidate forecasting values. On the other hand, since no further observations are assumed to be given after the observation period, the particles are generated by forecasting the latent process. As a result, the particles in the right panel do not need to stay in the same bin. Note that all of the particles seem to converge to zero for the longer forecasting horizon. As discussed in Section \ref{sse:convergence}, this behavior is expected when a stable VAR is used for forecasting. In addition, the particles are rather close to zero even for the first few horizons. This is a consequence of the interplay between the levels of signal (factors) and noise (errors) in the estimated model. The forecasted value naturally approaches 0 when forecasting noise, which in turn downweights the factor forecast as the noise magnitude increases, given that the variance of our latent process is 1 in each dimension.

Finally, Figure \ref{fig:Figure_forecast} shows the plots of the absolute differences between forecasts and the true values for each item. The items are arranged according to their categories, under the expectation that each factor primarily influences its corresponding category. Overall, the proposed forecasting approach slightly outperforms the reference methods, generally exhibiting small absolute differences. As discussed in Section \ref{sse:convergence}, the forecasting performance becomes identical to the Marginal for the longer horizon. But for shorter horizons, our approach performs better for 4 items, whereas the Marginal method performs better for 2 items. While this suggests a potential advantage, it is not sufficient to draw overarching conclusions based on this evidence.

\section{Conclusion}
\label{se:conclusion}

In this work, we considered a multivariate discrete-valued times series model, wherein component count series are obtained by binning the continuous values of latent Gaussian dynamic factor processes. We introduced an estimation method based on second-order properties of the count and latent processes, and PCA. We also suggested additional model selection approaches for determining the number of factor series and their lag orders through cross-validation. We provided the theoretical guarantees of the estimators by applying available concentration results for the considered model with general latent Gaussian processes. Facilitated by the state-space formulation of our model, we employed a sequential Monte Carlo method with resampling, for forecasting. Our estimation and forecasting methods were examined on simulated data and an empirical example. The R code used for the illustrations in Section \ref{sse:estimation-unknown}, the simulation study of Sections \ref{se:simulation}, and the data analysis of Section \ref{se:application} are available on GitHub at \href{https://github.com/yk748/latent_Gaussian_TS}{https://github.com/yk748/latent\_Gaussian\_TS}.

While our study advances a framework for latent Gaussian time series modeling of categorical observations collected over time, important questions remain. The question of how to employ time-varying covariates is discussed in Supplemental Material but left for future work. There are also potential improvements to make in terms of accuracy and computing time of our forecasting methods. Instead of employing standard particle filtering strategies, one could try other variants of sequential Monte Carlo sampling, for example, ensemble Kalman filtering in a high-dimensional regime \cite[e.g.,][]{katzfuss2020ensemble}.

\section*{Acknowledgment}

MD was supported partially by the FAU Emerging Talents Initiative. VP was supported partially by the grants NSF DMS-1712966, DMS-2113662, and DMS-2134107. The authors also thank two anonymous Reviewers for their many helpful comments.

\appendix

\section{Proofs}\label{ap:consistency}

\subsection{Proofs of results in Section \ref{se:main_results}}

An integral part of the proofs is Proposition 3.1 in \cite{duker2024high}. For completeness, we rephrase its statement here. 
Introduce the set $\mathcal{K}(2s) = \{ v \in \RR^{\Dim \Lag} : \| v \|_2 \leq 1, \| v \|_{0} \leq 2s \}$ with the Euclidean norm $\| v \|_2 = \left( \sum_{j =1}^{\Dim} |v_{j}|^2 \right)^{1/2}$, and the norm $\| v \|_{0}$ that counts all non-zero elements for a vector $v \in \RR^{\Dim}$. Define the mapping
\begin{align} 
\label{eq:matrixmapping}
A \mapsto \vertiii{ A }_{s} := \sup_{v \in \mathcal{K}(2s)} | v' A v|.
\end{align}
Besides submultiplicativity, the mapping in \eqref{eq:matrixmapping} satisfies all properties of a matrix norm. Set further $
\bmS = (\Sigma_X(r-s))_{r,s=1,\dots,\VO}$.
Under Assumptions \ref{ass:L1}--\ref{ass:L4}, for any $\delta >0 $, 
\begin{align} \label{eq:DLP_rephrased}
&
\Prob\left[
\vertiii{  \widehat{\ell}^{-1} ( \bmhS ) - \ell^{-1}(\bmS) }_s > Q(\bm{R}_Z) \delta
\right]
\nonumber
\\&\leq
\Prob\left[
\vertiii{ \bmhS - \bmS }_s > \delta \wedge \sqrt{\delta}
\right] + 
\Prob\left[ \| \widehat{\theta} - \E \widehat{\theta} \|_{\max} > \delta \wedge \sqrt{\delta} \right]
\end{align}
with a constant $Q(\bmR) := Q(\bmR, \delta)$ that depends on the quantities
\begin{equation} 
\label{eq:momentlikemk}
m^{(k)}_{i}(u) = \frac{1}{\sqrt{2\pi }} \sum_{n = 0}^{\infty} \exp\left(-\frac{1}{2 u} Q^{2}_{i,n} \right) | Q_{i,n} |^{k}, \hspace{0.2cm} k=0,2,3,
\end{equation}
and 
\begin{equation} 
\label{eq:momentlikemkderiv}
\mu_{i}^{(k)}(u) = 
\frac{1}{\sqrt{2\pi}} \sum_{n=0}^{\infty} \exp\left(-\frac{1}{2u} Q^{2}_{i,n} \right) |Q_{i,n}|^k 
\| \nabla_{\theta_{i}} Q_{i,n} \|_{1}, \hspace{0.2cm} k=0,3.
\end{equation}
The constant $Q(\bm{R}_Z)$ is defined precisely in Appendix A.2 in \cite{duker2024high}. Strictly speaking, the inequality in \eqref{eq:DLP_rephrased} is up to an absolute constant, which we do not write for notational simplicity.

\begin{proof}[Proof of Proposition \ref{prop1}]
Note that
\begin{align}
\| \bmhR - \bmL \bm{\Sigma}_Y \bmL' \|
&=
\| \bm{\widehat{R}}_Z - \bm{R}_Z + \bm{R}_Z - \bm{\Lambda} \bm{\Sigma}_Y \bm{\Lambda}' \|
\nonumber
\\&\leq
\| \widehat{\ell}^{-1} ( \bmhS ) - \ell^{-1}(\bmS) \| + \| \bmR - \bmL \bmSY \bmL' \|.
\label{eq:maineq1}
\end{align}
We consider the two terms in \eqref{eq:maineq1} separately. The second one satisfies
\begin{equation} \label{eq:rate_population}
\frac{1}{d} \| \bmR - \bmL \bmSY \bmL' \|
=
\frac{1}{d} \| \bmL \bmSY \bmL'+ \bm{\Sigma}_{\varepsilon} - \bmL \bmSY \bmL' \|
=
\frac{1}{d} \| \bm{\Sigma}_{\varepsilon} \| = \mathcal{O}(1/d)
\end{equation}
with $\bm{\Lambda} = I_p \otimes \Lambda$, $\bm{\Sigma}_{\varepsilon} = I_p \otimes \Sigma_{\varepsilon}$, by Assumption \ref{ass:2}.
For the first term in \eqref{eq:maineq1}, recall the constant $Q(\bm{R}_Z)$ from \eqref{eq:DLP_rephrased}.  
We further set
\begin{equation} \label{eq:delta_deltatilde}
    \delta_{dT} = \sqrt{\frac{\log(dT)}{T}}.
\end{equation}
Then, for all $\epsilon > 0$, there is $M>0$ such that
\begin{align}
&
\Prob\left[ \frac{1}{d}
\| \widehat{\ell}^{-1} ( \bmhS) - \ell^{-1}(\bmS) \| > 3pQ(\bmR) \delta_{dT} M
\right]
\nonumber
\\&\leq
\Prob\left[ 
\| \widehat{\ell}^{-1} ( \bmhS) - \ell^{-1}(\bmS) \|_{\max} > 3Q(\bmR) \delta_{dT} M
\right]
\label{al:main_00}
\\&\leq
\Prob\left[
\vertiii{ \widehat{\ell}^{-1} ( \bmhS) - \ell^{-1}(\bmS) }_{1} > Q(\bmR) \delta_{dT} M
\right]
\label{al:main_0}
\\&\leq
\Prob\left[
\vertiii{ \bmhS - \bmS }_1 > 
M \delta_{dT} \wedge \sqrt{M \delta_{dT} }
\right] + 
\Prob\left[ \| \widehat{\theta} - \E \widehat{\theta} \|_{\max} > M \delta_{dT} \wedge \sqrt{M \delta_{dT} } \right]
\label{al:main_1}
\\&\leq
\Prob\left[
\vertiii{ \bmhS - \bmS }_1 > 
(M \wedge \sqrt{M}) \delta_{dT}
\right] + 
\Prob\left[ \| \widehat{\theta} - \E \widehat{\theta} \|_{\max} > (M \wedge \sqrt{M}) \delta_{dT} \right] < \epsilon,
\label{al:main_2}
\end{align}
where \eqref{al:main_00} follows since $\| A \| \leq q \| A \|_{\max}$ for matrices $A \in \RR^{q \times q}$. Furthermore, \eqref{al:main_0} is due to the inequality 
$2|v' A w| \leq |v'Av| + |w'Aw| + |(v+w)' A (v+w)|$ and $A_{ij} = e_{q,i}' A e_{q,j}$ for $A = (A_{ij})_{i,j=1,\dots,q}$ with $e_{q,i}$ denoting the $i$th unit vector in $\RR^q$. 
Then, \eqref{al:main_1} follows from \eqref{eq:DLP_rephrased} with $s=1$. We can use the stochastic boundedness stated in Lemma \ref{le:le2} below and Assumption \ref{Ass:C1} for \eqref{al:main_2}.
\end{proof}

\begin{remark} \label{re}
    We pause here to comment on the proof of Proposition \ref{prop1} and the bound \eqref{al:main_00} in particular. Note that for the second summand in \eqref{eq:maineq1}, i.e., on the population level, it is crucial to consider the spectral norm to get \eqref{eq:rate_population}. However, for the first summand in \eqref{eq:maineq1}, we use that $\| A \| \leq q \| A \|_{\max}$ for matrices $A \in \RR^{q \times q}$ and argue with bounds on the maximum norm. Similar arguments but mostly based on Euclidean distances are
    used by \cite{doz2011two} when the series following a factor model is observed. Our use of the maximum norm results in the additional factor $\sqrt{\log(dT)}$ compared to \cite{doz2011two}. 
\end{remark}

\begin{proof}[Proof of Proposition \ref{prop2}]
We note first that there are a number of results which can be inferred from Proposition \ref{prop1}. 
Proposition \ref{prop1} is an analogue of Lemma 2(i) in \cite{doz2011two}. Following \cite{doz2011two}, one can infer the following asymptotics:

\begin{enumerate}[label=\textbf{DGR.\arabic*.},ref=DGR.\arabic*, align=left]
\item \label{item:Doz1}
$\widehat{U}_r^{'} U_r - I_r = \mathcal{O}_{\prob} \mleft( \delta_{dT} \mright) + \mathcal{O}_{\prob}\mleft( \frac{1}{d} \mright)$ by Lemma 4(i) in \cite{doz2011two}.
\item \label{item:Doz3}
$d \|\widehat{E}_{r}^{-1} - E_r^{-1} \| = 
\mathcal{O}_{\prob} \mleft( \delta_{dT} \mright) + \mathcal{O}_{\prob}\mleft( \frac{1}{d} \mright)$
 by Lemma 2(iii) in \cite{doz2011two}.
\item \label{item:Doz4}
$E_r \widehat{E}_{r}^{-1}-I_{r} = \mathcal{O}_{\prob} \mleft( \delta_{dT} \mright) + \mathcal{O}_{\prob}\mleft( \frac{1}{d} \mright)$ by Lemma 2(iv) in \cite{doz2011two}.
\end{enumerate}

Starting with the actual proof of Proposition \ref{prop2}, we follow the proof of Proposition 3 in \cite{doz2011two}. That is, in view of \eqref{eq:estimatorSigmaY},
\begin{align}
&
( I_p \otimes \widehat{E}^{-1/2}_r \widehat{U}_r' ) \bm{\widehat{R}}_Z (I_p \otimes \widehat{U}_r \widehat{E}_r^{-1/2} )
\nonumber
\\&=
( I_p \otimes \widehat{E}^{-1/2}_r \widehat{U}_r' ) 
\bm{\Lambda} \bm{\Sigma}_{Y} \bm{\Lambda}'
(I_p \otimes \widehat{U}_r \widehat{E}_r^{-1/2} ) 
+
( I_p \otimes \widehat{E}^{-1/2}_r \widehat{U}_r' ) ( \bm{\widehat{R}}_Z - 
\bm{\Lambda} \bm{\Sigma}_{Y} \bm{\Lambda}') (I_p \otimes \widehat{U}_r \widehat{E}_r^{-1/2} ).
\label{al:proofDozProp3al1}
\end{align}
We consider the two summands in \eqref{al:proofDozProp3al1} separately. The first summand in \eqref{al:proofDozProp3al1} satisfies 
\begin{align}
&
( I_p \otimes \widehat{E}^{-1/2}_r \widehat{U}_r' ) 
\bm{\Lambda} \bm{\Sigma}_{Y} \bm{\Lambda}'
(I_p \otimes \widehat{U}_r \widehat{E}_r^{-1/2} ) \nonumber
\\&=
( I_p \otimes \widehat{E}^{-1/2}_r \widehat{U}_r' ) 
\bm{\Lambda} \bm{Q} \bm{Q}' \bm{\Sigma}_{Y} \bm{Q} \bm{Q}' \bm{\Lambda}'
(I_p \otimes \widehat{U}_r \widehat{E}_r^{-1/2} )
\nonumber
\\&=
( I_p \otimes \widehat{E}^{-1/2}_r \widehat{U}_r' ) 
\bm{U} \bm{E}^{1/2} \bm{Q}' \bm{\Sigma}_{Y} \bm{Q} \bm{E}^{1/2} \bm{U}'
(I_p \otimes \widehat{U}_r \widehat{E}_r^{-1/2} )
\label{al:proofDozProp3al2}
\\&= 
\bm{Q}' \bm{\Sigma}_{Y} \bm{Q} + \mathcal{O}_{\prob} \mleft( \delta_{dT} \mright) + \mathcal{O}_{\prob}\mleft( \frac{1}{d} \mright),
\label{al:proofDozProp3al3}
\end{align}
where \eqref{al:proofDozProp3al2} follows by the relation $\Lambda Q_r = U_r E_r^{1/2}$ in \eqref{eq:other_relations} and $\bm{U} = I_p \otimes U_r$, $\bm{E} = I_p \otimes E_r$, \eqref{al:proofDozProp3al3} is due to 
\ref{item:Doz1} and \ref{item:Doz4} as well as $\bm{Q}' \bm{\Sigma}_{Y} \bm{Q} = \mathcal{O}(1)$.

For the second summand in \eqref{al:proofDozProp3al1} note that
\begin{align}
&   
    \|
    ( I_p \otimes \widehat{E}^{-1/2}_r \widehat{U}_r' ) ( \bm{\widehat{R}}_Z - 
    \bm{\Lambda} \bm{\Sigma}_{Y} \bm{\Lambda}') (I_p \otimes \widehat{U}_r \widehat{E}_r^{-1/2} ) 
    \|
    \nonumber
    \\&\leq
    \|( I_p \otimes \widehat{E}^{-1/2}_r \widehat{U}_r' ) \|^2 
    \| \bm{\widehat{R}}_Z - \bm{\Lambda} \bm{\Sigma}_{Y} \bm{\Lambda}' \|
    \label{al:proofDozProp3al4}
    \\&\leq
    \|(\widehat{E}_r/d)^{-1/2} \widehat{U}_r' \|^2 
    \frac{1}{d} \| \bm{\widehat{R}}_Z - \bm{\Lambda} \bm{\Sigma}_{Y} \bm{\Lambda}' \|
    \label{al:proofDozProp3al5}
    \\&= 
    \mathcal{O}_{\prob}\mleft( \delta_{dT} \mright) + \mathcal{O}\left(\frac{1}{d}\right)
    \label{al:proofDozProp3al5.1},
\end{align}
where \eqref{al:proofDozProp3al4} is due to submultiplicativity of the spectral norm, and \eqref{al:proofDozProp3al5} follows by the property of the eigenvalues of Kronecker products as stated in Theorem 1 (Section 3) in \cite{magnus2007matrix}. Finally, $\| \widehat{U}_r\|=1$, \ref{item:Doz3} and Proposition \ref{prop1} give \eqref{al:proofDozProp3al5.1}.
\end{proof}

\begin{proof}[Proof of Corollary \ref{cor1}]
Recall 
$\bm{\Sigma}_Y^{(p)} := (\Sigma_Y(r-s))_{r,s=1,\dots,\VO}$, $\Sigma_{Y}(h) = \E( Y_{t+h} Y_t')$ such that $\bm{\Sigma}_{QY}^{(p)} := (Q\Sigma_Y(r-s)Q')_{r,s=1,\dots,\VO} = \bm{Q} \bm{\Sigma}_{Y}^{(p)} \bm{Q}'$ and $\bm{\TM} = S_2 \bm{\Sigma}_{QY}^{(p+1)} S'_1(\bm{\Sigma}_{QY}^{(p)})^{-1}$. Then, 
\begin{align}
&
\| \widehat{\bm{\TM}} - \bm{\TM} \|
\nonumber
\\&=
\| S_2 \bm{\widehat{\Sigma}}_Y^{(p+1)} S'_1(\bm{\widehat{\Sigma}}_Y^{(p)})^{-1} - S_2 \bm{\Sigma}_{QY}^{(p+1)} S'_1(\bm{\Sigma}_{QY}^{(p)})^{-1} \|
\nonumber
\\&\leq
\| S_2 \| 
\| \bm{\widehat{\Sigma}}_Y^{(p+1)} S'_1(\bm{\widehat{\Sigma}}_Y^{(p)})^{-1} - \bm{\Sigma}_{QY}^{(p+1)} S'_1(\bm{\Sigma}_{QY}^{(p)})^{-1} \|
\label{al:proofDozProp3al6}
\\&\leq
\| \bm{\widehat{\Sigma}}_Y^{(p+1)} \|
\| (\bm{\widehat{\Sigma}}_Y^{(p)})^{-1} - (\bm{\Sigma}_{QY}^{(p)})^{-1} \|
+
\| \bm{\widehat{\Sigma}}_Y^{(p+1)} - \bm{\Sigma}_{QY}^{(p+1)} \| \| (\bm{\Sigma}_{QY}^{(p)})^{-1} \|
\label{al:proofDozProp3al7}
\\&\leq
\| \bm{\widehat{\Sigma}}_Y^{(p+1)} \|
\| (\bm{\widehat{\Sigma}}_Y^{(p)})^{-1} \|
\| \bm{\widehat{\Sigma}}_Y^{(p)} - \bm{\Sigma}_{QY}^{(p)} \|
\| (\bm{\Sigma}_{QY}^{(p)})^{-1} \|
+
\| \bm{\widehat{\Sigma}}_Y^{(p+1)} - \bm{\Sigma}_{QY}^{(p+1)} \| \| (\bm{\Sigma}_{QY}^{(p)})^{-1} \|
\label{al:proofDozProp3al8}
\\&= 
\mathcal{O}_{\prob} \mleft( \delta_{dT} \mright) + \mathcal{O}_{\prob}\mleft( \frac{1}{d} \mright),
\label{al:proofDozProp3al9}
\end{align}
where \eqref{al:proofDozProp3al6} is due to submultiplicativity, \eqref{al:proofDozProp3al7} can be inferred from $\| S_1 \| = \| S_2 \| = \| (0_{p}, I_{p})\| = 1 $ and \eqref{al:proofDozProp3al8} follows by $B^{-1} - A^{-1} = B^{-1}(A - B) A^{-1}$. 
For \eqref{al:proofDozProp3al9}, we apply Proposition \ref{prop2}, and also observe that
\begin{align*}
\| (\bm{\widehat{\Sigma}}_Y^{(p)})^{-1} \|
&\leq 
| \lambda_{\min}(\bm{\widehat{\Sigma}}_Y^{(p)}) - \lambda_{\min}(\bm{\Sigma}_{QY}^{(p)}) | + \lambda_{\min}(\bm{\Sigma}_{QY}^{(p)})
\\&\leq
\| \bm{\widehat{\Sigma}}_Y^{(p)} - \bm{\Sigma}_{QY}^{(p)} \| + \lambda_{\min}(\bm{\Sigma}_{QY}^{(p)})
\\&= \mathcal{O}_{\prob} \mleft( \delta_{dT} \mright) + \mathcal{O}_{\prob}\mleft( \frac{1}{d} \mright) + \lambda_{\min}(\bm{\Sigma}_{QY}^{(p)})
= \mathcal{O}_{\prob}(1)
\end{align*}
due to Weyl's Theorem (Theorem 4.3.1 in \cite{horn2012matrix}) and Proposition \ref{prop2}.
\end{proof}

\begin{proof}[Proof of Lemma \ref{lem1}]
Recall from \eqref{eq:Lambda_hat} that $\widehat{\Lambda} = \widehat{U}_{r}\widehat{E}_{r}^{1/2}$.
Then, with $\widehat{U}_{r} = \widehat{R}_Z(0) \widehat{U}_{r}\widehat{E}_{r}^{-1}$, we get
\begin{align} \label{eq:Doz_calcs200_le2}
    (\widehat{\Lambda} - \Lambda)_i
    &=
    e_{d,i}'(\widehat{R}_{Z}(0) - R_Z(0)) \widehat{U}_{r}\widehat{E}_{r}^{-1/2} + 
    e_{d,i}'\Sigma_{\varepsilon}\widehat{U}_{r}
    \widehat{E}_{r}^{-1/2}
    \nonumber
    \\&\hspace{2cm}+ 
    e_{d,i}'U_r E_r (U_r'\widehat{U}_{r} - E^{-1/2}_r \widehat{E}_{r}^{1/2} )\widehat{E}_{r}^{-1/2} 
\end{align}
with $e_{d,i}$ denoting the $i$th unit vector in $\RR^d$, $R_Z(0) = \Lambda \Sigma_Y(0) \Lambda'+ \Sigma_{\varepsilon} = \Lambda \Lambda'+ \Sigma_{\varepsilon}$ and 
by the same calculation as on p.\ 200 in \cite{doz2011two}. We consider the three summands in \eqref{eq:Doz_calcs200_le2} separately. 

For the first summand, note that
\begin{equation}
   \E \| e_{d,i}'(\widehat{R}_{Z}(0) - R_Z(0)) \|^2 = \mathcal{O} \mleft( d \delta^2_{dT} \mright),
\end{equation}
which can be shown following the same arguments as in the proof of Proposition \ref{prop1}. In particular, we get a $d \delta^2_{dT}$-rate instead of $d^2 \delta^2_{dT}$ as may be expected based on Proposition \ref{prop1} since we only consider the $i$th row of the $d \times d$ matrix $\widehat{R}_{Z}(0) - R_Z(0)$.

The second summand in \eqref{eq:Doz_calcs200_le2} satisfies
\begin{equation}
   \| e_{d,i}'\Sigma_{\varepsilon}\widehat{U}_{r} \widehat{E}_{r}^{-1/2} \| 
   \leq
   \| \Sigma_{\varepsilon} \| \| \widehat{U}_{r} \| \| \widehat{E}_{r}^{-1/2} \|
   = \mathcal{O}_{\prob}\left(\frac{1}{\sqrt{d}}\right)
\end{equation}
since $\| e_{d,i}'\Sigma_{\varepsilon} \| \leq \| \Sigma_{\varepsilon} \| = \mathcal{O}(1)$ by Assumption \ref{ass:2}, $\| \widehat{U}_{r} \| = 1$ and $\| \widehat{E}_{r}^{-1/2} \| = \mathcal{O}_{\prob}\left(\frac{1}{\sqrt{d}}\right)$ by \ref{item:Doz1} and \ref{item:Doz3}, respectively.

Finally, for the third summand in \eqref{eq:Doz_calcs200_le2},
\begin{align}
   \| e_{d,i}'U_r E_r (U_r'\widehat{U}_{r} - E^{-1/2}_r \widehat{E}_{r}^{1/2} )\widehat{E}_{r}^{-1/2} \| 
   &=
   \| e_{d,i}'\Lambda Q_r E_r^{1/2} (U_r'\widehat{U}_{r} - E^{-1/2}_r \widehat{E}_{r}^{1/2} )\widehat{E}_{r}^{-1/2} \|
   \label{al:proofDozProp3al10}
    \\&\leq 
   \| e_{d,i}'\Lambda Q_r E_r^{1/2} \|
   \| U_r'\widehat{U}_{r} - E^{-1/2}_r \widehat{E}_{r}^{1/2} \| \| \widehat{E}_{r}^{-1/2} \|
   \nonumber
    \\&=
   \mathcal{O}_{\prob} \mleft( \delta_{dT} \mright) + \mathcal{O}_{\prob}\mleft( \frac{1}{d} \mright),
   \label{al:proofDozProp3al12}
\end{align}
where for \eqref{al:proofDozProp3al10}, we used the representation $U_r = \Lambda Q_r E_r^{-1/2}$ from
\eqref{eq:other_relations}. Finally, \eqref{al:proofDozProp3al12} is due to \ref{item:Doz1}, \ref{item:Doz3} and
    $\| \widehat{E}_{r}^{-1/2} \| = \mathcal{O}_{\prob}\left(\frac{1}{\sqrt{d}}\right)$ by \ref{item:Doz3}. Furthermore, note that $\| e_{d,i}'\Lambda Q_r E_r^{1/2} \| \leq \| e_{d,i}'\Lambda \| \| Q_r \| \| E_r^{1/2} \| = \mathcal{O}(\sqrt{d})$.
\end{proof}

\subsection{Auxiliary result and its proof} \label{se:auxandproof}

In the following, we aim to illustrate that the assumptions made in Section \ref{se:assumptions} are satisfied in quite general settings. The next auxiliary result shows that it suffices to impose Assumption \ref{Ass:C2} on the maximum distance between the sample and true autocovariances of the observed series. For simplicity, We replace the sample mean by its population counterpart in $\bmhS$, i.e., consider the centered random variables $\widetilde{X}_{t} = X_{t} - \E X_{t}$, so that
\begin{equation} \label{eq:unbiased_estimator}
\begin{aligned}
    \bmhS 
    &= \left( \frac{1}{T-p} \sum_{t=p+1}^{T} \widetilde{X}_{t-i} \widetilde{X}_{t-j}' \right)_{i,j =1,\dots, p}   
    \\&=
    \left( \frac{1}{T-p} \sum_{t=p+1}^{T} \left( G(Z_{t-i} ) - \E G(Z_{t-i} ) \right)
    \left( G( Z_{t-j} ) - \E G( Z_{t-j} ) \right)' \right)_{i,j =1,\dots, p},
\end{aligned}
\end{equation}
which is an unbiased estimator for $\bmS$, i.e., 
\begin{equation} \label{al:qwqwqw1.3}
\begin{aligned}
\E \bmhS
&=
\left(\frac{1}{T-p} \E \sum_{t=p+1}^{T} \widetilde{X}_{t-i} \widetilde{X}_{t-j}' \right)_{i,j =1,\dots, p}
\\&=
\left( \frac{1}{T-p} \sum_{t=p+1}^{T} ( \E ( G(Z_{t-i}) G(Z_{t-j})' ) - \E G(Z_{t-i}) \E G(Z_{t-j})' ) \right)_{i,j =1,\dots, p}
= \bmS.
\end{aligned}
\end{equation}
We also set 
$$
\widetilde{G}(z) = G(z)- \E G(Z_{t} ).
$$
When the population mean is estimated, one can argue for concentration as in the verification of Assumption \ref{Ass:C1} below. 
\begin{lemma} \label{le:le2}
Suppose Assumptions \ref{Ass:C2}, \ref{ass:2}--\ref{ass:7}. Then, 
\begin{equation*}
\vertiii{ \bmhS - \bmS }_1
=
\mathcal{O}_{\prob}\left( \delta_{dT} \right).
\end{equation*}
\end{lemma}

\begin{proof}
Recall the definition of $\vertiii{ \cdot }_s$ from \eqref{eq:matrixmapping}. For $v \in \RR^{dp}$, we set $v = \vecop( [v_{1} : \dots : v_{\VO} ])$ with $v_{r} \in \RR^{\Dim}$ and $N=T-p$. Then, by \eqref{eq:unbiased_estimator} and \eqref{al:qwqwqw1.3},
\begin{align}
&
\vertiii{ \bmhS - \bmS }_1
\nonumber
\\&=
\BIGvertiii{ \frac{1}{N} \sum_{t=p+1}^{T} \Big( \widetilde{G}(Z_{t-i} ) \widetilde{G}( Z_{t-j} )' - \E \widetilde{G}(Z_{t-i} ) \widetilde{G}( Z_{t-j} )' \Big)_{i,j=1,\dots,\VO} }_1
\nonumber
\\&=
\sup_{v \in \mathcal{K}(2)} \left| \frac{1}{N} \sum_{t=p+1}^{T} \sum_{i,j=1}^{p} v'_i \Big( \widetilde{G}(Z_{t-i} ) \widetilde{G}( Z_{t-j} )' - \E \widetilde{G}(Z_{t-i} ) \widetilde{G}( Z_{t-j} )' \Big) v_j \right|
\nonumber
\\&\leq
\sup_{v \in \mathcal{K}(2)} \sum_{i,j=1}^{p} |v'_i| \left\| \frac{1}{N} \sum_{t=p+1}^{T} \Big( \widetilde{G}(Z_{t-i} ) \widetilde{G}( Z_{t-j} )' - \E \widetilde{G}(Z_{t-i} ) \widetilde{G}( Z_{t-j} )' \Big) \right\|_{\max} | v_j |
\nonumber
\\&\leq
\max_{i,j=1,\dots,p} \max_{r,s=1,\dots,d} \Big| \frac{1}{N} \sum_{t=p+1}^{T} \Big( \widetilde{G}_{r}( Z_{r,t-i}  ) \widetilde{G}_{s}( Z_{s,t-j}  ) 
\nonumber
\\& \hspace{1cm}  - 
 \E ( \widetilde{G}_{r}( Z_{r,t-i}  ) \widetilde{G}_{s}( Z_{s,t-j}  ) ) \Big) \Big| \sup_{v \in \mathcal{K}(2)} \sum_{i,j=1}^{p} \sum_{r,s=1}^{d} |v_{i,r} | |  v_{j,s} |
 \nonumber
 \\&=
\max_{i,j=1,\dots,p} \max_{r,s=1,\dots,d} \Big| \frac{1}{N} \sum_{t=p+1}^{T} \Big( \widetilde{G}_{r}( Z_{r,t-i}  ) \widetilde{G}_{s}( Z_{s,t-j} ) 
 - 
 \E ( \widetilde{G}_{r}( Z_{r,t-i}  ) \widetilde{G}_{s}( Z_{s,t-j}  ) ) \Big) \Big|
 \label{eq:HDLFMeq6}
 \\&=
 \| \bmhS - \bmS \|_{\max},
  \nonumber
 \end{align}
 where \eqref{eq:HDLFMeq6} can be inferred since $\mathcal{K}(2) = \{ v \in \RR^{\Dim \Lag} : \| v \|_2 \leq 1, \| v \|_{0} \leq 2 \}$.
 The result then follows by Assumption \ref{Ass:C2}.
\end{proof}

\subsection{Proof of Lemma \ref{le:justifyAss}} \label{se:proofs_verif_ass}
We aim here to rewrite the latent factor model as a Markov chain. 
We then use the following concentration inequality for bounded functions of general-state-space Markov chains derived in \cite{fan2021hoeffding}. The result is expressed in terms of $\lambda_{\rig} := \lambda_{\rig}(P)$, the rightmost value of the spectrum $[-\lambda, \lambda]$ of the corresponding Markov operator $P$. We refer to $1-\lambda_{\rig}$ as the right spectral gap of the Markov chain.

\begin{theorem}[Theorem 3 in \cite{fan2021hoeffding}] \label{theorem:CIMarkovchain}
Let $\{ Y_{t} \}_{t \geq 1}$ be a Markov chain on $\mathcal{X}$ with Markov operator $P$ and right spectral gap $1-\lambda_{\rig} >0$. For any $\varepsilon > 0$ and bounded function $f: \mathcal{X} \to [a,b]$, 
\begin{equation*}
\Prob \left[ \left| \frac{1}{T} \sum_{t =1}^{T} f(Y_{t}) - \frac{1}{T} \sum_{t =1}^{T} \E[f(Y_{t})] \right| > \varepsilon \right]
\leq
2\exp\left(- \frac{1-\max\{0,\lambda_{\rig}\}}{1+\max\{0,\lambda_{\rig}\}} \frac{T \varepsilon^{2}}{ (b-a)^{2}/2} \right).
\end{equation*}
\end{theorem}

For the remainder of the section we assume without loss of generality that $\frac{1}{d} \Lambda' \Lambda =  I_r$ such that Assumptions \ref{ass:4}--\ref{ass:7} are satisfied. Relation \eqref{e:dynamic_factor} then implies 
$Y_t = \frac{1}{d} \Lambda'Z_t - \frac{1}{d} \Lambda' \varepsilon_t$, and hence by \eqref{e:VAR}, 
\begin{equation} \label{eq:VARMA_rep}
    Z_t = \Lambda Y_t + \varepsilon_t
    =
    \sum_{i=1}^{p} \Lambda \TM_i Y_{t-i} 
    + \Lambda \eta_t + \varepsilon_t
    =
    \sum_{i=1}^{p} \Phi_i Z_{t-i} + \Lambda \eta_t + \varepsilon_t - \sum_{i=1}^{p} \Phi_i \varepsilon_{t-i}
\end{equation}
with
\begin{equation*}
    \Phi_i = \frac{1}{d} \Lambda \TM_i \Lambda'.
\end{equation*}
Similar arguments are used in \cite{bhamidi2023dynamic}, who show additionally that $\{Z_t\}$ is a VARMA($p,p$) series.
In view of the representation \eqref{eq:VARMA_rep}, we write
\begin{equation} \label{eq:VAR_rep_VARMA}
    \begin{pmatrix}
        \mathcal{Z}_{t} \\
        \mathcal{E}_{t}
    \end{pmatrix}
    =
    \begin{pmatrix}
        \bm{\Phi} & \widetilde{\bm{\Phi}} \\
        0_{dp \times dp} & 
        \begin{matrix}
            0_{d \times d(p-1)} & 0_{d \times d} \\
            I_{d(p-1)} & 0_{d(p-1) \times d}
        \end{matrix}
    \end{pmatrix}
    \begin{pmatrix}
        \mathcal{Z}_{t-1} \\
        \mathcal{E}_{t-1}
    \end{pmatrix}  
    +
    \begin{pmatrix}
        \Lambda \eta_t + \varepsilon_t\\
        0_{d(p-1)} \\
        \varepsilon_{t} \\
        0_{d(p-1)}
    \end{pmatrix}   
    \hspace{0.2cm}
    \text{ or }
    \hspace{0.2cm}
    \mathcal{V}_{t} = \bm{\TM}_{\mathcal{V}} \mathcal{V}_{t-1} + \xi_{t}
\end{equation}
with 
\begin{equation*}
    \mathcal{Z}_{t} = (Z_t', \dots, Z_{t-p+1}')',
    \hspace{0.2cm}
    \mathcal{E}_{t} = (\varepsilon_t', \dots, \varepsilon_{t-p+1}')'
\end{equation*}
and 
\begin{equation*}
    \bm{\Phi} =
    \begin{pmatrix}
    \Phi_{1}	&	\cdots	& \Phi_{\VO-1}	& \Phi_{\VO} \\
    I_{d}		&	\cdots	& 0			&	0 \\
    \vdots	&	\ddots	&	\vdots	&	\vdots \\
    0		&	\cdots	&	I_{d}		&	0\\
    \end{pmatrix},
    \hspace{0.2cm}
    \widetilde{\bm{\Phi}} =
    \begin{pmatrix}
    -\Phi_{1}	&	\cdots	& -\Phi_{\VO} \\
    0_{d(p-1) \times d}&	\cdots	&	0_{d(p-1) \times d} \\
    \end{pmatrix}.
\end{equation*}
The process $\{ \mathcal{V}_{t} \}$ in \eqref{eq:VAR_rep_VARMA} has a VAR($1$) representation. Note that the characteristic polynomial of its transition matrix $\bm{\TM}_{\mathcal{V}}$ satisfies 
\begin{equation*}
    p_{\bm{\TM}_{\mathcal{V}}}(\omega) := \det(\omega I_{2dp} - \bm{\TM}_{\mathcal{V}}) = 
    \det( \omega I_{dp} - \bm{\Phi})
\end{equation*}
by employing the Schur complement. Then, $\{ \mathcal{V}_{t} \}$ is stable due to Assumption \ref{ass:2}. Therefore, $\{ \mathcal{V}_{t} \}$ satisfies the Markov property and is also geometrically ergodic; see p.\ 944 in \cite{an1996geometrical}. Under geometric ergodicity, Theorem 2.1 in \cite{roberts1997geometric} implies that there is a spectral gap $\lambda$ with $1-\lambda>0$.
We start with the verification of Assumption \ref{Ass:C1}. 

\textit{Verification of Assumption \ref{Ass:C1}:}
First, note that by Assumption \ref{Ass:E1},
\begin{equation*}
\E[\widehat{\theta}_{i}]
= 
\E\left[ \frac{1}{T}
\sum_{t =1}^{T} X_{i,t} \right]
=
\theta_{i}.
\end{equation*}
Then, 
\begin{align}
\Prob[ \max_{i =1, \dots, \Dim} | \widehat{\theta}_{i} - \theta_{i} | > \varepsilon ]
&
=
\Prob[ \max_{i =1, \dots, \Dim} | \widehat{\theta}_{i} - \E[\widehat{\theta}_{i}] | > \varepsilon ]
\nonumber
\\&
\leq
\sum_{i=1}^{\Dim} \Prob[ | \widehat{\theta}_{i} - \E[\widehat{\theta}_{i}] | > \varepsilon ] \nonumber
\\&
\leq
\sum_{i=1}^{\Dim} \Prob\left[ \left| \frac{1}{T} \sum_{t =1}^{T} X_{i,t} - \E[ X_{i,t} ] \right| > \varepsilon \right] \nonumber
\\&
=
\sum_{i=1}^{\Dim} \Prob\left[ \left| \frac{1}{T} \sum_{t =1}^{T} f_i(\mathcal{V}_{t}) - \frac{1}{T} \sum_{t =1}^{T} \E[f_i(\mathcal{V}_{t})] \right| > \varepsilon \right] \label{al:CIthetaineq1}
\\&
\leq
2 \Dim
\exp\left(- \frac{1-\max\{0,\lambda_{\rig} \}}{1+\max\{0,\lambda_{\rig} \}} \frac{T \varepsilon^{2}}{ 2 b^2} \right), \label{al:CIthetaineq2}
\end{align}
where \eqref{al:CIthetaineq1} follows by choosing the function $f_i: \RR^{2dp} \to \RR$ in terms of $\mathcal{G}_1: \RR^{2dp} \to \RR^{dp}$ such that 
\begin{equation*}
    f_i(y) = e_{d \VO,i}'\mathcal{G}_1(y), \hspace{0.2cm}
    \mathcal{G}_1(y) = \vecop( [G(y_1): \dots: G(y_p)]),
\end{equation*}
where $y = \vecop([y_1: \cdots: y_p: y_{p+1}: \cdots : y_{2p}])$ with $y_{i} \in \RR^d$ for $i=1,\dots, 2p$ and $G$ as in \eqref{eq:subo}. 
Then, it remains to observe that $f_i$ is bounded which follows since $f_i(y) = e_{d \VO,i}' \mathcal{G}_1(y) \leq b e_{d \VO,i}' j_{\Dim \VO} = b$, where $j_{\Dim \VO}$ denotes a $\Dim \VO$-dimensional column vector with all entries equal to one and by Assumption \ref{Ass:E2}. The last inequality \eqref{al:CIthetaineq2} is a consequence of applying Theorem \ref{theorem:CIMarkovchain}.

Finally, we choose $\varepsilon = \delta_{dT} M$ for some $M >0$ in \eqref{al:CIthetaineq2}.
We can then get that for any $\epsilon>0$, there are finite $M>0$ and $K_0$ such that
\begin{align} 
&
    \Prob\left[ \| \widehat{\theta} - \E \widehat{\theta} \|_{\max} > \delta_{dT} M\right]
    \nonumber
    \\&\leq
    2 \Dim
    \exp\left(- \frac{1-\max\{0,\lambda_{\rig} \}}{1+\max\{0,\lambda_{\rig} \}} \frac{ \log(dT) }{ 2 b^2} \right) < \epsilon \label{eq:new1_1}
\end{align}
for all $d,T > K_0$, where we also need Assumption \ref{Ass:E3}.

\textit{Verification of Assumption \ref{Ass:C2}:}
Following the discussion in Section \ref{se:auxandproof}, we consider here the centered random variables $\widetilde{X}_{t} = X_{t} - \E X_{t}$ to estimate $\bmS$.
In particular, the estimator $\bmhS$ is then an unbiased estimator for $\bmS$ such that $\E \bmhS = \bmS$; see \eqref{eq:unbiased_estimator}.
We now aim to apply Theorem \ref{theorem:CIMarkovchain}. Recalling \eqref{eq:unbiased_estimator} and \eqref{al:qwqwqw1.3}, we write
\begin{align}
\Prob[ \| \bmhS - \bmS \|_{\max} > \delta ]
&
=
\Prob[ \| \bmhS - \E \bmhS \|_{\max} > \delta ] \nonumber
\\&
=
\Prob\left[ \max_{r,s=1,\dots,dp} \Big| e_{dp,r}' \Big( \bmhS - \E \bmhS \Big) e_{dp,s} \Big| > \delta \right] \nonumber
\\&
\leq
\sum_{r,s=1}^{dp}
\Prob\left[ \Big| \frac{1}{N} \sum_{t =p+1}^{T} f_{rs}(\mathcal{V}_{t}) - \frac{1}{N} \sum_{t =p+1}^{T} \E[f_{rs}(\mathcal{V}_{t})] \Big| > \delta \right]
\label{al:CIGammaXineq1}
\\&
\leq
2(dp)^2 \exp\left(- \frac{1-\max\{0,\lambda_{\rig} \}}{1+\max\{0,\lambda_{\rig}\}} \frac{N \delta^{2}}{ 8 b^4} \right). \label{al:CIGammaXineq2}
\end{align}
The function $f_{rs}$ in \eqref{al:CIGammaXineq1} is specified below and satisfies $|f_{rs}(y)| \leq b^2 4$. Finally, \eqref{al:CIGammaXineq2} is a consequence of applying Theorem \ref{theorem:CIMarkovchain}.

The function $f_{rs}: \RR^{2dp} \to \RR$ in \eqref{al:CIGammaXineq1} is given in terms of $\mathcal{G}_2: \RR^{2dp} \to \RR^{dp}$ with
\begin{equation*}
    f_{rs}(y) = e_{dp,r}' \mathcal{G}_2(y ) \mathcal{G}_2(y)' e_{dp,s}
\end{equation*}
and
\begin{equation*}
    \mathcal{G}_2(y) = \vecop([
    G(y_1) - \E G(Z_{1}), \dots, G(y_p) - \E G(Z_{1})])
    = \vecop([
    \widetilde{G}(y_1) , \dots, \widetilde{G}(y_p)]).
\end{equation*}
Then, it remains to verify that $f_{rs}$ is bounded. Let $J_{\Dim}$ be a $\Dim \times \Dim$-matrix with all entries equal to one and $j_{\Dim}$ be a $\Dim$-dimensional column vector with all entries equal to one. Then, with explanations given below,
\begin{align}
| f_{rs}(y) |
&= 
\bigg|
e_{dp,r}' \mathcal{G}_2(y ) \mathcal{G}_2(y)' e_{dp,s}
\bigg|
\nonumber
\\&\leq
4b^{2}
e_{dp,r}' J_{\Dim \VO} e_{dp,s}
\label{al:fbounded1}
\\& =
4b^{2},
\nonumber
\end{align}
where \eqref{al:fbounded1} follows since we have by Assumption \ref{Ass:E2} that $|\widetilde{G}_{i}(y_{i})| = |G_{i}(y_{i}) - \E G_{i}(y_{i}) | \leq 2b$.

Finally, we choose $\delta = \delta_{dT} M$ for some $M >0$ in \eqref{al:CIGammaXineq2}.
We can then show that for any $\epsilon>0$, there are finite $M>0$ and $K_0$ such that
\begin{align*}
&
    \Prob\left[ \| \bmhS - \bmS \|_{\max} > \delta_{dT} M\right]
    \\&\leq
    2(dp)^2 \exp\left(- \frac{1-\max\{0,\lambda_{\rig} \}}{1+\max\{0,\lambda_{\rig}\}} \frac{N \log(dT) \delta^{2}}{ T 8 b^4} \right) < \epsilon
\end{align*}
for all $d,T > K_0$, by Assumption \ref{Ass:E3}.

\subsection{Discussion concerning Assumption \ref{Ass:E3}} \label{se:discussion_Ass_E3}

The purpose of this section is to verify Assumption \ref{Ass:E3}. 
Recall that this assumption was critical in obtaining, for example, the relation \eqref{eq:new1_1}. 
Rephrasing the assumption, the question is whether we can expect the spectral gap of the Markov chain \eqref{eq:VAR_rep_VARMA} underlying our latent dynamic factor model to be bounded away from $1$ as the dimension $d$ is increasing.
The following lemma formally states one possible quite general scenario under which Assumption \ref{Ass:E3} is satisfied. Recall $\Psi_1 = \Psi$ and $\{\eta_t\}$ from Assumption \ref{e:VAR2}.

\begin{lemma} \label{le:spectral_gap}
Suppose $p=1$, $\Psi$ is symmetric and $\Psi$ and $\Sigma_{\eta} = \E \eta_t \eta_t'$ have common eigenspaces (i.e. $\Psi$ and $\Sigma_{\eta}$ commute). Then, Assumption \ref{Ass:E3} is satisfied.
\end{lemma}

The proof of Lemma \ref{le:justifyAss} is developed over Appendices \ref{se:A.4.1}--\ref{se:A.4.3} and could be of independent interest as a way to analyze the spectral gap of a Gaussian dynamic factor model as its dimension increases. What happens with spectral gaps of Markov chains as their state space dimensions increase seems to be attracting some attention only recently; see, e.g., \cite{negrea2021approximations,yang2023complexity,herve2024explicit}.

\subsubsection{Preliminaries: Hermite polynomials} \label{se:A.4.1}
While we gave a brief introduction of Hermite polynomials in Section \ref{sse:Gaussian_correlation}, we need here a more general definition allowing to expand functions in multiple variables in a Hermite basis. In addition, we introduce a way of handling Gaussian subordination with non-standardized latent variables. For the content of this section, we refer to \cite{arcones1994limit} and \cite{nourdin2011quantitative} for Hermite expansions of functions in multiple variables and to Section 9.2.2 in \cite{DaPr06} for standardizing the latent Gaussian process within the Hermite polynomials.

Let $\{X_t\}_{t\in\mathbb{Z}}$ be a $D-$dimensional Gaussian process with zero mean and positive-definite covariance matrix $\Sigma_{X}$. We further introduce the eigenvectors $v_1, \dots, v_D$ and eigenvalues $\lambda_1,\dots, \lambda_D$ of $\Sigma_X$ such that $\Sigma_X v_i = \lambda_i v_i$. 
Define the linear functionals 
\begin{equation} \label{eq:linearfunctionalsW}
    W_{v_i}(x) = \langle x, \Sigma_X^{-\frac{1}{2}} v_i \rangle;
\end{equation}
see Section 1.7 in \cite{DaPr06}.

Let $\Lambda$ denote the set of all vectors
$\alpha=(\alpha_1,  \dots, \alpha_D)$ with $\alpha_i\in \NN \cup \{0\}$. For any multi-index
$\alpha \in \Lambda $, we introduce
the notation $|\alpha | = \sum_{i= 1}^D \alpha _i$ and $\alpha ! = \prod_{i= 1}^D \alpha _i!$. Given a function $f: \RR^D \to \RR$ with $\E[f^2(X_1)]<\infty $,
$f$ admits a Hermite expansion
\begin{equation} \label{hermite}
f(x)= \sum_{\alpha \in \Lambda} a_{\alpha} \prod_{i=1}^D H_{\alpha_i} (W_{v_i}(x)), \qquad
a_{\alpha} = (\alpha !)^{-1} \E\Big[f(X_1) \prod_{i=1}^D H_{\alpha_i} (W_{v_i}(X_1))\Big],
\end{equation}
where $\{H_j\}_{j\geq 0}$ is the sequence of Hermite polynomials.
We say that the function $f$
has Hermite rank $q\geq 1$ if and only if $a_{\alpha}=0$ for all
$\alpha \in \Lambda $ with $|\alpha |<q$ and $a_{\alpha}\not =0$ for some $\alpha \in \Lambda $ with $|\alpha |=q$.
Then, a function $f: \RR^D \to \RR$ with Hermite rank $q$ can also be written as
\begin{equation} \label{fm}
f(x)= \sum_{m=q}^\infty f_m(x), \qquad f_m(x)= \sum_{\alpha \in \Lambda: |\alpha|=m} a_{\alpha} \prod_{i=1}^D H_{\alpha_i} (W_{v_i}(x));
\end{equation}
see Section 2 in 
\cite{nourdin2011quantitative} and also Section 9.2.2 in \cite{DaPr06}.

\subsubsection{Preliminaries: Isonormal Gaussian processes} \label{se:WienerChaos}

We give a brief overview of the construction of isonormal Gaussian processes and refer the reader to Section 3.1 in \cite{nourdin2011quantitative} and also to Appendix B in \cite{nourdin2012normal} for more information.

Let $\{X_t\}_{t\in\mathbb{Z}}$ be a $D-$dimensional Gaussian process with  zero mean and positive definite covariance matrix $\Sigma_X$. Let $v_1,\dots, v_D$ denote the eigenvectors of $\Sigma_X$  and recall the linear functionals $W_{v_i}(\cdot)$ from \eqref{eq:linearfunctionalsW}. Then, the autocorrelations can be written as $\E W_{v_i}(X_{t}) W_{v_j}(X_{s}) = \rho_{ij}(t-s) = \corr(X_{i,t}, X_{j,s})$.
A Gaussian process $\{X_t\}_{t\in\mathbb{Z}}$ can always be regarded as a subset of an isonormal Gaussian process
$\{W(u):~u\in \mfh\}$, where $\mfh$ is a separable Hilbert space with scalar product $\langle \cdot,\cdot \rangle_\mfh$. 
For every $m\geq 1$, we write $\mfh^{\otimes m}$ to indicate the $m$th tensor power of $\mfh$; $\mfh^{\odot m}$ indicates the $m$th symmetric tensor power of $\mfh$, equipped with the norm $\sqrt{m} \| \cdot \|_{\mfh^{\otimes m}}$. 
We can assume that there exists $u_{i,t} \in \mfh$ such that
\begin{equation*}
X_{i,t} = W(u_{i,t}), \qquad \mbox{ and } \qquad \langle u_{i,t}, u_{j,s} \rangle_{\mfh}=\rho_{ij}(t-s),
\end{equation*}
for every $t,s \in \ZZ$ and every $1\leq i,j \leq D$. 
Using the Hermite expansion \eqref{fm} of $f$ with $\E f(X_t) = 0$ and $\E [f^2(X_t) ]=1$, we obtain the representation 
\begin{equation} \label{eq:chaos}
    f(X_t) = \sum_{m=q}^\infty I_m(h_m^t),
\end{equation}
where $I_m$ denotes the isometry between $\mfh^{\odot m}$ and the $m$th Wiener chaos of $X_t$ with kernels
\begin{equation} \label{kernel}
h_m^t= \sum_{i \in \{1, \dots, D\}^m} b_{i} \,u_{i_1,t} \otimes \cdots
\otimes u_{i_m,t}
\end{equation}
for certain coefficients $b_{i}$ such that the mapping $i \mapsto b_i$ is symmetric on $\{1,\dots,D\}^m$. One also has the identities
\begin{equation} \label{eq:second_mom_contr}
\E[f_m(X_1)]^2 = m! \sum_{i\in \{1, \dots, D\}^m} b_{i}^2, \quad m\geq q, \qquad
\E[f(X_1)]^2 = \sum_{m=q}^\infty m!\sum_{i\in \{1, \dots, D\}^m} b_{i}^2
\end{equation}
for the functions in \eqref{fm}; see Section 4.1 in \cite{nourdin2011quantitative}.

\subsubsection{Proof of Lemma \ref{le:spectral_gap}}  \label{se:A.4.3}
Recall the representation of our latent factor model in terms of a VAR(1) model from \eqref{eq:VAR_rep_VARMA}. From the representation \eqref{eq:VAR_rep_VARMA}, we know that $\{\mathcal{V}_t\}_{t\in\mathbb{Z}}$ is a Markov chain. Then, setting $p=1$ and using \eqref{e:dynamic_factor}, we can also write $\mathcal{V}_t$ as
\begin{equation} \label{eq:diff_MC}
    \begin{pmatrix}
    Z_t \\
    \varepsilon_t
    \end{pmatrix}
    =
    \begin{pmatrix}
    \Lambda Y_t + \varepsilon_t \\
    \varepsilon_t
    \end{pmatrix}  
    =
    \begin{pmatrix}
    \Lambda & I_d \\
    0_{d \times r} & I_d
    \end{pmatrix}
    \begin{pmatrix}
    Y_t \\
    \varepsilon_t
    \end{pmatrix}.
\end{equation}
The Markov chain \eqref{eq:diff_MC} is therefore an affine map of the process
\begin{equation} \label{eq:mathcalW}
    \begin{pmatrix}
    Y_t \\
    \varepsilon_t
    \end{pmatrix}
=
    \begin{pmatrix}
    \Psi & 0_{r \times d} \\
    0_{d \times r} & 0_{d \times d}
    \end{pmatrix}
    \begin{pmatrix}
    Y_{t-1} \\
    \varepsilon_{t-1}
    \end{pmatrix}
    +
    \begin{pmatrix}
    \eta_{t} \\
    \varepsilon_t
    \end{pmatrix}
    \text{ or }
    \hspace{0.2cm}
    \mathcal{W}_{t} = \bm{\TM}_{\mathcal{W}} \mathcal{W}_{t-1} + \zeta_{t},
\end{equation}
that itself is a $(r+d)$-dimensional Markov chain with $\E \zeta_{t} \zeta_{t}' := \diag( \Sigma_{\eta}, \Sigma_{\varepsilon}) $. For simplicity, we assume throughout the remainder of the proof that $\Sigma_{\varepsilon} = I_{d}$. Recall the definition of the right spectral gap from \eqref{def:spectral_gap}. Therefore, it suffices to bound the lag one correlation for functionals of the Markov chain \eqref{eq:diff_MC}. We use the following relationship:
\begin{align} \label{eq:lag1_cor_G}
    \langle Ph, h \rangle_{\pi}
=
    \E \left[ h(\mathcal{V}_1) h(\mathcal{V}_2) \right]
=
    \E \left[ g(\mathcal{W}_1) g(\mathcal{W}_2) \right],
\end{align}
where
\begin{equation*}
    g(\cdot) = (h \circ h_{\Lambda}) (\cdot),
    \hspace{0.2cm}
    \text{ with }
    \hspace{0.2cm}
    h_{\Lambda}: \RR^{r+d} \to \RR^{2d}, 
    \hspace{0.2cm}
    h_{\Lambda}(w) = 
    \begin{pmatrix}
    \Lambda & I_d \\
    0_{d \times r} & I_d
    \end{pmatrix}
    w.
\end{equation*}
In order to bound \eqref{eq:lag1_cor_G}, we study the autocovariance matrices at lags $0$ and $1$ of $\{ \mathcal{W}_t \}$ in \eqref{eq:mathcalW}.
The autocovariance matrices of $\{\mathcal{W}_t\}_{t\in\mathbb{Z}}$ for general lag $h$ are given by 
\begin{equation}
    \Gamma_{\mathcal{W}}(h) = \sum_{i=0}^{\infty} \bm{\TM}_{\mathcal{W}}^{i+h} \Sigma_{\zeta} \bm{\TM}_{\mathcal{W}}^{i}
\end{equation}
since $\Psi$ is assumed to be symmetric (and hence so is $\bm{\TM}_{\mathcal{W}}$). We can infer further that
\begin{align*}
    \Gamma_{\mathcal{W}}(h) 
    &= \sum_{i=0}^{\infty} \bm{\TM}_{\mathcal{W}}^{i+h} \Sigma_{\zeta} \bm{\TM}_{\mathcal{W}}^{i}
    \\&=
    \sum_{i=1}^{\infty}      
    \begin{pmatrix}
        \Psi^{i+h} \Sigma_{\eta} \Psi^{i} & 0_{r \times d} \\
        0_{d \times r} & 0_{d\times d}
    \end{pmatrix}    
    +
    \bm{\TM}_{\mathcal{W}}^{h} \Sigma_{\zeta}
    \\&=  
    \diag( \Psi^{h} , 0_{d \times d}) 
    \begin{pmatrix}
        \sum_{i=1}^{\infty}   \Psi^{i} \Sigma_{\eta} \Psi^{i} & 0_{r \times d} \\
        0_{d \times r} & 0_{d\times d}
    \end{pmatrix}   
    +
    \bm{\TM}_{\mathcal{W}}^{h} \Sigma_{\xi}.
\end{align*}
In particular,
\begin{align} \label{eq:not_Sigma}
    \Gamma_{\mathcal{W}}(0) 
= 
    \begin{pmatrix}
        \sum_{i=0}^{\infty}   \Psi^{i} \Sigma_{\eta} \Psi^{i} & 0_{r \times d} \\
        0_{d \times r} & I_d
    \end{pmatrix}   
    =
    \begin{pmatrix}
        \Sigma_Y & 0_{r \times d} \\
        0_{d \times r} & I_d
    \end{pmatrix}
    = \Sigma.
\end{align}
such that
\begin{equation} \label{eq:acfhaczero}
    \Gamma_{\mathcal{W}}(h) = \bm{\TM}_{\mathcal{W}} \Gamma_{\mathcal{W}}(0) = \bm{\TM}_{\mathcal{W}} \Sigma.  
\end{equation}
Define eigenvectors corresponding to the covariance matrix $\Sigma = \diag(\Sigma_Y, I_{d})$ with eigenvectors $v_i' = (\widetilde{v}_i', 0_d') $ for $i=1,\dots,r$ and $v_i' = (0_r', e_{d,i}') $ for $i=r+1,\dots,2d$ with $\widetilde{v}_i$, $i=1,\dots,r$, denoting the eigenvectors of $\Sigma_Y$ and 
$e_{d,i}$ being the $i$th unit vector in $\RR^d$. Then, 
\begin{align} 
    \rho_{ij}(0) 
    &:= \E \left( \langle \mathcal{W}_1, \Sigma^{-\frac{1}{2}} v_i \rangle \langle \mathcal{W}_{2}, \Sigma^{-\frac{1}{2}} v_j \rangle \right)
    = v_i' \Sigma^{-\frac{1}{2}} \Gamma_{\mathcal{W}}(0) \Sigma^{-\frac{1}{2}} v_j
    = 
    \langle v_i, v_j \rangle;
    \label{eq:cor_scores}
\\    
    \rho_{ij}(1) 
    &:= \E \left( \langle \mathcal{W}_1, \Sigma^{-\frac{1}{2}} v_i \rangle \langle \mathcal{W}_{2}, \Sigma^{-\frac{1}{2}} v_j \rangle \right)
    = v_i' \Sigma^{-\frac{1}{2}} \Gamma_{\mathcal{W}}(1) \Sigma^{-\frac{1}{2}} v_j
    \nonumber
    \\&= v_i' \Sigma^{-\frac{1}{2}}  \bm{\TM}_{\mathcal{W}} \Sigma  \Sigma^{-\frac{1}{2}} v_j
    =
    \langle \bm{\TM}_{\mathcal{W}} \Sigma^{-\frac{1}{2}} v_i, \Sigma^{\frac{1}{2}} v_j \rangle
    =
    \langle \Psi \widetilde{v}_i, \widetilde{v}_j \rangle;
    \label{eq:cor_scores_1}
\\    
    \rho_{ij}(h) &:=
    \langle \Psi^h \widetilde{v}_i, \widetilde{v}_j \rangle, \hspace{0.2cm} h \geq 1, \label{eq_rho_h}
\end{align}
where \eqref{eq:cor_scores} follows since $\Sigma = \Gamma_{\mathcal{W}}(0)$ as introduced in \eqref{eq:not_Sigma}. For \eqref{eq:cor_scores_1}, we use \eqref{eq:acfhaczero}, as well as that, by assumption, $\Sigma_{\eta}$ and $\Psi$ commute and the definition of $\bm{\TM}_{\mathcal{W}} $ in \eqref{eq:mathcalW}.
Finally, \eqref{eq_rho_h} can be inferred by the same arguments as \eqref{eq:cor_scores_1}.

Then, following Section \ref{se:WienerChaos}, we define an isonormal Gaussian process
$\{W(u):~u\in \mfh\}$, where $\mfh$ is a separable Hilbert space with the inner product defined through \eqref{eq_rho_h} as 
\begin{equation} \label{eq:inner_iso}
\langle u_{i,t}, u_{j,s} \rangle_{\mfh}=\rho_{ij}(t-s).
\end{equation}
Denote by $\lambda_{\bm{\TM},i}$ and $\lambda_{\Psi,i}$ respectively the $i$th eigenvalues of $\bm{\TM}$ and $\Psi$. Since by assumption $\Psi$ is symmetric and has common eigenspaces with $\Sigma_{\eta}$, we have $\Psi \widetilde{v}_i = \lambda_{\Psi,i} \widetilde{v}_i$. Then, for $t-s = 1$, we get by \eqref{eq:cor_scores}--\eqref{eq_rho_h},
\begin{align}
    \langle u_{i,t}, u_{j,s} \rangle_{\mfh}
    =\rho_{ij}(t-s)
    =\rho_{ij}(1)
    = \langle \Psi \widetilde{v}_i, \widetilde{v}_j \rangle
    = \lambda_{\Psi,i} \langle \widetilde{v}_i, \widetilde{v}_j \rangle
    = \lambda_{\bm{\TM},i} \langle v_i, v_j \rangle
    = \lambda_{\bm{\TM},i} \langle u_{i,t}, u_{j,t} \rangle_{\mfh}
\end{align}
with $\lambda_{\bm{\TM},i} = \lambda_{\Psi,i}$ for $i=1,\dots, r$ and $\lambda_{\bm{\TM},i} = 0$ for $i = r+1, \dots, d$. 
Further applying \eqref{eq:chaos} and the isometry property (see Proposition 2.7.5 in \cite{nourdin2011quantitative}), 
\begin{align} \label{eq:gap_contr}
    \langle Ph, h \rangle_{\pi}
=
    \E \left[ h(\mathcal{V}_1) h(\mathcal{V}_2) \right]
=
    \E \left[ g(\mathcal{W}_1) g(\mathcal{W}_2) \right]
=
    \sum_{m=q}^\infty m!\langle  h^1_m, h^2_m \rangle_{\mathfrak{H}^{\otimes m}}.
\end{align}
With explanations given below, we have
\begin{align}
&
    \langle  h^1_m, h^2_m \rangle_{\mathfrak{H}^{\otimes m}} 
    \nonumber
    \\&\le
    \left|
    \sum_{i,j \in \{1, \dots, r+d \}^m} b_{i} b_{j} \prod_{l=1}^m \rho_{i_l j_l}(1) \right| \nonumber
    \\&=
    \left|
    \sum_{i,j \in \{1, \dots, r+d \}^m} b_{i} b_{j} \prod_{l=1}^m \lambda_{\bm{\TM},i_l} \langle u_{i_l,1}, u_{j_l,1} \rangle_{\mfh}  \right| \nonumber
    \\&=
    \left|
    \left\langle \sum_{i \in \{1, \dots, r+d\}^m} b_i (u_{i_1,1} \otimes \cdots \otimes u_{i_m,1}), 
    \sum_{j \in \{1, \dots, r+d \}^m} b_j (\lambda_{\bm{\TM},j_1} u_{j_1,1} \otimes \cdots \otimes \lambda_{\bm{\TM},j_m} u_{j_m,1})
    \right\rangle_{\mathfrak{H}^{\otimes m}}
    \right|
    \nonumber 
    \\&
    \leq
    \left|
    \left(
    \sum_{i \in \{1, \dots, r+d\}^m} b_{i}^2
    \sum_{j \in \{1, \dots, r+d\}^m} 
    b_{j}^2 \prod_{l=1}^m \langle \bm{\TM} v_{j_l}, \bm{\TM} v_{j_l} \rangle 
    \right)^{\frac{1}{2}}
    \right|
    \label{eq:2}
    \\&
    \leq
    \sum_{i \in \{1, \dots, r+d\}^m} b_{i}^2 \| \bm{\TM} \|,
    \label{eq:3}
\end{align}
where \eqref{eq:2} is due to Cauchy-Schwarz inequality and using that 
$\{ u_{i_1,1} \otimes \cdots \otimes u_{i_m,1} ~|~ i_1, \dots, i_m \geq 1\}$ is an orthonormal basis in $\mathfrak{H}^{\otimes m}$
as well as \eqref{eq:cor_scores} and \eqref{eq:inner_iso}; see also Appendix B.1 in \cite{nourdin2012normal}.

Finally, combining \eqref{eq:gap_contr} and \eqref{eq:3}, we get
\begin{align} \label{eq:last_ineq}
    \E \left[ g(\mathcal{W}_1) g(\mathcal{W}_2) \right]
\leq 
    \sum_{m=q}^\infty m! 
    \sum_{i \in \{1, \dots, r+m\}^m} b_{i}^2 \| \bm{\TM} \| 
    =
    \| \Psi \| < 1,
\end{align}
since, by \eqref{eq:second_mom_contr} and \eqref{eq:lag1_cor_G}, we have
$\sum_{m=q}^\infty m! \sum_{i \in \{1, \dots, r+d\}^m} b_{i}^2 = \E[g(\mathcal{W}_1)]^2 = \E[h(\mathcal{V}_1)]^2=\| h \|^2_{\pi}=1$. The last inequality in \eqref{eq:last_ineq}, i.e., $\| \Psi \| < 1$ follows since $\Psi $ is the transition matrix of the latent VAR(1) model $\{Y_t\}$ that is stationary by 
Assumption \ref{ass:2}.

\section{Kalman recursions and forecasting for SIS/R algorithm}
\label{ap:kalman}

This section describes how the Kalman recursions are used to obtain the $h$-step-ahead linear prediction of $Z_t$, $\widehat{Z}_{t+h|t}=H_{t1}^{(h)}Z_t+\ldots+H_{tt}^{(h)}Z_1$, and how they enter into the SIS/R algorithm. Having the autocovariance function of $\{Z_t\}$, the predictor $\widehat{Z}_{t+h|t}$ can naturally be computed through e.g. the Durbin-Levinson algorithm, but the Kalman recursions route provides computational benefit for higher dimension $d$. Related technical details can be found in \cite{durbin2012time}, \cite{douc2014nonlinear} but this list is not exhaustive.

We consider below the case $p=1$ only for simplicity. But for higher $p$, one can convert the VAR($p$) structure of the factor series into an augmented VAR(1) model by using a companion form of the VAR transition matrix. We define the one-step-ahead prediction of $Y_t$ by $\widehat{Y}_{t|t-1}$ when $Z_{1},\ldots,Z_{t-1}$ are given. Also, we denote the corresponding covariance matrix of prediction error by $\widehat{Q}_{t|t-1}:=\mathbb{E}[(Y_{t}-\widehat{Y}_{t|t-1})(Y_{t}-\widehat{Y}_{t|t-1})']$. Also, let $\widetilde{Y}_{t|t}$ be the filtered estimate and denote the corresponding error covariances $\widetilde{Q}_{t|t}$. By convention of Kalman recursions, we let $\widetilde{Y}_{0|0} \sim \mathcal{N}(0,\widetilde{Q}_{0|0})$ where $\widetilde{Q}_{0|0}=\textrm{Var}(Y_0)$.

The forecast step, which generates the filtering distribution conditioned on the previous information up to $t-1$, is 
\begin{equation}\label{e:prediction_Y}
    \widehat{Y}_{t|t-1} = \Psi \widetilde{Y}_{t-1|t-1}
\end{equation}
and the corresponding covariance matrix of prediction error is
\begin{equation}\label{e:prediction_Q}
    \widehat{Q}_{t|t-1} = \Psi \widetilde{Q}_{t-1|t-1} \Psi' + \Sigma_{\eta}.
\end{equation}
As a consequence, $\widehat{Z}_{t|t-1}=\Lambda\widehat{Y}_{t|t-1}$ and
$\widehat{R}_{t|t-1}  = \Lambda \widehat{Q}_{t|t-1} \Lambda' +\Sigma_{\varepsilon}$. The joint distribution of the forecast $Y_{t}$ and $Z_t$ conditioning on $Z_{1:t-1}$ is
\begin{equation}
    \begin{pmatrix}
        Y_{t} \\
        Z_{t}
    \end{pmatrix}\Bigg| Z_{1:t-1} \sim \mathcal{N}_{r+d}\left(
    \begin{pmatrix}
        \widehat{Y}_{t|t-1} \\
        \Lambda \widehat{Y}_{t|t-1}
    \end{pmatrix},
    \begin{pmatrix}
        \widehat{Q}_{t|t-1} & \widehat{Q}_{t|t-1}\Lambda' \\
        \Lambda\widehat{Q}_{t|t-1} & \Lambda\widehat{Q}_{t|t-1}\Lambda' + \Sigma_{\varepsilon}
    \end{pmatrix}
    \right).
\end{equation}

From this perspective, one can interpret the update step as sampling $Y_t$ conditioned on $Z_{1:t}$. That is,  $Y_t|Z_{1:t}\sim\mathcal{N}_{r}(\widetilde{Y}_{t|t},\widetilde{Q}_{t|t})$ where
\begin{eqnarray} 
    \widetilde{Y}_{t|t} &=& \widehat{Y}_{t|t-1} + K_t(Z_{t}-\widehat{Z}_{t|t-1}), \label{e:kalman_update_F} \\
    \widetilde{Q}_{t|t} &=& (I_{r} - K_t \Lambda) \widehat{Q}_{t|t-1}, \label{e:kalman_update_cov} 
\end{eqnarray}
where $K_t = \widehat{Q}_{t|t-1}\Lambda'(\Lambda\widehat{Q}_{t|t-1}\Lambda'+\Sigma_{\varepsilon})^{-1} = \widehat{Q}_{t|t-1}\Lambda'\widehat{R}_{t|t-1}^{-1}$ is called the Kalman gain. The relations \eqref{e:kalman_update_F} and \eqref{e:kalman_update_cov} form the update equations for Kalman recursions. One can apply the Sherman-Morisson-Woodbury formula for the matrix inversion of $\widehat{R}_{t|t-1}^{-1}$ when the dimension $d$ is high. These two equations are used to update filtered estimators given $Z_{1:t-1}$ when new information about $Z_t$ is added.

Hence, the Kalman recursions for the SIS/R algorithm suggested in Section \ref{se:forecast} to obtain one-step-ahead linear prediction $\widehat{Z}_{t|t-1}$ are as follows: At each time $t=1,\ldots,T$, carry out the following two steps for $k=1,\ldots,N$,
\begin{enumerate}
    \item[1.] Forecasting step: 
    \begin{eqnarray}
        \widehat{Y}_{t|t-1}^{(k)} &=& \Psi \widetilde{Y}_{t|t}^{(k)}, \\
        \widehat{Q}_{t|t-1} &=& \Psi \widetilde{Q}_{t-1|t-1} \Psi' + \Sigma_{\eta}, \\
        \widehat{Z}_{t|t-1}^{(k)} &=& \Lambda \widehat{Y}_{t|t-1}^{(k)}, \\
        \widehat{R}_{t|t-1} &=& \Lambda \widehat{Q}_{t|t-1} \Lambda' +\Sigma_{\varepsilon}.
    \end{eqnarray}
    
    \item[4.] Updating step: 
    \begin{eqnarray}
        K_t &=& \widehat{Q}_{t|t-1}\Lambda'(\Lambda \widehat{Q}_{t|t-1} \Lambda' + \Sigma_{\varepsilon})^{-1}, \\
        \widetilde{Y}_{t|t}^{(k)} &=& \widehat{Y}_{t|t-1}^{(k)} + K_t(\widetilde{Z}_{t}^{(k)}-\widehat{Z}_{t|t-1}^{(k)}), \label{e:kalman_update_Y_k}\\
        \widetilde{Q}_{t|t} &=& (I_r - K_t\Lambda) \widehat{Q}_{t|t-1}.
    \end{eqnarray}
    
\end{enumerate}
Note that $Z_t$ in \eqref{e:kalman_update_F} is replaced by $\widetilde{Z}_t$ in \eqref{e:kalman_update_Y_k} because this is the notation used in the SIS/R algorithm.

Forecasting $h$-step-ahead linear prediction after $T$ observations in the algorithm is straightforward. Since the latent factor series $\{Y_t\}$ follows a VAR model, the prediction of $Y_{T+h}$ with the information only up to $T$ is 
\begin{equation}\label{e:forecast_Y_after_T}
    \widehat{Y}_{T+h|T} = \Psi \widehat{Y}_{T+h-1|T} 
    = \ldots = \Psi^{h} \widetilde{Y}_{T|T},
 \end{equation}
and the corresponding covariance matrix of prediction error is
\begin{equation}
    \widehat{Q}_{T+h|T} = \Psi \widehat{Q}_{T+h-1|T} \Psi' + \Sigma_{\eta} = \ldots = \Psi^{h} \widetilde{Q}_{T|T} \Psi^{h'} + \sum_{s=1}^{h}\Psi^{s-1}\Sigma_{\eta}\Psi^{s-1'}
\end{equation}
from \eqref{e:prediction_Q}.
\begin{enumerate}
    \item[5.] Prediction step: 
    \begin{eqnarray}
        \widehat{Z}_{T+h|T}^{(k)} &=& \Lambda \widehat{Y}_{T+h|T}^{(k)}, \label{e:forecast_Z_after_T}\\
        \widehat{R}_{T+h|T} &=& \Lambda \widehat{Q}_{T+h|T} \Lambda' + \Sigma_{\varepsilon}. \label{e:forecast_Z_after_T_cov}
    \end{eqnarray}
\end{enumerate}
Hence, for each $k=1,\ldots,N$, we compute $\{\widehat{Y}_{T+h|T}^{(k)}\}_h$ as above and then $\widehat{Z}_{T+h|T}^{(k)}$ as \eqref{e:forecast_Z_after_T}. Likewise, the covariance of the prediction error is also computed by \eqref{e:forecast_Z_after_T_cov}. Note that unlike within the observation period, computing \eqref{e:sampling_error} is impossible beyond the period. So rather than following Forecasting and Updating steps, we directly compute \eqref{e:forecast_Z_after_T} and \eqref{e:forecast_Z_after_T_cov}.


\clearpage

\begin{figure}[th]
\centering
    \begin{subfigure}[]{0.49\textwidth}
    \centering
        \includegraphics[width=0.65\linewidth,height=0.65\linewidth]{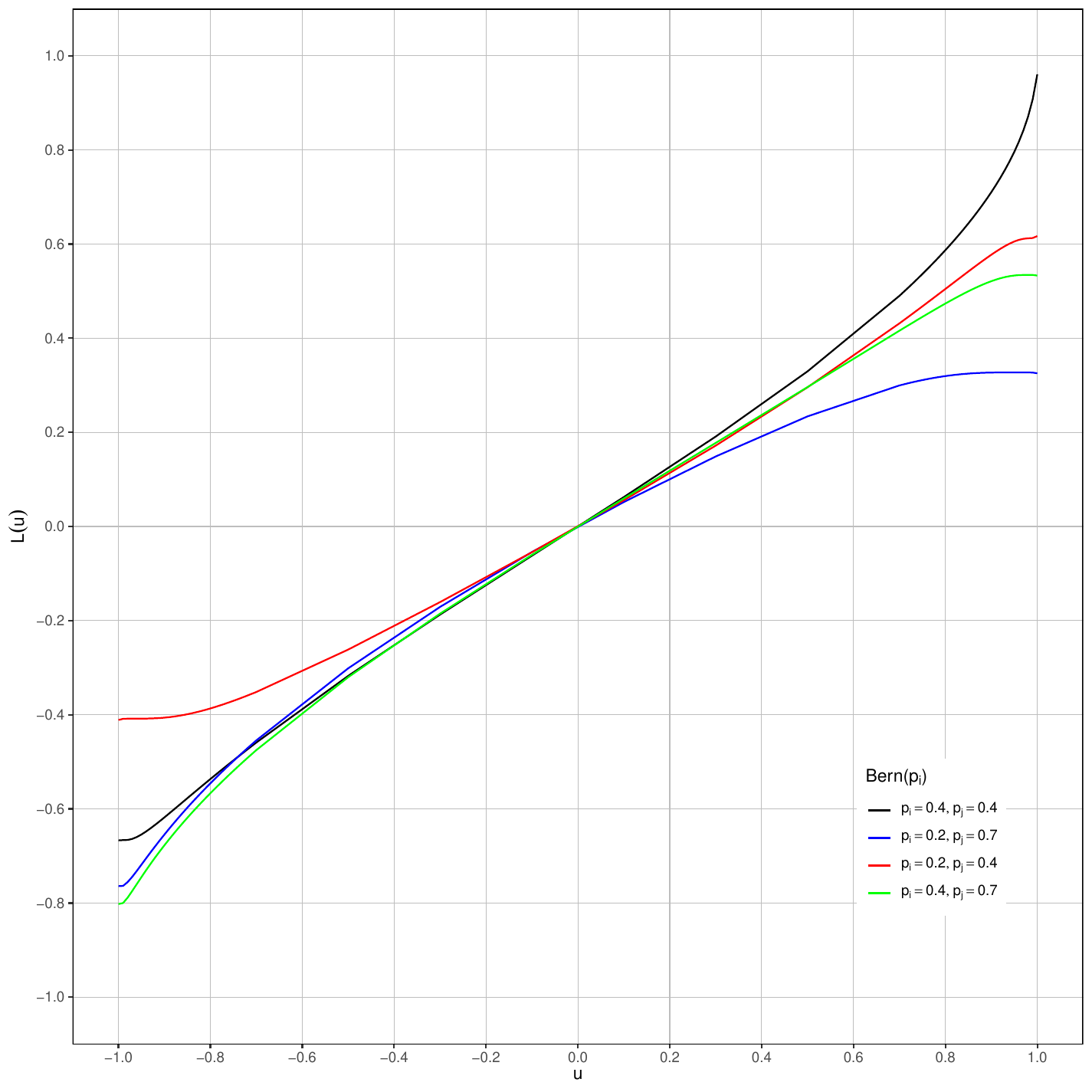} 
    \end{subfigure}
    \begin{subfigure}[]{0.49\textwidth}
    \centering
        \includegraphics[width=0.65\linewidth,height=0.65\linewidth]{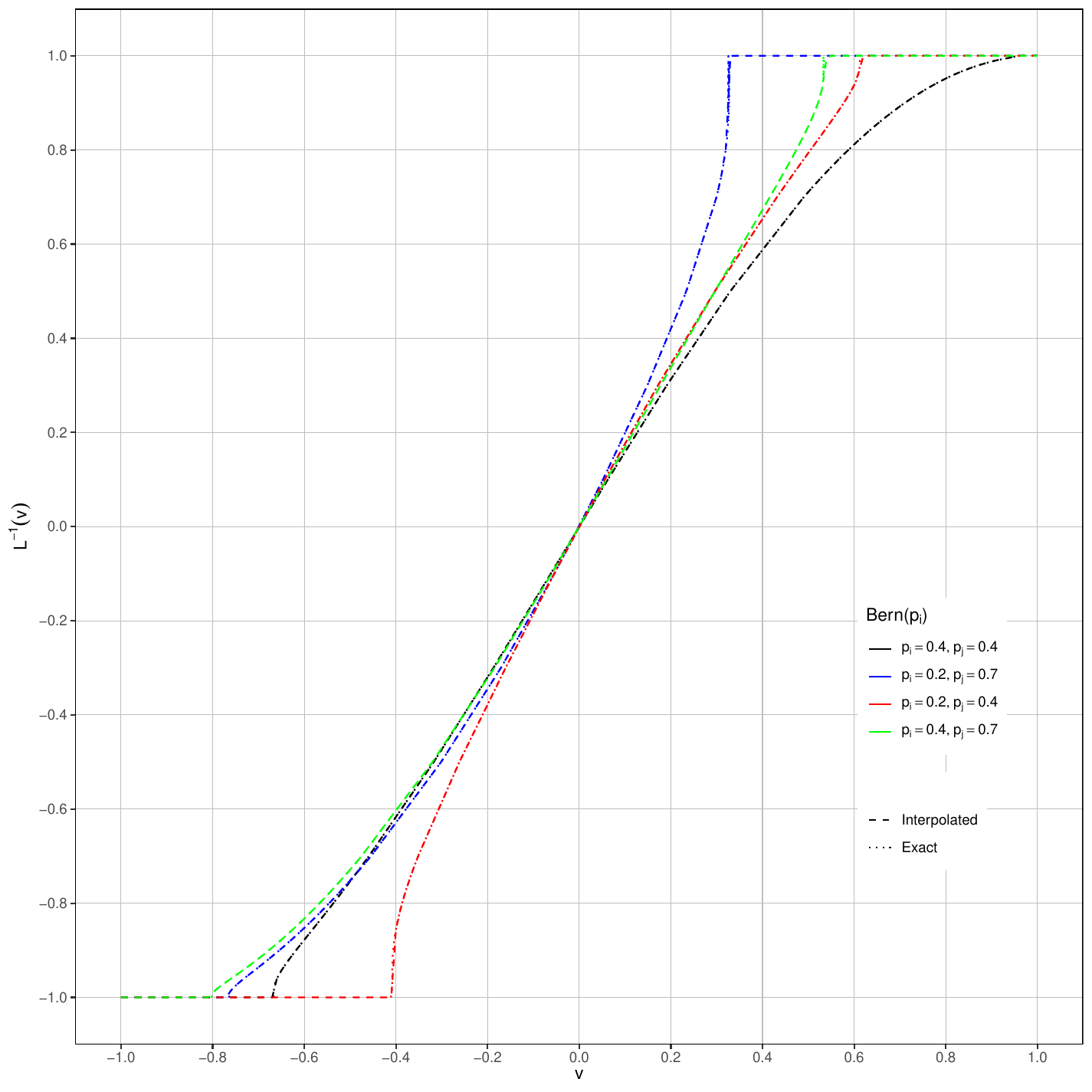}
    \end{subfigure}

    \smallskip
    \begin{subfigure}[]{0.49\textwidth}
    \centering
        \includegraphics[width=0.65\linewidth,height=0.65\linewidth]{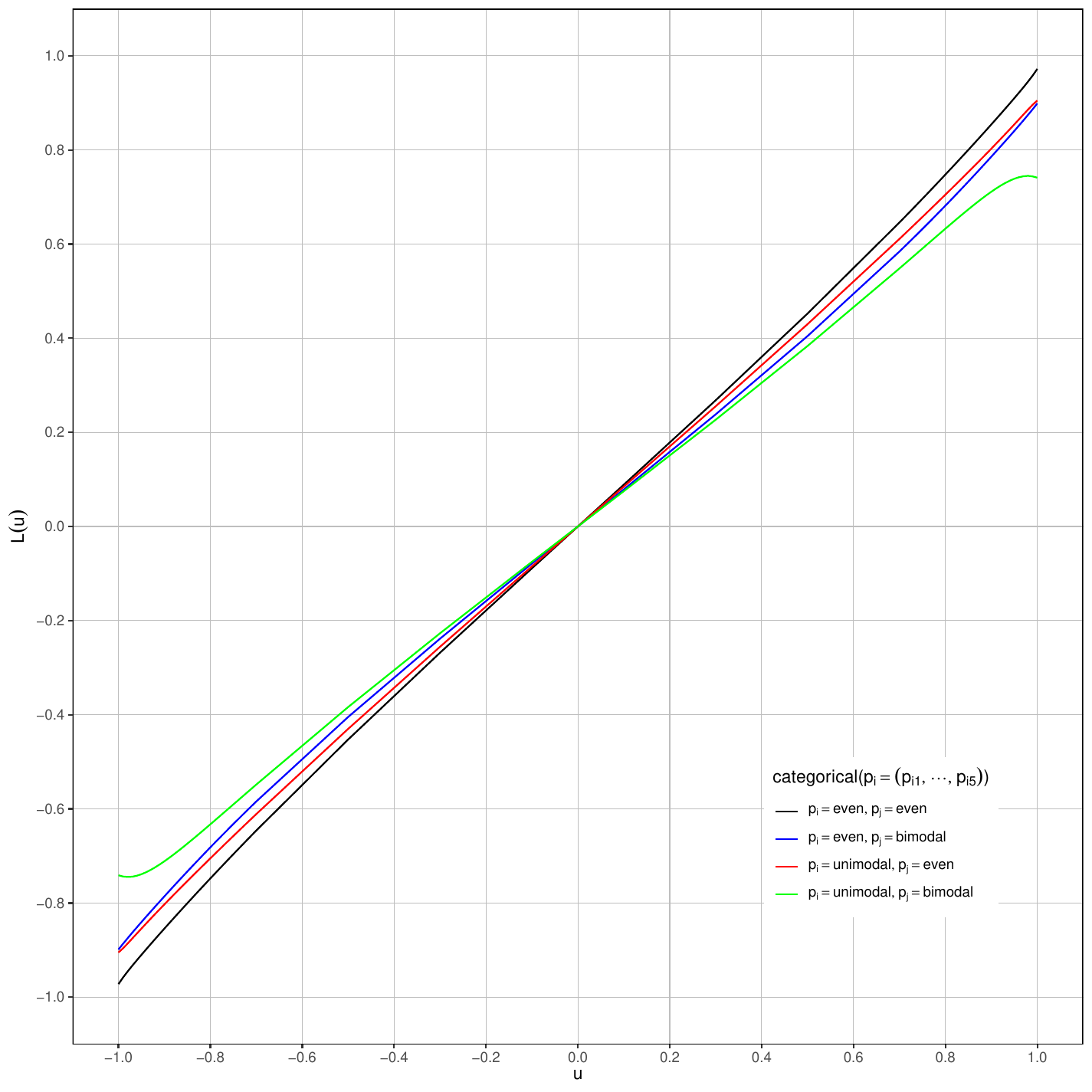}
    \end{subfigure}
    \begin{subfigure}[]{0.49\textwidth}
    \centering
        \includegraphics[width=0.65\linewidth,height=0.65\linewidth]{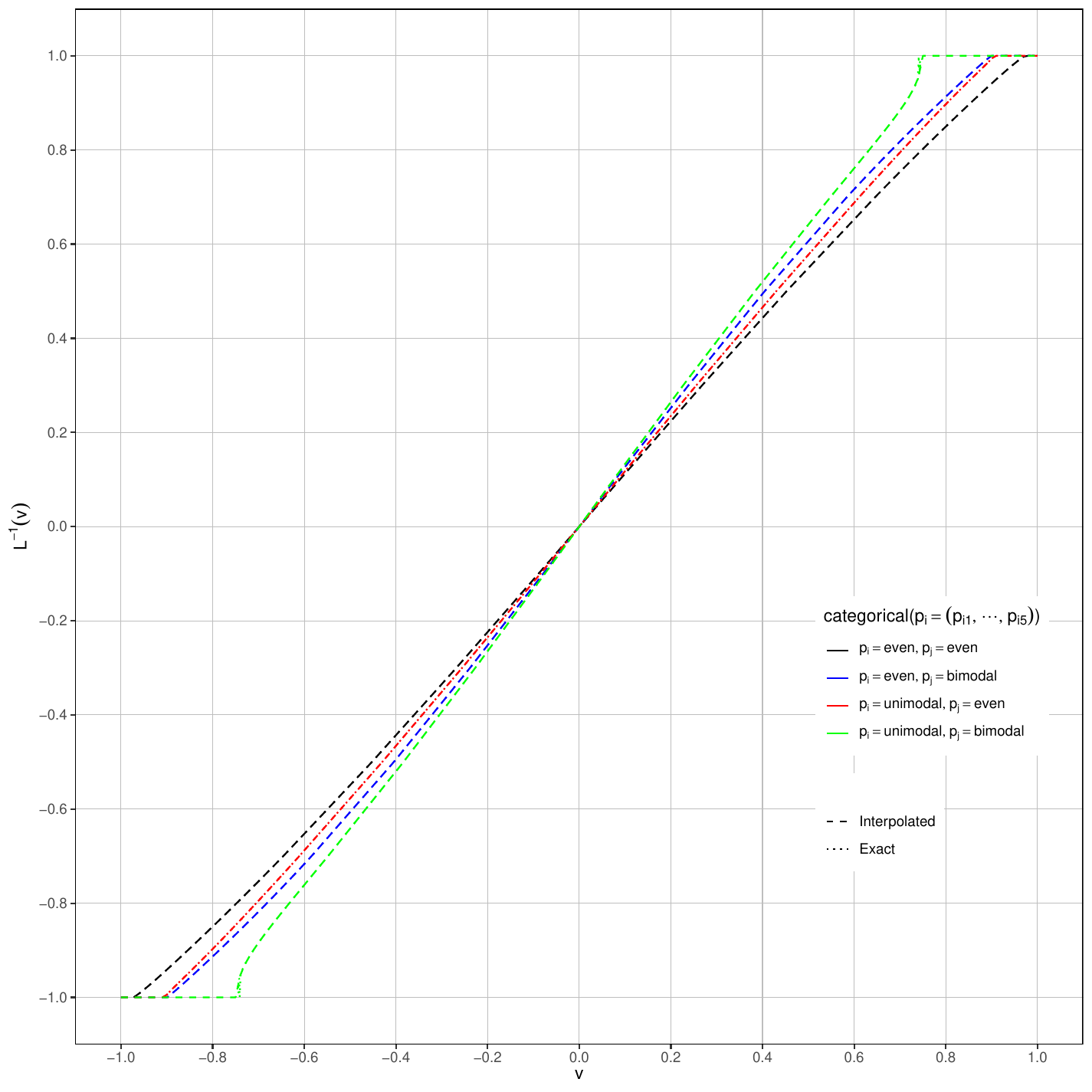}
    \end{subfigure}
    
    \smallskip
    \begin{subfigure}[]{0.49\textwidth}
    \centering
        \includegraphics[width=0.65\linewidth,height=0.65\linewidth]{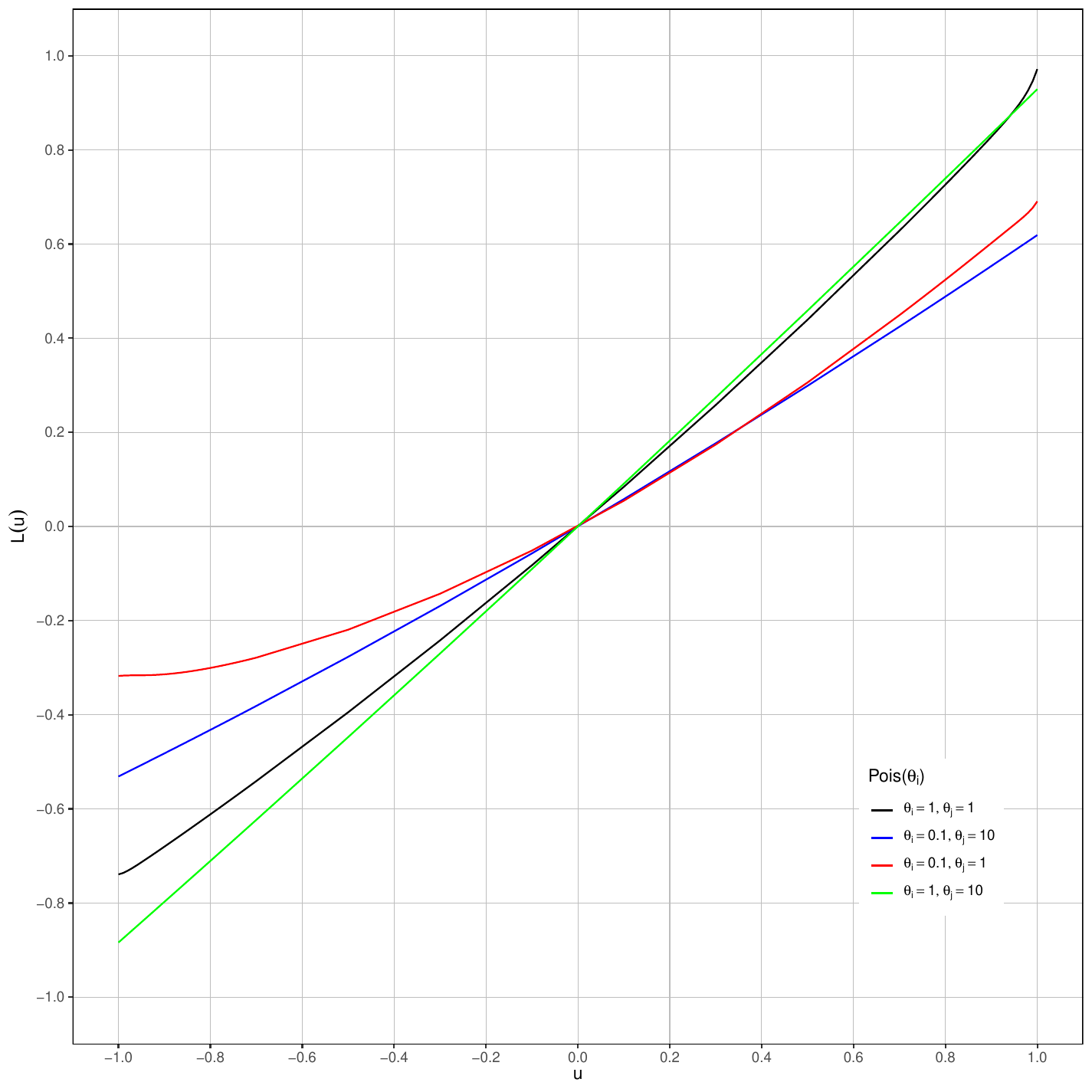}
    \end{subfigure}
    \begin{subfigure}[]{0.49\textwidth}
    \centering
        \includegraphics[width=0.65\linewidth,height=0.65\linewidth]{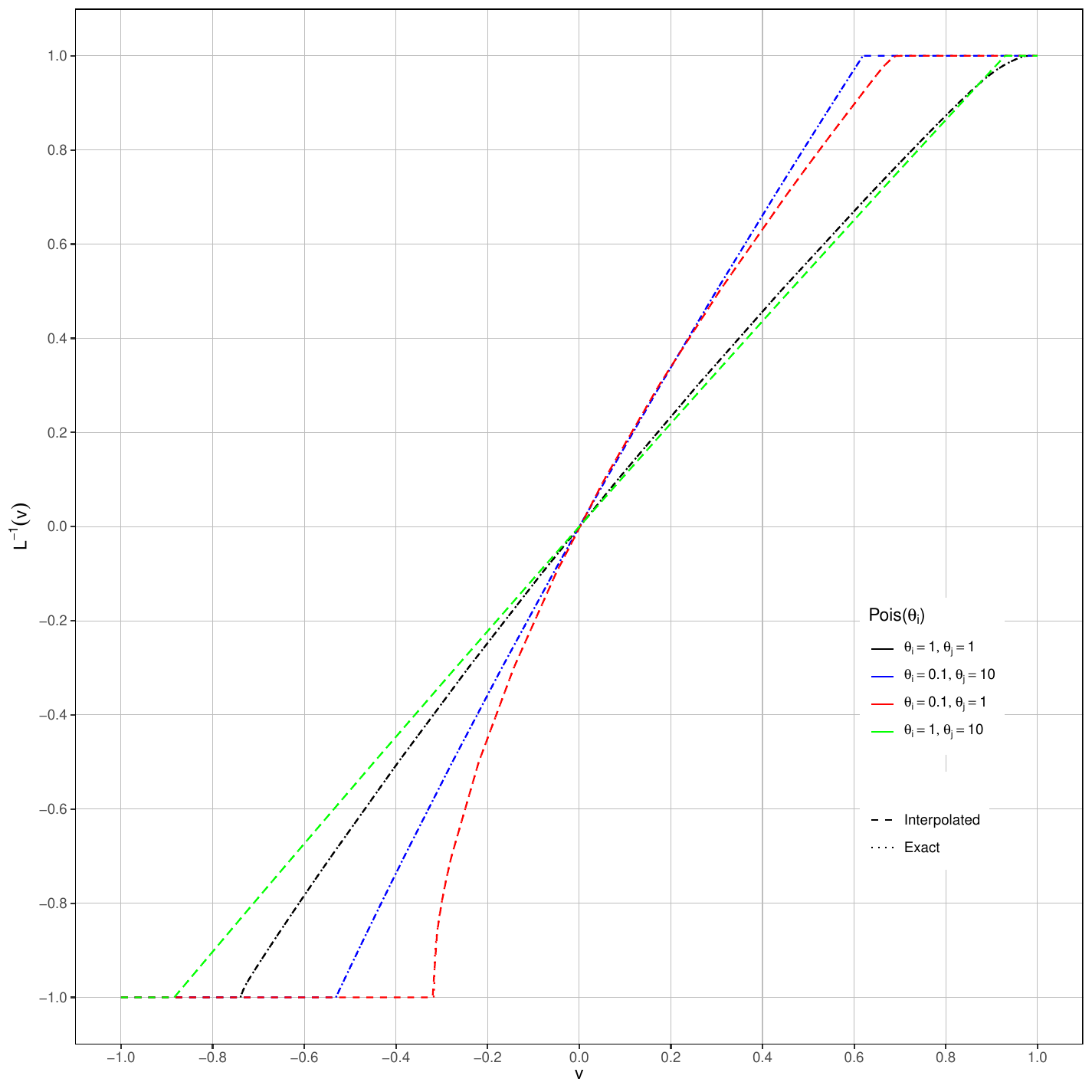}
    \end{subfigure}
    
    \smallskip
    \begin{subfigure}[]{0.49\textwidth}
    \centering
        \includegraphics[width=0.65\linewidth,height=0.65\linewidth]{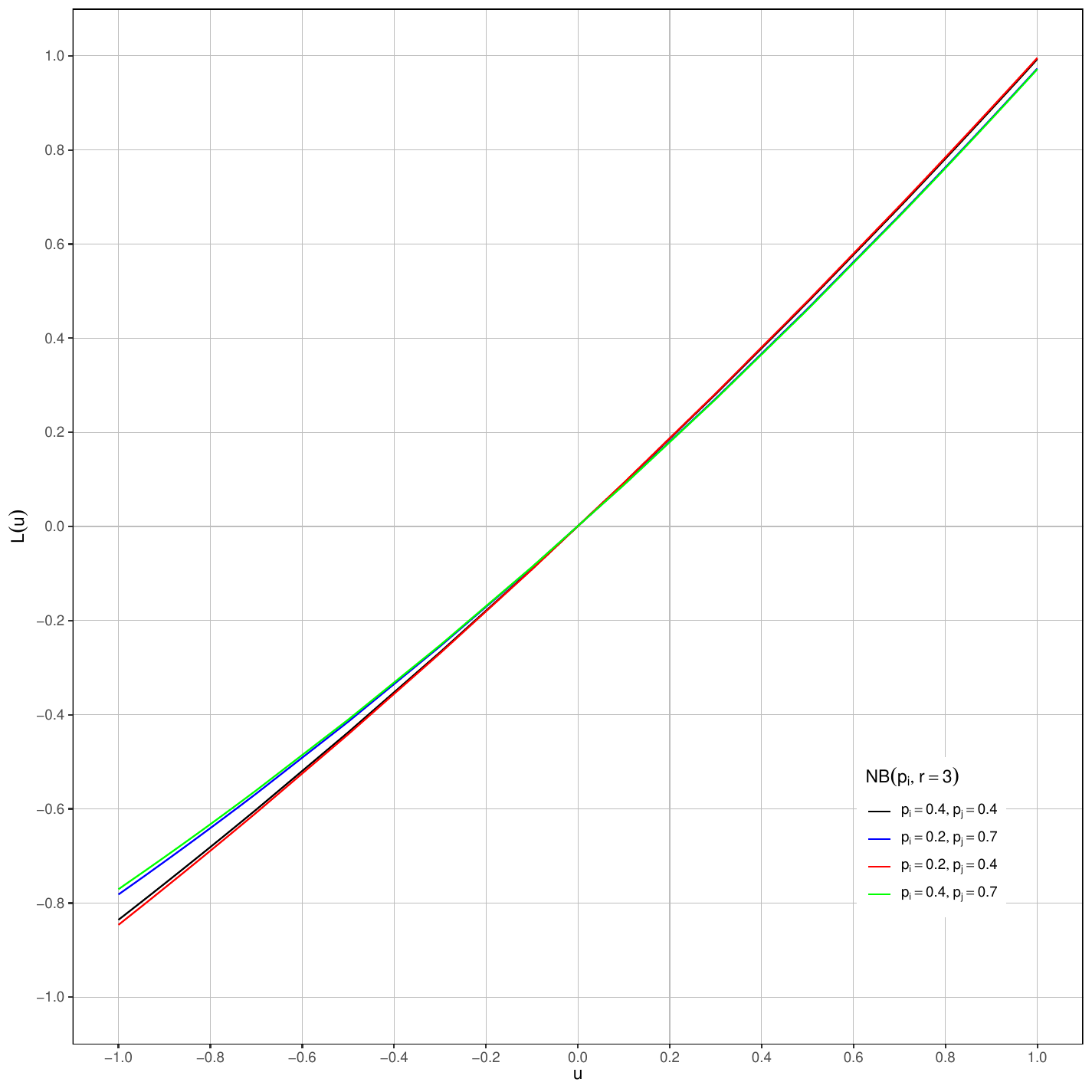}
        \caption{}
        \label{fig:Figure_link_1}
    \end{subfigure}
    \begin{subfigure}[]{0.49\textwidth}
    \centering
        \includegraphics[width=0.65\linewidth,height=0.65\linewidth]{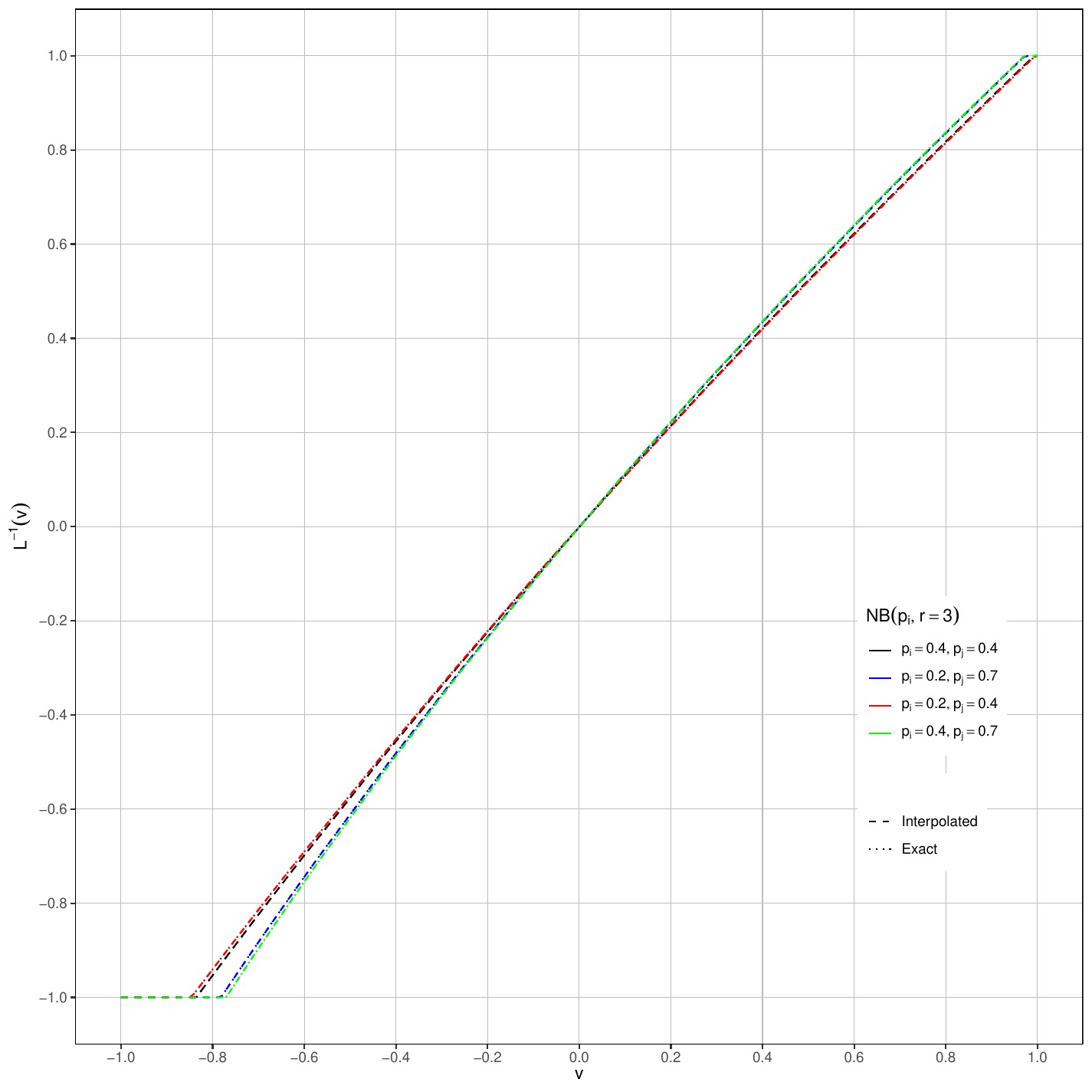}
        \caption{}
        \label{fig:Figure_link_2}
    \end{subfigure}
    \caption{(a) The link function $L_{ij}(u)$ for several combinations of CDFs. (b) The inverse link function $L^{-1}_{ij}(v)$ and its interpolation $\widetilde{L}_{ij}^{-1}(v)$ for chosen combinations of CDFs.}
    \label{fig:Figure_link}
\end{figure}

\clearpage

\begin{table}[]
\resizebox{\columnwidth}{!}{%
\begin{tabular}{cclcccccccccccccccc}
\cline{4-19}
 &  &  & \multicolumn{4}{c}{$d=15$} & \multicolumn{4}{c}{$d=30$} & \multicolumn{4}{c}{$d=60$} & \multicolumn{4}{c}{$d=90$} \\ \cline{4-19} 
 &  &  & \multicolumn{2}{c}{$T=100$} & \multicolumn{2}{c}{$T=200$} & \multicolumn{2}{c}{$T=100$} & \multicolumn{2}{c}{$T=200$} & \multicolumn{2}{c}{$T=100$} & \multicolumn{2}{c}{$T=200$} & \multicolumn{2}{c}{$T=100$} & \multicolumn{2}{c}{$T=200$} \\ \hline
\multirow{10}{*}{Bern} & \multirow{5}{*}{$r=2$} & $\widehat{\theta}$ & 0.0745 & (0.0251) & 0.0582 & (0.0206) & 0.0789 & (0.0272) & 0.0568 & (0.0185) & 0.0718 & (0.0216) & 0.0629 & (0.0239) & 0.0772 & (0.0260) & 0.0555 & (0.0189) \\
 &  & $\widehat{\Lambda}$ & 0.9018 & (0.2387) & 0.8877 & (0.3312) & 0.9960 & (0.2834) & 0.9920 & (0.2753) & 0.9979 & (0.2686) & 0.9571 & (0.2541) & 0.9584 & (0.2684) & 0.8739 & (0.2779) \\
 &  & $\widehat{\Sigma}_{\varepsilon}$ & 0.2541 & (0.0988) & 0.1591 & (0.0338) & 0.1938 & (0.0473) & 0.1246 & (0.0256) & 0.1672 & (0.0346) & 0.1080 & (0.0215) & 0.1502 & (0.0215) & 0.0972 & (0.0135) \\
 &  & $\widehat{\Psi}$ & 0.6067 & (0.1340) & 0.4856 & (0.0949) & 0.5422 & (0.1341) &  0.3817 & (0.0830) & 0.4515 & (0.1437) & 0.3487 & (0.0851) & 0.4153 & (0.1188) & 0.3188 & (0.0730) \\
 &  & $\widehat{\Sigma}_{\eta}$ & 0.7778 & (0.0720) & 0.7189 & (0.0726) & 0.7237 & (0.0877) & 0.6276 & (0.0840) & 0.6544 & (0.1119) & 0.5946 & (0.0896) & 0.6257 & (0.0998) & 0.5606 & (0.0794) \\ \cline{2-19} 
 & \multirow{5}{*}{$r=5$} & $\widehat{\theta}$ & 0.0807 & (0.0228) & 0.0560 & (0.0172) & 0.0819 & (0.0202) & 0.0590 & (0.0155) & 0.0798 & (0.0202) & 0.0608 & (0.0158) & 0.0789 & (0.0178) & 0.0576 & (0.0146) \\
 &  & $\widehat{\Lambda}$ & 1.1765 & (0.1661) & 1.0724 & (0.1308) & 1.1211 & (0.1029) & 1.0413 & (0.1360) & 1.0755 & (0.1063) & 1.1076 & (0.1158) & 1.0496 & (0.1163) & 1.0097 & (0.1366) \\
 &  & $\widehat{\Sigma}_{\varepsilon}$ & 0.4569 & (0.0480) & 0.3738 & (0.0234) & 0.3588 & (0.0581) & 0.2265 & (0.0243) & 0.2986 & (0.0622) & 0.1639 & (0.0205) & 0.2734 & (0.0521) & 0.1394 & (0.0169) \\
 &  & $\widehat{\Psi}$ & 0.8330 & (0.0736) & 0.7593 & (0.0536) & 0.7874 & (0.0626) &  0.6542 & (0.0531) & 0.7145 & (0.0667) & 0.5409 & (0.0554) & 0.6955 & (0.0595) & 0.5143 & (0.0697) \\
 &  & $\widehat{\Sigma}_{\eta}$ & 0.7858 & (0.0370) & 0.8233 & (0.0282) & 0.7797 & (0.0362) & 0.7657 & (0.0301) & 0.7551 & (0.0358) & 0.7040 & (0.0376) & 0.7467 & (0.0397) & 0.6844 & (0.0471) \\ \hline
\multirow{10}{*}{Categ} & \multirow{5}{*}{$r=2$} & $\widehat{\theta}$ & 0.1271 & (0.0356) & 0.0828 & (0.0255) & 0.1317 & (0.0397) & 0.0922 & (0.0282) & 0.1275 & (0.0353) & 0.0919 & (0.0265) & 0.1274 & (0.0375) & 0.0907 & (0.0253) \\
 &  & $\widehat{\Lambda}$ & 1.0327 & (0.3267) & 0.8321 & (0.3107) & 0.9170 & (0.2959) & 0.9391 & (0.3003) & 1.0572 & (0.2231) & 0.7869 & (0.2769) & 1.0111 & (0.3130) & 0.8437 & (0.2940) \\
 &  & $\widehat{\Sigma}_{\varepsilon}$ & 0.1753 & (0.0336) & 0.1518 & (0.0268) & 0.1314 & (0.0296) & 0.1028 & (0.0226) & 0.1236 & (0.0308) & 0.0898 & (0.0216) & 0.1161 & (0.0231) & 0.0850 & (0.0233) \\
 &  & $\widehat{\Psi}$ & 0.5485 & (0.1370) & 0.5129 & (0.0839) & 0.4560 & (0.1095) &  0.3543 & (0.0749) & 0.4202 & (0.1172) & 0.3184 & (0.0746) & 0.3734 & (0.1038) & 0.3022 & (0.0684) \\
 &  & $\widehat{\Sigma}_{\eta}$ & 0.7396 & (0.0871) & 0.7477 & (0.0625) & 0.6748 & (0.0877) & 0.6010 & (0.0753) & 0.6341 & (0.1006) & 0.5630 & (0.0809) & 0.5957 & (0.1107) & 0.5450 & (0.0804) \\ \cline{2-19} 
 & \multirow{5}{*}{$r=5$} & $\widehat{\theta}$ & 0.1285 & (0.0291) & 0.0910 & (0.0240) & 0.1315 & (0.0302) & 0.0958 & (0.0198) & 0.1315 & (0.0244) & 0.0917 & (0.0172) & 0.1302 & (0.0255) & 0.0890 & (0.0188) \\
 &  & $\widehat{\Lambda}$ & 1.1377 & (0.1570) & 1.0889 & (0.1857) & 1.1187 & (0.1399) & 1.1164 & (0.1476) & 1.0427 & (0.1170) & 0.9799 & (0.1138) & 1.0800 & (0.1036) & 1.0259 & (0.0990) \\
 &  & $\widehat{\Sigma}_{\varepsilon}$ & 0.3675 & (0.0253) & 0.3061 & (0.0223) & 0.2294 & (0.0274) & 0.1744 & (0.0199) & 0.1778 & (0.0236) & 0.1234 & (0.0187) &  0.1590 & (0.0258) & 0.1053 & (0.0176) \\
 &  & $\widehat{\Psi}$ & 0.7930 & (0.0584) & 0.6895 & (0.0465) & 0.6949 & (0.0643) &  0.5801 & (0.0609) & 0.6444 & (0.0693) & 0.5095 & (0.0541) & 0.6029 & (0.0644) & 0.4606 & (0.0560) \\
 &  & $\widehat{\Sigma}_{\eta}$ & 0.8132 & (0.0293) & 0.8033 & (0.0263) & 0.7599 & (0.0362) & 0.7423 & (0.0365) & 0.7211 & (0.0402) & 0.6961 & (0.0386) & 0.6938 & (0.0403) & 0.6537 & (0.0435) \\ \hline
\multirow{10}{*}{Pois} & \multirow{5}{*}{$r=2$} & $\widehat{\theta}$ & 0.4095 & (0.1904) & 0.2542 & (0.1168) & 0.3701 & (0.1709) & 0.2775 & (0.1233) & 0.3827 & (0.1538) & 0.2742 & (0.1174) & 0.3687 & (0.1428) & 0.2731 & (0.1089) \\
 &  & $\widehat{\Lambda}$ & 1.0918 & (0.2642) & 0.7946 & (0.3217) & 1.0475 & (0.2319) & 0.9560 & (0.3129) & 0.9204 & (0.2424) & 0.9346 & (0.3276) & 0.9445 & (0.2596) & 0.9930 & (0.2364) \\
 &  & $\widehat{\Sigma}_{\varepsilon}$ & 0.3971 & (0.0786) & 0.2171 & (0.1085) & 0.3118 & (0.0880) & 0.2133 & (0.0966) & 0.3183 & (0.4118) & 0.2075 & (0.2179) & 0.2353 & (0.1325) & 0.1684 & (0.1645) \\
 &  & $\widehat{\Psi}$ & 0.7011 & (0.1526) & 0.5854 & (0.1636) & 0.6071 & (0.1774) &  0.4731 & (0.1337) & 0.5071 & (0.1651) & 0.4001 & (0.1067) & 0.4346 & (0.1455) & 0.3265 & (0.0960) \\
 &  & $\widehat{\Sigma}_{\eta}$ & 0.7840 & (0.0744) & 0.7531 & (0.0738) & 0.7759 & (0.2761) & 0.6899 & (0.0879) & 0.7085 & (0.2209) & 0.6351 & (0.0886) & 0.6323 & (0.0999) & 0.5652 & (0.0929)  \\ \cline{2-19} 
 & \multirow{5}{*}{$r=5$} & $\widehat{\theta}$ & 0.4048 & (0.1553) & 0.2705 & (0.1035) & 0.3852 & (0.1274) & 0.2810 & (0.0871) & 0.3955 & (0.1147) & 0.2994 & (0.0884) & 0.3988 & (0.1041) & 0.2953 & (0.0823) \\
 &  & $\widehat{\Lambda}$ & 1.1810 & (0.1466) & 1.0689 & (0.1496) & 1.1508 & (0.1116)  & 1.0766 & (0.1476) & 1.1222 & (0.0991) & 1.0756 & (0.1236) & 1.0964 & (0.0801) & 1.0604 & (0.1281) \\
 &  & $\widehat{\Sigma}_{\varepsilon}$ & 0.5581 & (0.3447) & 0.4201 & (0.0677) & 0.5472 & (0.0824) & 0.4426 & (0.2551) & 0.5448 & (0.2331) & 0.3782 & (0.1069) & 0.5474 & (0.2924) & 0.3506 & (0.1045) \\
 &  & $\widehat{\Psi}$ & 0.8575 & (0.0978) & 0.7631 & (0.0879) & 0.8100 & (0.0860) &  0.7451 & (0.0800) & 0.7542 & (0.0867) & 0.6375 & (0.0905) & 0.7201 & (0.1027) & 0.5931 & (0.0682) \\
 &  & $\widehat{\Sigma}_{\eta}$ & 0.7660 & (0.0557) & 0.7890 & (0.0363) & 0.7449 & (0.0381) & 0.7692 & (0.0325) & 0.7734 & (0.2479) & 0.7385 & (0.1532) & 0.7968 & (0.2761) & 0.7122 & (0.0519) \\ \hline
\multirow{10}{*}{Negbin} & \multirow{5}{*}{$r=2$} & $\widehat{\theta}$ & 0.0355 & (0.0171) & 0.0235 & (0.0087) & 0.0980 & (0.0106) & 0.0966 & (0.0069) & 0.1739 & (0.0061) & 0.1742 & (0.0059) & 0.2336 & (0.0082) & 0.2356 & (0.0041) \\
 &  & $\widehat{\Lambda}$ & 0.6957 & (0.3128) & 0.9873 & (0.2703) & 0.9215 & (0.2655) & 0.8582 & (0.3072) & 0.7562 & (0.2894) & 0.9888 & (0.2931) & 0.8651 & (0.2541) & 0.9536 & (0.2015) \\
 &  & $\widehat{\Sigma}_{\varepsilon}$ & 0.1542 & (0.0275) & 0.1324 & (0.0229) & 0.1114 & (0.0191) & 0.0832 & (0.0196) & 0.0993 & (0.0151) & 0.0752 & (0.0142) & 0.0960 & (0.0133) & 0.0730 & (0.0105) \\
 &  & $\widehat{\Psi}$ & 0.5812 & (0.1133) & 0.5151 & (0.0887) & 0.4889 & (0.1091) & 0.3590 & (0.0768) & 0.4228 & (0.1028) & 0.3237 & (0.0719) & 0.3846 & (0.0997) & 0.3107 & (0.0768) \\
 &  & $\widehat{\Sigma}_{\eta}$ & 0.7736 & (0.0708) & 0.7525 & (0.0648) & 0.7043 & (0.0869) & 0.6138 & (0.0803) & 0.6442 & (0.0979) & 0.5709 & (0.0785) & 0.6097 & (0.1021) & 0.5533 & (0.0875) \\ \cline{2-19} 
 & \multirow{5}{*}{$r=5$} & $\widehat{\theta}$ & 0.0400 & (0.0133) & 0.0269 & (0.0092) & 0.1008 & (0.0075) & 0.0960 & (0.0058) & 0.1736 & (0.0056) & 0.1742 & (0.0032) & 0.2352 & (0.0068) & 0.2353 & (0.0037)  \\
 &  & $\widehat{\Lambda}$ & 1.1473 & (0.1920) & 1.0099 & (0.1826) & 1.0702 & (0.1368) & 1.1149 & (0.1526) & 1.0519 & (0.1225) & 0.9496 & (0.1385) & 1.0308 & (0.1147) & 0.9765 & (0.1215) \\
 &  & $\widehat{\Sigma}_{\varepsilon}$ & 0.3055 & (0.0245) & 0.2922 & (0.0175)  & 0.1944 & (0.021) & 0.1509 & (0.0173) & 0.1307 & (0.0153) & 0.0991 & (0.0128) & 0.1148 & (0.0161) & 0.0852 & (0.0123) \\
 &  & $\widehat{\Psi}$ & 0.7353 & (0.0576) & 0.6921 & (0.0408) & 0.6872 & (0.0591) &  0.5563 & (0.0481) & 0.6291 & (0.0571) & 0.5031 & (0.0504) & 0.6046 & (0.0606) & 0.4568 & (0.0447) \\
 &  & $\widehat{\Sigma}_{\eta}$ & 0.8046 & (0.0303) & 0.8184 & (0.0213) & 0.7715 & (0.0321) & 0.7391 & (0.0329) &  0.7341 & (0.0322) & 0.6978 & (0.0356) & 0.7075 & (0.0418) & 0.6600 & (0.0384)  \\ \hline
\end{tabular}%
}
\caption{The average and standard deviation (in the parenthesis) of losses of the estimators for various combinations of model parameters including number of factors $r$, dimension $d$, number of time points $T$, and several marginal distributions.}
\label{tab:table_estimation}
\end{table}

\clearpage

\begin{figure}[h]
    \centering
    \includegraphics[width=1\textwidth,height=1\textwidth]{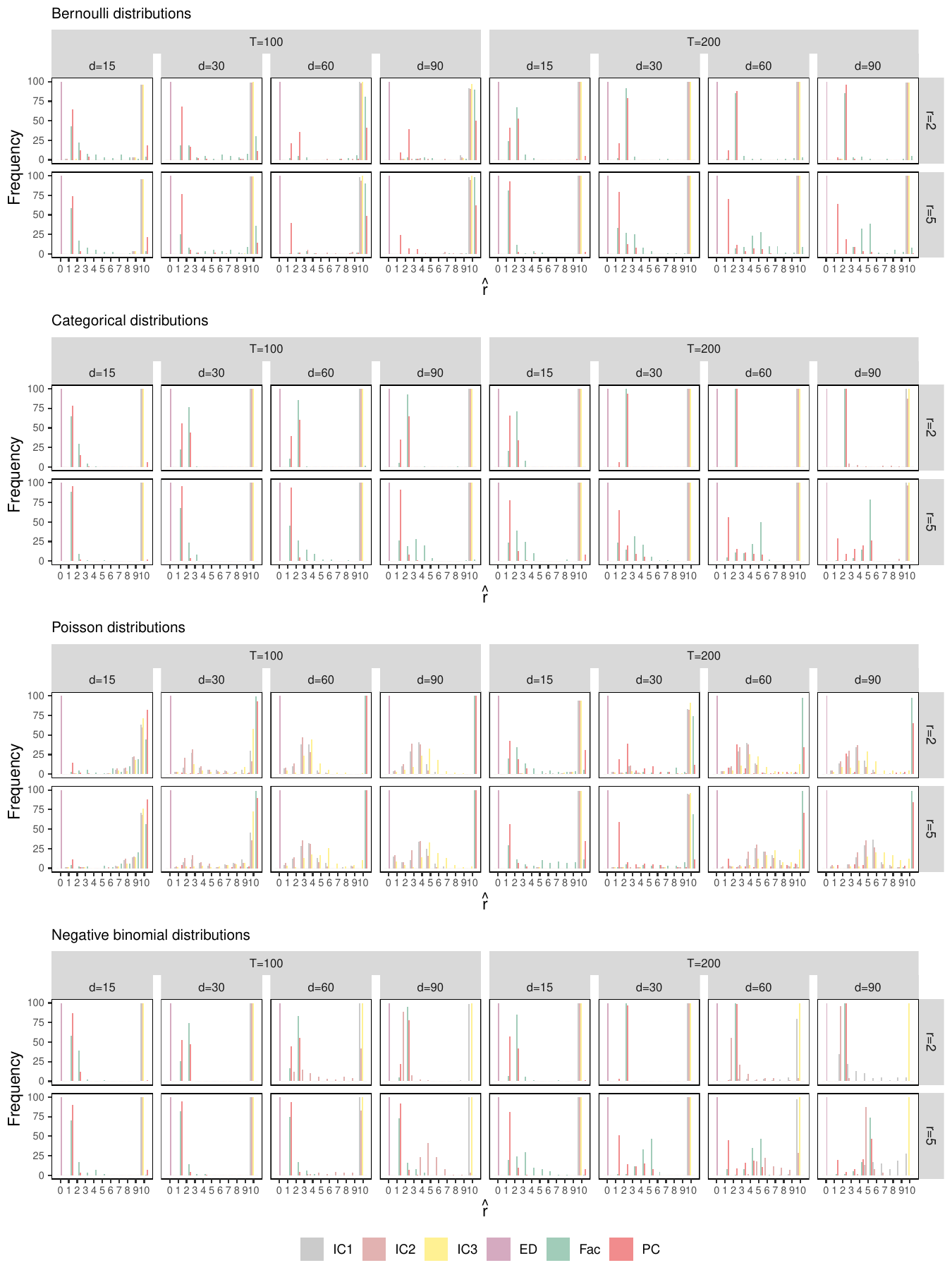}
    \caption{Estimated number of factors from simulated data for various combinations of model parameters including number of factors $r$, dimension $d$, number of time points $T$, and several marginal distributions.}
    \label{fig:Figure_rank}
\end{figure}


\clearpage

\begin{table}[t]
\resizebox{\textwidth}{!}{%
\begin{tabular}{ccccccccccccccccccccccc}
\cline{4-23}
 &  &  & \multicolumn{10}{c|}{$\{Y_{t}^{(k)}\}$} & \multicolumn{10}{c}{$\{Z_{t}^{(k)}\}$} \\ \cline{4-23} 
 &  &  & \multicolumn{2}{c}{$H=1$} & \multicolumn{2}{c}{$H=2$} & \multicolumn{2}{c}{$H=3$} & \multicolumn{2}{c}{$H=6$} & \multicolumn{2}{c|}{$H=12$} & \multicolumn{2}{c}{$H=1$} & \multicolumn{2}{c}{$H=2$} & \multicolumn{2}{c}{$H=3$} & \multicolumn{2}{c}{$H=6$} & \multicolumn{2}{c}{$H=12$} \\ \hline
\multirow{8}{*}{Bern} & \multirow{4}{*}{$r=2$} & $d=15$ & 0.3807 & (0.1370) & 0.3827 & (0.1664) & 0.4062 & (0.2019) & 0.4600 & (0.2304) & 0.5218 & (0.2746) & 0.9276 & (0.1482) & 0.9077 & (0.1589) & 0.9386 & (0.1838) & 0.9485 & (0.1923) & 0.9812 & (0.2102) \\
 &  & $d=30$ & 0.3402 & (0.1467) & 0.3882 & (0.1971) & 0.4180 & (0.2175) & 0.4522 & (0.2373) & 0.5126 & (0.2648) & 0.9067 & (0.1283) & 0.9366 & (0.1374) & 0.9606 & (0.1462) & 0.9451 & (0.1443) & 0.9724 & (0.1728) \\
 &  & $d=60$ & 0.3175 & (0.1338) & 0.3446 & (0.1515) & 0.4063 & (0.1863) & 0.4375 & (0.2128) & 0.5015 & (0.2767) & 0.8932 & (0.1015) & 0.9126 & (0.0987) & 0.9191 & (0.1139) & 0.9389 & (0.1207) & 0.9873 & (0.1519) \\
 &  & $d=90$ & 0.2855 & (0.1324) & 0.3269 & (0.1654) & 0.3603 & (0.1853) & 0.4532 & (0.2211) & 0.5030 & (0.2567) & 0.9019 & (0.0781) & 0.9227 & (0.0849) & 0.9249 & (0.0980) & 0.9587 & (0.1162) & 0.9787 & (0.1419) \\ \cline{3-23} 
 & \multirow{4}{*}{$r=5$} & $d=15$ & 0.3263 & (0.1618) & 0.3710 & (0.1598) & 0.3899 & (0.1700) & 0.4647 & (0.2175) & 0.4497 & (0.2360) & 0.8897 & (0.1715) & 0.9071 & (0.1580) & 0.9117 & (0.1804) & 0.9370 & (0.1701) & 0.9227 & (0.1977)  \\
 &  & $d=30$ & 0.3210 & (0.1238) & 0.3503 & (0.1843) & 0.4077 & (0.1897) & 0.4294 & (0.2105) & 0.5062 & (0.2710) & 0.8938 & (0.1194) & 0.8981 & (0.1238) & 0.9206 & (0.1500) & 0.9407 & (0.1633) & 0.9714 & (0.1752) \\
 &  & $d=60$ & 0.2759 & (0.1487) & 0.3435 & (0.1692) & 0.3825 & (0.1831) & 0.4255 & (0.2058) & 0.4432 & (0.2415) & 0.8883 & (0.0934) & 0.9137 & (0.0967) & 0.9363 & (0.1055) & 0.9437 & (0.1079) & 0.9641 & (0.1338)   \\
 &  & $d=90$ & 0.2473 & (0.1214) & 0.3049 & (0.1686) & 0.3779 & (0.1785) & 0.4534 & (0.2348) & 0.5577 & (0.2928) & 0.8785 & (0.0768) & 0.9062 & (0.0904) & 0.9333 & (0.0965) & 0.9641 & (0.1113) & 1.0265 & (0.1588)   \\ \hline
\multirow{8}{*}{Categ} & \multirow{4}{*}{$r=2$} & $d=15$ & 0.3263 & (0.1618) & 0.3710 & (0.1598) & 0.3899 & (0.1700) & 0.4647 & (0.2175) & 0.4497 & (0.2360) & 0.8897 & (0.1715) & 0.9071 & (0.1580) & 0.9117 & (0.1804) & 0.9370 & (0.1701) & 0.9227 & (0.1977)   \\
 &  & $d=30$ & 0.3210 & (0.1238) & 0.3503 & (0.1843) & 0.4077 & (0.1897) & 0.4294 & (0.2105) & 0.5062 & (0.271) & 0.8938 & (0.1194) & 0.8981 & (0.1238) & 0.9206 & (0.1500) & 0.9407 & (0.1633) & 0.9714 & (0.1752)    \\
 &  & $d=60$ & 0.2759 & (0.1487) & 0.3435 & (0.1692) & 0.3825 & (0.1831) & 0.4255 & (0.2058) & 0.4432 & (0.2415) & 0.8883 & (0.0934) & 0.9137 & (0.0967) & 0.9363 & (0.1055) & 0.9437 & (0.1079) & 0.9641 & (0.1338)  \\
 &  & $d=90$ & 0.2473 & (0.1214) & 0.3049 & (0.1686) & 0.3779 & (0.1785) & 0.4534 & (0.2348) & 0.5577 & (0.2928) & 0.8785 & (0.0768) & 0.9062 & (0.0904) & 0.9333 & (0.0965) & 0.9641 & (0.1113) & 1.0265 & (0.1588)  \\ \cline{2-23} 
 & \multirow{4}{*}{$r=5$} & $d=15$ &  0.4326 & (0.1257) & 0.4578 & (0.1436) & 0.4883 & (0.1756) & 0.5104 & (0.1699) & 0.5349 & (0.1648) & 0.8967 & (0.1537) & 0.8978 & (0.1672) & 0.9219 & (0.1849) & 0.9322 & (0.1829) & 0.9757 & (0.1996)  \\
 &  & $d=30$ & 0.3538 & (0.0918) & 0.3974 & (0.1188) & 0.4382 & (0.1325) & 0.4918 & (0.1606) & 0.5515 & (0.1601) & 0.8710 & (0.1042) & 0.8863 & (0.1199) & 0.9405 & (0.1304) & 0.9505 & (0.1383) & 0.9958 & (0.1572) \\
 &  & $d=60$ & 0.3261 & (0.0953) & 0.3708 & (0.1195) & 0.4126 & (0.1132) & 0.4966 & (0.1731) & 0.5535 & (0.1791) & 0.9066 & (0.0829) & 0.9185 & (0.0837) & 0.9511 & (0.0916) & 0.9699 & (0.1027) & 0.9882 & (0.1281) \\
 &  & $d=90$ & 0.3175 & (0.0853) & 0.3745 & (0.1071) & 0.4364 & (0.1346) & 0.4843 & (0.1703) & 0.5268 & (0.1618) & 0.9147 & (0.0632) & 0.9221 & (0.075) & 0.9385 & (0.0958) & 0.9703 & (0.1059) & 0.9727 & (0.0920) \\ \hline
\multirow{8}{*}{Pois} & \multirow{4}{*}{$r=2$} & $d=15$ & 0.3372 & (0.1420) & 0.3877 & (0.1793) & 0.4279 & (0.1886) & 0.4660 & (0.2142) & 0.5504 & (0.2893) & 0.8517 & (0.1353) & 0.9493 & (0.1863) & 0.9342 & (0.1838) & 0.9521 & (0.1913) & 0.9996 & (0.2231)   \\
 &  & $d=30$ & 0.3140 & (0.1364) & 0.3608 & (0.1643) & 0.4047 & (0.1684) & 0.4471 & (0.2131) & 0.4911 & (0.2389) & 0.9296 & (0.1337) & 0.9148 & (0.1367) & 0.9340 & (0.1335) & 0.9433 & (0.1373) & 0.9710 & (0.1467)   \\
 &  & $d=60$ & 0.2967 & (0.1306) & 0.3464 & (0.1669) & 0.4042 & (0.1865) & 0.4657 & (0.2338) & 0.5359 & (0.2694) & 0.9017 & (0.0938) & 0.9163 & (0.0917) & 0.9298 & (0.1115) & 0.9544 & (0.1303) & 0.9887 & (0.1382)  \\
 &  & $d=90$ & 0.2564 & (0.1388) & 0.3358 & (0.1848) & 0.3732 & (0.2033) & 0.4377 & (0.2101) & 0.4939 & (0.2580) & 0.8884 & (0.0734) & 0.9123 & (0.0845) & 0.9213 & (0.1152) & 0.9383 & (0.0951) & 0.9665 & (0.1449)  \\ \cline{2-23} 
 & \multirow{4}{*}{$r=5$} & $d=15$ & 0.4368 & (0.1205) & 0.4827 & (0.1450) & 0.4934 & (0.1580) & 0.5064 & (0.1448) & 0.5435 & (0.1591) & 0.9276 & (0.1759) & 0.9454 & (0.1679) & 0.9231 & (0.1688) & 0.9995 & (0.1927) & 0.9849 & (0.2040)  \\
 &  & $d=30$ & 0.3662 & (0.1045) & 0.3899 & (0.1288) & 0.4157 & (0.1418) & 0.4791 & (0.1537) & 0.5105 & (0.1791) & 0.8869 & (0.1184) & 0.9145 & (0.1151) & 0.9470 & (0.1247) & 0.9484 & (0.1434) & 0.9777 & (0.1605)   \\
 &  & $d=60$ & 0.3382 & (0.1021) & 0.3914 & (0.1373) & 0.4137 & (0.1312) & 0.4772 & (0.1510) & 0.5313 & (0.1514) & 0.9008 & (0.0852) & 0.9300 & (0.1048) & 0.9332 & (0.1008) & 0.9523 & (0.1011) & 0.9919 & (0.1175) \\
 &  & $d=90$ & 0.3033 & (0.0878) & 0.3718 & (0.1325) & 0.4189 & (0.1292) & 0.4744 & (0.1531) & 0.5537 & (0.1694) & 0.8938 & (0.0655) & 0.9200 & (0.0855) & 0.9368 & (0.0804) & 0.9545 & (0.0947) & 0.9938 & (0.1081)   \\ \hline
\multirow{8}{*}{Negbin} & \multirow{4}{*}{$r=2$} & $d=15$ & 0.3100 & (0.1602) & 0.3791 & (0.1933) & 0.4103 & (0.2139) & 0.4470 & (0.2029) & 0.4969 & (0.2546) & 0.9091 & (0.1744) & 0.8973 & (0.1755) & 0.9212 & (0.1897) & 0.9491 & (0.2008) & 0.9857 & (0.2079)  \\
 &  & $d=30$ & 0.2914 & (0.1546) & 0.3491 & (0.1796) & 0.3926 & (0.1935) & 0.4476 & (0.2028) & 0.4917 & (0.2631) & 0.8853 & (0.1391) & 0.9495 & (0.1361) & 0.9365 & (0.1531) & 0.9521 & (0.1429) & 0.9703 & (0.1637) \\
 &  & $d=60$ & 0.2705 & (0.1431) & 0.3237 & (0.1645) & 0.3578 & (0.1861) & 0.4261 & (0.2440) & 0.4790 & (0.2505) & 0.9156 & (0.0893) & 0.9159 & (0.0975) & 0.9365 & (0.1092) & 0.9522 & (0.1240) & 0.9632 & (0.1382)   \\
 &  & $d=90$ & 0.2493 & (0.1373) & 0.2992 & (0.1438) & 0.3464 & (0.1676) & 0.4680 & (0.2351) & 0.5498 & (0.3010) & 0.8902 & (0.0826) & 0.8900 & (0.0795) & 0.9172 & (0.0821) & 0.9607 & (0.1386) & 1.0021 & (0.1756)  \\ \cline{2-23} 
 & \multirow{4}{*}{$r=5$} & $d=15$ & 0.3997 & (0.1264) & 0.4330 & (0.1367) & 0.4620 & (0.1458) & 0.4976 & (0.1646) & 0.5445 & (0.1903) & 0.8984 & (0.1598) & 0.9312 & (0.1528) & 0.9123 & (0.1940) & 0.9689 & (0.1900) & 0.9310 & (0.1792)  \\
 &  & $d=30$ & 0.3501 & (0.0994) & 0.4033 & (0.1237) & 0.4276 & (0.1577) & 0.4644 & (0.1501) & 0.5479 & (0.1775) & 0.9008 & (0.1171) & 0.9284 & (0.1243) & 0.9174 & (0.1426) & 0.9485 & (0.1258) & 1.0020 & (0.1428)  \\
 &  & $d=60$ & 0.3106 & (0.0989) & 0.3787 & (0.1232) & 0.4221 & (0.1243) & 0.4986 & (0.1768) & 0.5423 & (0.1913) & 0.8925 & (0.0853) & 0.9035 & (0.0973) & 0.9404 & (0.0995) & 0.9547 & (0.1068) & 0.9861 & (0.1247)   \\
 &  & $d=90$ & 0.2923 & (0.0924) & 0.3639 & (0.0978) & 0.4185 & (0.1097) & 0.4689 & (0.1384) & 0.5073 & (0.1752) & 0.9010 & (0.0628) & 0.9155 & (0.0723) & 0.9370 & (0.0736) & 0.9672 & (0.0864) & 0.9622 & (0.1087)  \\ \hline
\end{tabular}%
}
\caption{Means and standard deviations (in the parentheses) of RMSEs of the $H$-step forecasting errors at the two latent process levels $\{Y_{t}^{(k)},Z_{t}^{(k)}\}$ for various combinations of model parameters including number of factors $r$, dimension $d$, number of time points $T$, and several marginal distributions.}
\label{tab:table_forecast1}
\end{table}

\begin{table}[]
\resizebox{\columnwidth}{!}{%
\begin{tabular}{cccccccccccccccccl}
\cline{4-18}
 &  &  & \multicolumn{5}{c|}{$\{X_t\}$} & \multicolumn{10}{c}{ACC$(H)$} \\ \cline{4-18} 
 &  &  & $H=1$ & $H=2$ & $H=3$ & $H=6$ & \multicolumn{1}{c|}{$H=12$} & \multicolumn{2}{c}{$H=1$} & \multicolumn{2}{c}{$H=2$} & \multicolumn{2}{c}{$H=3$} & \multicolumn{2}{c}{$H=6$} & \multicolumn{2}{c}{$H=12$} \\ \hline
\multirow{8}{*}{Bern} & \multirow{4}{*}{$r=2$} & $d=15$ & 0.3618 & 0.3595 & 0.3555 & 0.3657 & 0.3614 & 0.7273 & (0.6633,0.6793,0.3067) & 0.7293 & (0.6707,0.700,0.2913) & 0.7360 & (0.6513,0.7060,0.2920) & 0.7093 & (0.6413,0.6920,0.3020) & 0.7180 & (0.6233,0.6927,0.2920)  \\
 &  & $d=30$ & 0.3035 & 0.3074 & 0.3118 & 0.3140 & 0.3148 & 0.7347 & (0.6583,0.6930,0.2903) & 0.7247 & (0.6550,0.6943,0.2970) & 0.7067 & (0.6383,0.6763,0.313) & 0.7007 & (0.6283,0.6857,0.3070) & 0.6960 & (0.6140,0.6847,0.3000) \\
 &  & $d=60$ & 0.2566 & 0.2600 & 0.2588 & 0.2596 & 0.2643 & 0.7340 & (0.6673,0.6922,0.3020) & 0.7198 & (0.6427,0.6878,0.2997) & 0.7240 & (0.6432,0.6987,0.2932) & 0.7233 & (0.6480,0.6985,0.2863) & 0.7030 & (0.6147,0.6862,0.2990)  \\
 &  & $d=90$ & 0.2330 & 0.2340 & 0.2353 & 0.2392 & 0.2395 & 0.7306 & (0.6543,0.6889,0.3052) & 0.7268 & (0.6486,0.7009,0.2983) & 0.7208 & (0.6388,0.7013,0.2966) & 0.7019 & (0.6182,0.6906,0.3007) & 0.7006 & (0.6093,0.6934,0.2998) \\ \cline{2-18} 
 & \multirow{4}{*}{$r=5$} & $d=15$ & 0.3580 & 0.3713 & 0.3620 & 0.3702 & 0.3678 & 0.7280 & (0.6600,0.6980,0.2947) & 0.6953 & (0.6287,0.6867,0.3140) & 0.7160 & (0.6360,0.6980,0.2880) & 0.6967 & (0.6027,0.6820,0.3053) & 0.6980 & (0.6193,0.6853,0.3033)  \\
 &  & $d=30$ & 0.3085 & 0.3097 & 0.3092 & 0.3140 & 0.3116 & 0.7173 & (0.6560,0.6950,0.3037) & 0.7147 & (0.642,0.7017,0.2963) & 0.7163 & (0.6373,0.6950,0.3003) & 0.7010 & (0.6237,0.6893,0.3020) & 0.7077 & (0.6037,0.7020,0.2907) \\
 &  & $d=60$ & 0.2605 & 0.2625 & 0.2632 & 0.2655 & 0.2656 & 0.7183 & (0.6438,0.6843,0.3052) & 0.7107 & (0.6350,0.6805,0.3127) & 0.7080 & (0.6275,0.6830,0.3088) & 0.6977 & (0.6265,0.6760,0.3108) & 0.6963 & (0.6147,0.6885,0.3040)  \\
 &  & $d=90$ & 0.2319 & 0.2353 & 0.2368 & 0.2382 & 0.2392 & 0.7352 & (0.6604,0.6998,0.2946) & 0.7209 & (0.6398,0.6904,0.3032) & 0.7143 & (0.6382,0.6958,0.2981) & 0.7077 & (0.6278,0.6967,0.2977) & 0.7023 & (0.6087,0.6900,0.3003) \\ \hline
\multirow{8}{*}{Categ} & \multirow{4}{*}{$r=2$} & $d=15$ & 0.6382 & 0.6653 & 0.6718 & 0.6778 & 0.7004 & 0.4813 & (0.4073,0.3987,0.2560) & 0.4620 & (0.3853,0.3940,0.2620) & 0.4247 & (0.3740,0.3740,0.2527) & 0.4473 & (0.3747,0.3567,0.2680) & 0.3940 & (0.3573,0.3913,0.2740)  \\
 &  & $d=30$ & 0.5426 & 0.5591 & 0.5543 & 0.5640 & 0.5937 & 0.4707 & (0.4053,0.3820,0.2647) & 0.4557 & (0.3797,0.3837,0.2667) & 0.4513 & (0.3843,0.3913,0.2630) & 0.4427 & (0.3653,0.3863,0.2720) & 0.4080 & (0.3550,0.3967,0.2667)  \\
 &  & $d=60$ & 0.4584 & 0.4583 & 0.4663 & 0.4769 & 0.5056 & 0.4718 & (0.3818,0.3950,0.2725) & 0.4622 & (0.3948,0.3898,0.2682) & 0.4572 & (0.3768,0.3885,0.2667) & 0.4413 & (0.3663,0.3837,0.2680) & 0.4060 & (0.3495,0.3833,0.2705)  \\
 &  & $d=90$ & 0.4180 & 0.4209 & 0.4238 & 0.4329 & 0.4464 & 0.4690 & (0.3943,0.3954,0.2698) & 0.4650 & (0.3912,0.3968,0.2627) & 0.4581 & (0.3754,0.3980,0.2658) & 0.4280 & (0.3647,0.3818,0.2573) & 0.4184 & (0.3560,0.3804,0.2513)  \\ \cline{2-18} 
 & \multirow{4}{*}{$r=5$} & $d=15$ & 0.6417 & 0.6540 & 0.6600 & 0.6996 & 0.7136 & 0.4673 & (0.4087,0.4040,0.2567) & 0.4553 & (0.3907,0.3920,0.2667) & 0.4380 & (0.3887,0.3873,0.2667) & 0.4187 & (0.3573,0.4067,0.2860) & 0.4113 & (0.3220,0.3767,0.2740)  \\
 &  & $d=30$ & 0.5451 & 0.5439 & 0.5589 & 0.5875 & 0.6017 & 0.4757 & (0.4053,0.3947,0.2607) & 0.4730 & (0.3997,0.3950,0.2657) & 0.4413 & (0.3843,0.3930,0.2540) & 0.4117 & (0.3633,0.389,0.2740) & 0.4140 & (0.3500,0.3900,0.2627)  \\
 &  & $d=60$ & 0.4618 & 0.4743 & 0.4799 & 0.4957 & 0.5150 & 0.4807 & (0.3975,0.3820,0.2565) & 0.4513 & (0.3710,0.3927,0.2817) & 0.4393 & (0.3632,0.3850,0.2705) & 0.4260 & (0.3493,0.3847,0.2608) & 0.4000 & (0.3335,0.3903,0.2673) \\
 &  & $d=90$ & 0.4191 & 0.4215 & 0.4311 & 0.4428 & 0.4582 & 0.4662 & (0.3848,0.4020,0.2609) & 0.4627 & (0.3834,0.3926,0.2641) & 0.4483 & (0.3757,0.3918,0.2671) & 0.4228 & (0.3548,0.3819,0.2668) & 0.4072 & (0.3411,0.3892,0.2637) \\ \hline
\multirow{8}{*}{Pois} & \multirow{4}{*}{$r=2$} & $d=15$ & 0.6406 & 0.6675 & 0.6759 & 0.6860 & 0.6885 & 0.4973 & (0.4347,0.4773,0.4860) & 0.4973 & (0.4333,0.4813,0.4687) & 0.4873 & (0.4280,0.4753,0.4687) & 0.4827 & (0.4313,0.4740,0.4780) & 0.4593 & (0.3793,0.4493,0.4447) \\
 &  & $d=30$ & 0.5628 & 0.5646 & 0.5723 & 0.5789 & 0.5820 & 0.4900 & (0.4440,0.4650,0.4630) & 0.4907 & (0.4470,0.4660,0.4703) & 0.4873 & (0.4247,0.4757,0.4637) & 0.4793 & (0.4143,0.4697,0.4707) & 0.4733 & (0.4253,0.4543,0.4680)  \\
 &  & $d=60$ & 0.4716 & 0.4740 & 0.4810 & 0.4843 & 0.5028 & 0.4985 & (0.4345,0.4560,0.4673) & 0.4932 & (0.4363,0.4655,0.4688) & 0.4860 & (0.4307,0.4642,0.4705) & 0.478 & (0.4205,0.4573,0.4662) & 0.4695 & (0.4095,0.4593,0.4650)  \\
 &  & $d=90$ & 0.4309 & 0.4338 & 0.4369 & 0.4386 & 0.4481 & 0.4921 & (0.4432,0.4464,0.4659) & 0.4918 & (0.4311,0.4448,0.4619) & 0.4923 & (0.4351,0.4502,0.4627) & 0.4932 & (0.4276,0.4474,0.4620) & 0.4817 & (0.4197,0.4487,0.4722) \\ \cline{2-18} 
 & \multirow{4}{*}{$r=5$} & $d=15$ & 0.6746 & 0.6865 & 0.6674 & 0.6923 & 0.7043 & 0.4960 & (0.4340,0.4613,0.4773) & 0.4667 & (0.4187,0.4473,0.4680) & 0.4833 & (0.4300,0.4560,0.4633) & 0.4753 & (0.4187,0.4460,0.4520) & 0.4567 & (0.4207,0.4480,0.4533)  \\
 &  & $d=30$ & 0.5557 & 0.5630 & 0.5754 & 0.5762 & 0.5880 & 0.5030 & (0.4427,0.4750,0.4687) & 0.4870 & (0.4247,0.4637,0.4563) & 0.4813 & (0.4227,0.4623,0.4513) & 0.4800 & (0.4223,0.4583,0.4677) & 0.4923 & (0.4193,0.4677,0.4717)  \\
 &  & $d=60$ & 0.4735 & 0.4812 & 0.4836 & 0.4889 & 0.4990 & 0.5032 & (0.4338,0.4603,0.4650) & 0.4910 & (0.4278,0.4528,0.4620) & 0.4890 & (0.4313,0.4600,0.4647) & 0.4840 & (0.4218,0.4578,0.4690) & 0.4772 & (0.4112,0.4583,0.4697)  \\
 &  & $d=90$ & 0.4294 & 0.4326 & 0.4393 & 0.4436 & 0.4539 & 0.4980 & (0.4422,0.4570,0.4599) & 0.4936 & (0.4367,0.4544,0.4647) & 0.4919 & (0.4278,0.4551,0.4663) & 0.4787 & (0.4229,0.4561,0.4637) & 0.4631 & (0.4114,0.4443,0.4582) \\ \hline
\multirow{8}{*}{Negbin} & \multirow{4}{*}{$r=2$} & $d=15$ & 1.0689 & 1.0438 & 1.0944 & 1.1393 & 1.1390 & 0.2087 & (0.1727,0.1853,0.1640) & 0.2060 & (0.1753,0.1947,0.1500) & 0.2033 & (0.1667,0.1833,0.1460) & 0.1840 & (0.1627,0.1747,0.1653) & 0.1833 & (0.1447,0.1773,0.1540)  \\
 &  & $d=30$ & 0.9406 & 0.9601 & 0.9538 & 0.9644 & 0.9954 & 0.2087 & (0.1550,0.1783,0.1587) & 0.2030 & (0.1497,0.1710,0.1617) & 0.1923 & (0.1467,0.1660,0.1503) & 0.1953 & (0.1433,0.1717,0.1623) & 0.1897 & (0.1377,0.1767,0.1763) \\
 &  & $d=60$ & 0.8059 & 0.8112 & 0.8086 & 0.8225 & 0.8392 & 0.2058 & (0.1518,0.1768,0.1577) & 0.2053 & (0.1532,0.1748,0.1610) & 0.2005 & (0.1425,0.1795,0.1562) & 0.1942 & (0.1522,0.1767,0.1660) & 0.1883 & (0.1400,0.1750,0.1607)  \\
 &  & $d=90$ & 0.7017 & 0.7025 & 0.7187 & 0.7481 & 0.7589 & 0.2110 & (0.1651,0.1770,0.1588) & 0.2040 & (0.1548,0.1778,0.1640) & 0.1966 & (0.1573,0.1743,0.1546) & 0.1912 & (0.1460,0.1768,0.1703) & 0.1843 & (0.1441,0.1633,0.1544) \\ \cline{2-18} 
 & \multirow{4}{*}{$r=5$} & $d=15$ & 1.1232 & 1.1295 & 1.1030 & 1.1550 & 1.1567 & 0.2040 & (0.1413,0.1727,0.1620) & 0.1947 & (0.1727,0.1727,0.1567) & 0.1800 & (0.1407,0.1660,0.1787) & 0.2027 & (0.1560,0.1767,0.1473) & 0.1840 & (0.1413,0.1820,0.1693) \\
 &  & $d=30$ & 0.9302 & 0.9331 & 0.9469 & 0.9501 & 1.0094 & 0.2027 & (0.1530,0.1757,0.1610) & 0.1907 & (0.1483,0.1810,0.1640) & 0.2040 & (0.1457,0.1653,0.1563) & 0.1883 & (0.1443,0.1803,0.1690) & 0.1820 & (0.1303,0.1713,0.1513) \\
 &  & $d=60$ & 0.7887 & 0.8058 & 0.8136 & 0.8145 & 0.828 & 0.2118 & (0.1558,0.1777,0.1648) & 0.2037 & (0.1505,0.1755,0.1590) & 0.1905 & (0.1463,0.1635,0.1568) & 0.1862 & (0.1433,0.1730,0.1640) & 0.1763 & (0.1422,0.1677,0.1587) \\
 &  & $d=90$ & 0.7014 & 0.7182 & 0.7201 & 0.7509 & 0.7583 & 0.2006 & (0.1560,0.1737,0.1612) & 0.1958 & (0.1538,0.1728,0.1601) & 0.1956 & (0.1459,0.1701,0.1571) & 0.1907 & (0.1419,0.1752,0.1669) & 0.1809 & (0.1519,0.1697,0.1630) \\ \hline
\end{tabular}%
}
\caption{Means of RMSEs of the $H$-step forecasting errors at the observation level $\{X_{t}\}$ and means of accuracy of $H$-step ahead forecasting (ACC$(H)$) with three benchmarks (Last, Marginal, and Null, in the parentheses) for various combinations of model parameters including number of factors $r$, dimension $d$, number of time points $T$, and several marginal distributions.}
\label{tab:table_forecast2}
\end{table}

\begin{figure}[th]
    \centering
    \includegraphics[width=1\textwidth,height=0.7\textwidth]{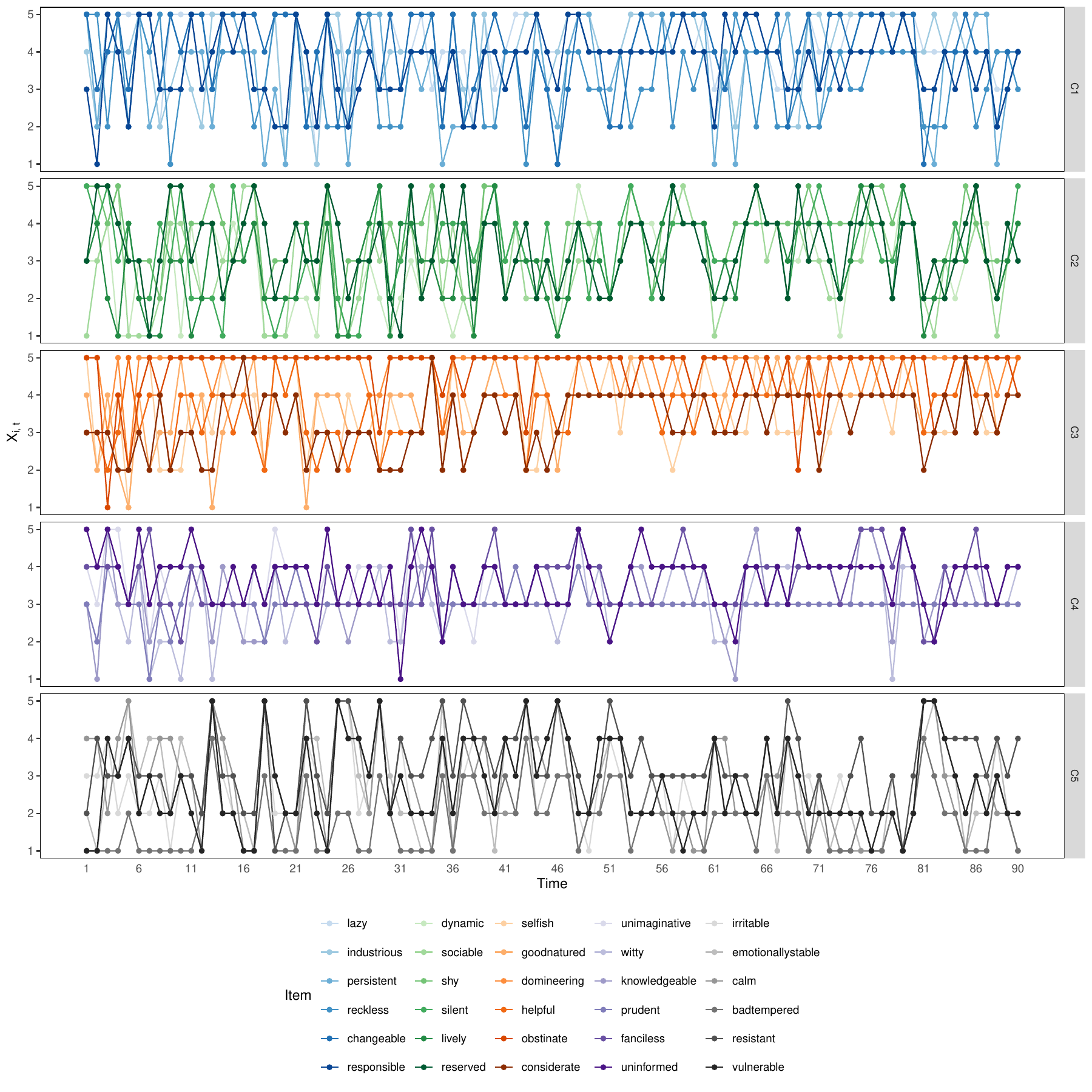}
    \caption{Time plots of 90-day observations for 30 items.}
    \label{fig:Figure_observation}
\end{figure}

\begin{figure}[th]
    \centering
    \includegraphics[width=1\textwidth,height=1\textwidth]{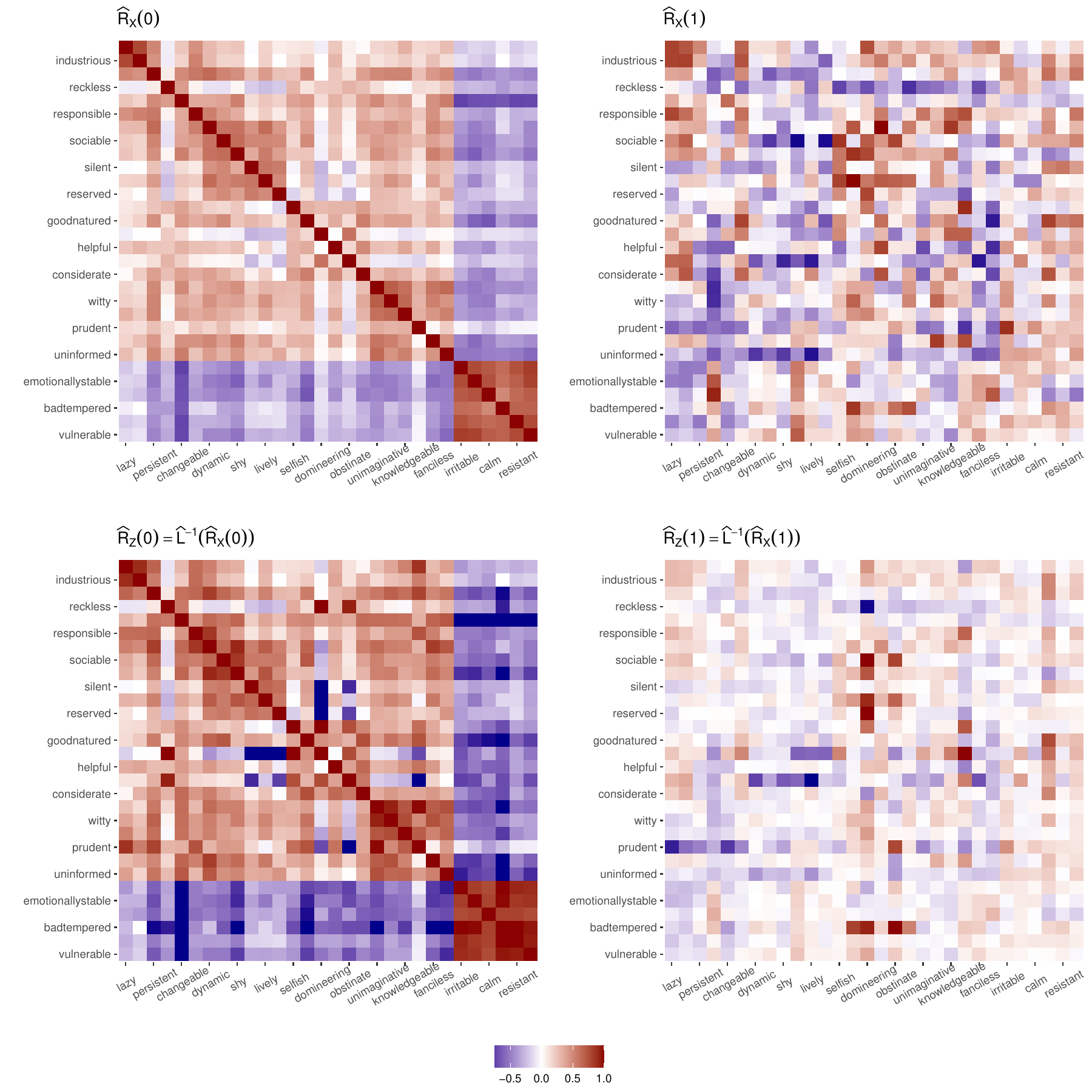}
    \caption{The sample autocorrelation matrices of the observations at lag 0 (top left) and lag 1 (top right), $\widehat{R}_{X}(h)$, $h=0,1$. The estimated autocorrelation matrices of the latent Gaussian series at lag 0 (bottom left) and lag 1 (bottom right), $\widehat{R}_{Z}(h)$, $h=0,1$.}
    \label{fig:Figure_covariance}
\end{figure}

\begin{figure}[th]
    \centering
    \includegraphics[width=1\textwidth,height=0.6\textwidth]{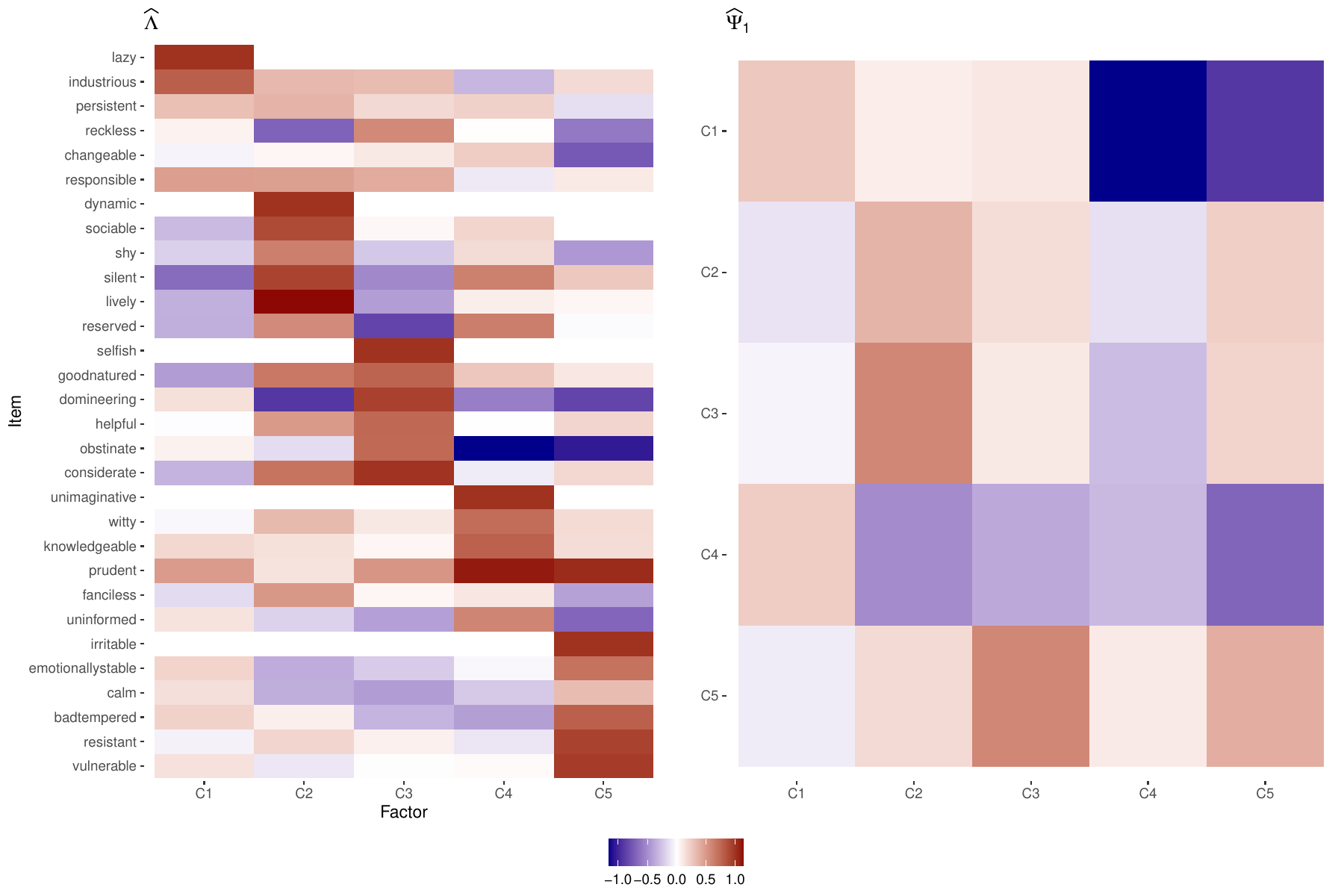}
    \caption{Estimate of the loadings matrix $\widehat{\Lambda}$ (left). Estimate of the VAR(1) transition matrix $\widehat{\Psi}_1$ (right).}
    \label{fig:Figure_param}
\end{figure}

\begin{figure}[th]
    \centering
    \includegraphics[width=1\textwidth,height=0.6\textwidth]{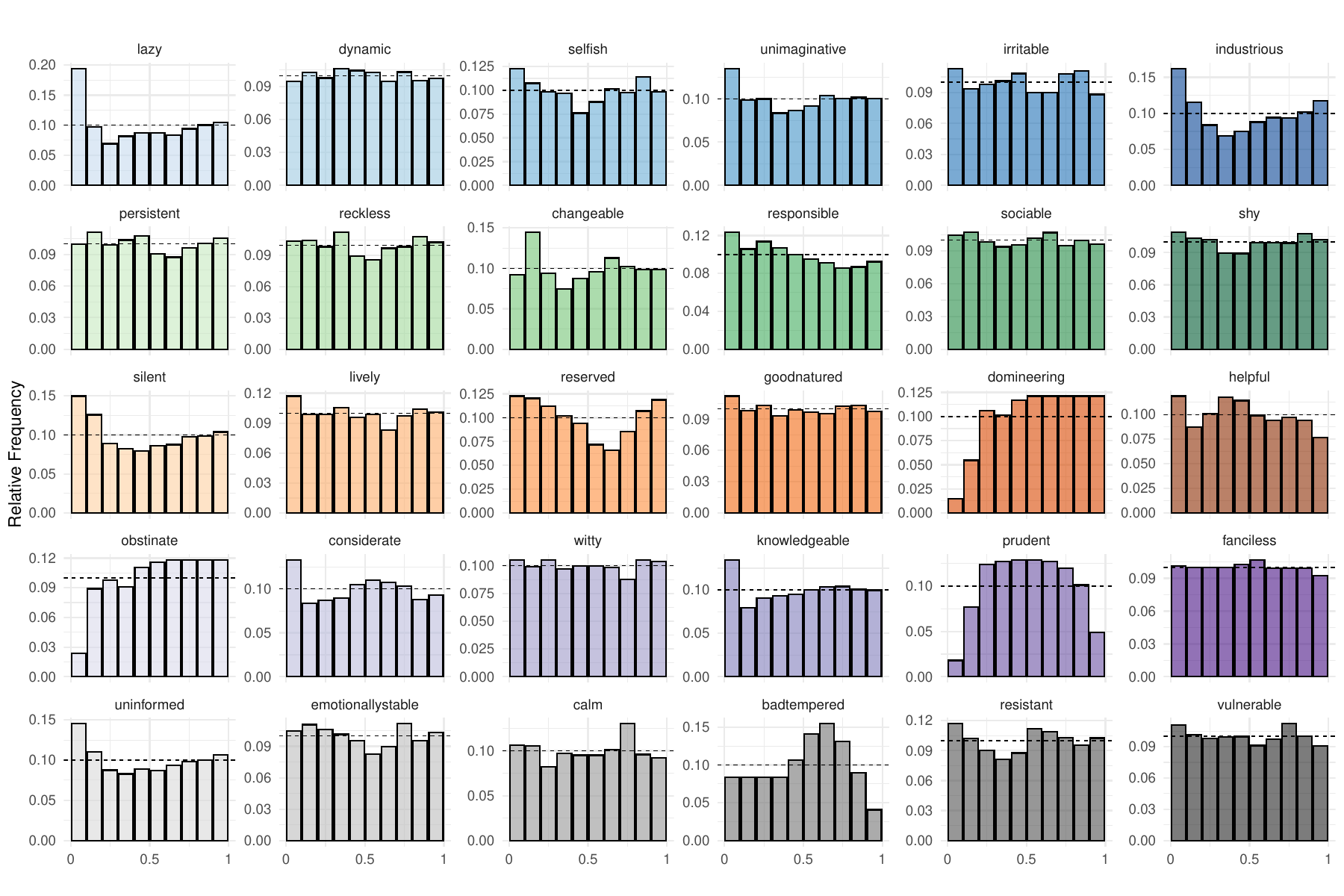}
    \caption{Sample PIT histograms for marginal distributions of each 30 items.}
    \label{fig:Figure_pit}
\end{figure}

\begin{figure}[th]
    \centering
    \includegraphics[width=1\textwidth,height=1\textwidth]{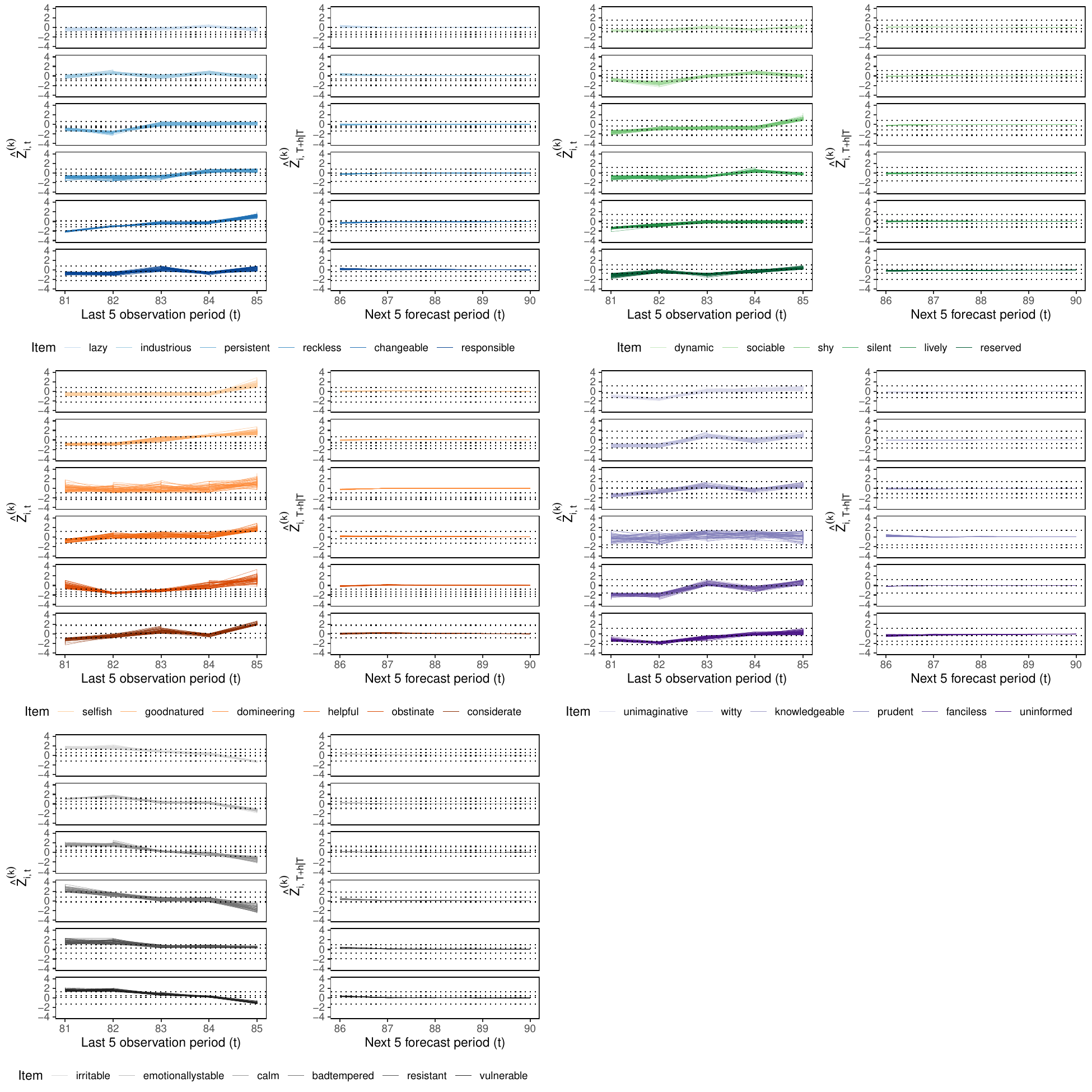}
    \caption{The simulated particles for the 30 items in the observation period (left of each panel) and forecasting period (right of each panel). Two consecutive horizontal lines with different line types form the bin for discrete observations of each item.}
    \label{fig:Figure_particle}
\end{figure}

\begin{figure}[th]
\centering
    \includegraphics[width=1\textwidth,height=0.6\textwidth]{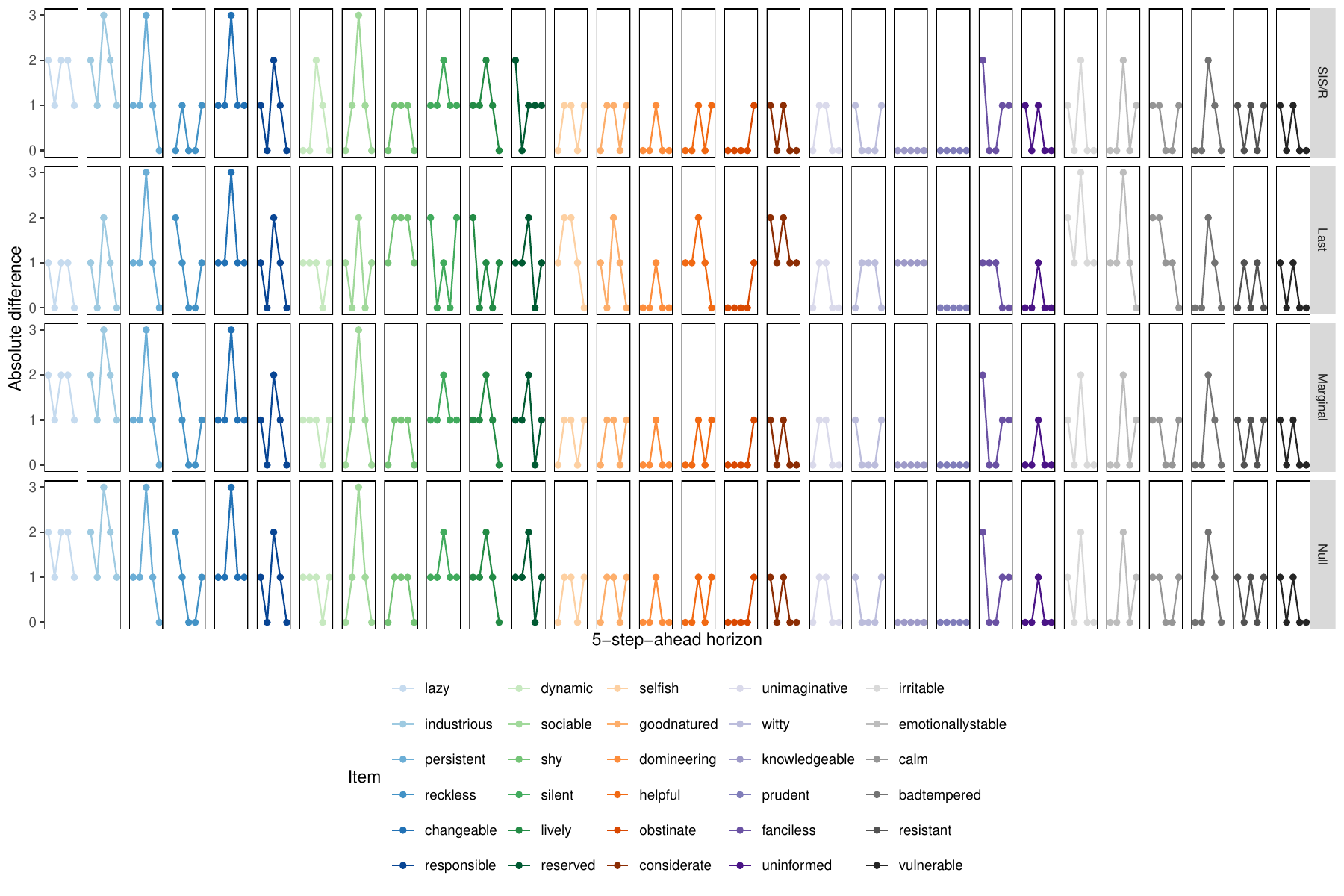}
    \caption{Result of 5-step-ahead prediction of each item with three baseline approaches. Each of 6 items in order are known to be affected by the same category.}
    \label{fig:Figure_forecast}
\end{figure}

\clearpage
\small
\bibliography{latent-DFM}

\appendix
\newpage
\setcounter{section}{0}
\setcounter{page}{1}
\setcounter{figure}{0}
\setcounter{equation}{0}
\renewcommand{\thesection}{S\arabic{section}}
\renewcommand{\thepage}{\arabic{page}}
\renewcommand{\thetable}{S\arabic{table}}
\renewcommand{\thefigure}{S\arabic{figure}}

\begin{center}
\textbf{\LARGE {Supplemental Material to \\ ``Latent Gaussian dynamic factor modeling and forecasting for multivariate count time series}''}
\end{center}

\begin{center}
Younghoon Kim$^1$, Marie-Christine D\"uker$^2$, Zachary F. Fisher$^3$, and Vladas Pipiras$^{*4}$
\end{center}

\begin{center}
$^1$Cornell University \\
$^2$Friedrich-Alexander-Universit\"at Erlangen-N\"urnberg \\
$^3$The Pennsylvania State University \\
$^4$University of North Carolina at Chapel Hill
\end{center}

\def\thefootnote{$*$}\footnotetext{Corresponding author. Email: pipiras@email.unc.edu}


\section{Nonstationary models with covariates and differencing}
\label{se:covariate}

As discussed in \cite{jia2023latent}, covariates can be incorporated into the model through its marginal parameters $\theta_i$ as follows. Suppose that $\theta_i(t)$ varies over time and is driven by $J-$dimensional deterministic covariate vector $M_t$ as 
\begin{equation}\label{e:covariate}
    \theta_i(t) = f_i(\beta_i' M_{t}),
\end{equation}
where $\beta_i' M_{t}$ is a linear combination of a coefficient vector and the covariate, and $f_i$ is a suitable function in the spirit of GLM for $i$th marginal distribution. Then, the model \eqref{e:generalized_inverse} becomes
\begin{equation}\label{e:X_nonstationary}
    X_{i,t} = F_{i,\theta_i(t)}^{-1}(\Phi(Z_{i,t})) := G_{i,\theta_i(t)}(Z_{i,t}).
\end{equation}
This time-varying parameter leads to the cumulative distribution of $X_{i,t}$ as $C_{i,n}(t) =\mathbb{P}(X_{i,t}\leq n)=F_{i,\theta_i(t)}(n)$ and the computation of the Hermite coefficients in \eqref{e:hermite_coefficient}, now denoted by $g_{\theta_i(t),k}$, is still valid. The cross-correlation between $X_{i,t_1}$ and $X_{j,t_2}$ is represented by the cross-correlation between $Z_{i,t_1}$ and $Z_{j,t_2}$ as
\begin{equation}\label{e:cov_X_nonstationary}
    R_{X,ij}(t_1,t_2) = L_{\theta_i(t_1),\theta_j(t_2)}(R_{Z,ij}(t_1-t_2)),
\end{equation}
where, from \eqref{e:link_function_expansion},
\begin{equation}\label{e:link_function_nonstationary}
   L_{\theta_i(t_1),\theta_j(t_2)}(u) 
   = \sum_{k=1}^{\infty} \frac{k!g_{\theta_i(t_1),k}g_{\theta_j(t_2),k}}{\textrm{Var}(X_{i,t})^{1/2}\textrm{Var}(X_{j,t})^{1/2}} u^k.
\end{equation}
As in \eqref{e:inverse_link_matrix}, one then has
\begin{equation}\label{e:cov_Z_nonstationary}
    R_{Z,ij}(h) = L_{\theta_{i}(t+h),\theta_{j}(t)}^{-1}(R_{X,ij}(t+h,t)).
\end{equation}
The estimation strategy outlined following \eqref{e:inverse_link_matrix} presents additional challenges for \eqref{e:cov_Z_nonstationary}. While parameters $\beta_i$ in \eqref{e:covariate} can, in principle, be estimated by considering $i$th marginal data leading to $\theta_i(t)$, the link function $L_{\theta_i(t_1),\theta_j(t_2)}$ now needs to be computed potentially for many values of $\theta_i,\theta_j$. A bigger challenge is estimation of $R_{X,ij}(t+h,t)$ which now depends on $t$ and $h$. Some averaging locally over time might lead to valid estimation in a suitable framework but this is beyond the scope of this work and will be considered elsewhere.

Another simpler possibility to accommodate nonstationarity with our model is to difference the count series $X_{i,t}$. That is, consider the series $\Delta X_{i,t} = X_{i,t} - X_{i,t-1}$, whose range now lies in $\mathbb{Z}$. As noted above, the considered model \eqref{e:generalized_inverse} may as well be defined on $\mathbb{Z}$, with a suitable choice of marginal distribution $F_i$ (e.g.\ the Skellam distribution). Differencing of count time series and subsequent modeling of $\mathbb{Z}-$valued time series was considered by many researchers, including \cite{kim2008non},\cite{zhang2010inference},\cite{kachour2011p}, and others, but their models are generally non-trivial extension of the counterpart models for counts.

\section{Estimation of parameters of marginal distributions}
\label{se:estimation-marginal}

The parameters $\theta_{i}$ of the marginal CDF $F_{i}$ can be estimated through maximum likelihood (ML), or another method applicable to the parametric CDF $F_i$ of interest. We describe here the estimators for the parameters of the distributions $F_i$ considered in this paper. When $F_{i}=\textrm{Bern}(p_{i})$, $\widehat{p}_{i}$ is just the sample proportion of $X_{i,t}=1$. That is, $\widehat{p}_i = \#\{X_{i,t}=1,t=1,\ldots,T\}/T$. Similarly, the parameters $\theta_{i}=(p_{1,i},\ldots,p_{k,i})$ of categorical marginal distribution with $k$ number of categories is estimated by $\widehat{p}_{j,i} = \#\{X_{i,t}=j,t=1,\ldots,T\}/T$, $j=1,\ldots,k$. If the $i$th marginal count series follows the Poisson distribution with parameter $\theta_i$, we take $\widehat{\theta}_i = \sum_{t=1}^T X_{i,t}/T$. Estimating the parameters of a negative binomial marginal distribution is more subtle. Recall that the probability of a negative binomial random variable $X_{i,t}$ with success probability $p$ and number of successes $r$ is given by
\begin{displaymath}
    \mathbb{P}(X_{i,t} = n) = \frac{\Gamma(n+r)}{n!\Gamma(r)}(1-p)^n p^r,\quad n=0,1,2,\ldots.
\end{displaymath}
By introducing the mean $\mu$ such that $p=r/(r+\mu)$, the log-likelihood function can be expressed as a function of $\mu$ and $r$:
\begin{displaymath}
    \ell(\mu,r) = \sum_{t=1}^T \left(\log\Gamma(X_{i,t}+r) - \log\Gamma(r) - \log(X_{i,t}!) + r\log \frac{r}{r+\mu} + n_t\log\frac{\mu}{r+\mu}\right).
\end{displaymath}
The ML estimation is then as follows. First, estimate $\mu$ using $\widehat{\mu}=\sum_{t=1}^T X_{i,t}/T$. Then, estimate $r$ by solving the following equation numerically:
\begin{equation}\label{e:negbin_likelihood}
    \sum_{t=1}^T\left( \psi(X_{i,t}+r) -\psi(r) + \log\frac{r}{r+\widehat{\mu}} - \frac{X_{i,t} + r}{r +\widehat{\mu}} \right) = 0,
\end{equation}
where $\psi(x)=\Gamma'(x)/\Gamma(x)$ is digamma function. Finally, estimate $p$ by $\widehat{p} = \widehat{r}/(\widehat{r} + \widehat{\mu})$. Compared to the computational cost of the numerical procedure above, the advantage gained by the ML estimation is not substantial. For this reason, \cite{jia2023latent} used instead the method of moments to estimate the parameters of negative binomial marginal distributions. In the same spirit, we assume that $r$ is known throughout the simulation and estimate $\mu$ and $p$ directly, thereby skipping the need to solve \eqref{e:negbin_likelihood}.

\section{Selection of lag order}
\label{se:selection-p}

One common approach to select the lag order $p$ of the factor series \eqref{e:VAR} is through an information criterion,
\begin{equation}\label{e:IC_lag_selection}
    \widehat{p} = \argmin_{l} \left\{ \ln(|\widehat{\Sigma}_{\eta}(l)|) + g_{i,l}(r,T)\right\},
\end{equation}
where $\widehat{\Sigma}_{\eta}(l)$ is the estimate of the covariance matrix of the innovations of the factor series for fixed lag order $l$. The possible penalty functions $g_{i,l}$ include
\begin{equation} \label{e:penalties_lag}
    \begin{gathered}
    g_{1,l}(r,T) = \frac{2}{T}lr^2, \quad g_{2,l}(r,T) = \frac{2\log(\log(T))}{T}lr^2, \\
    g_{3,l}(r,T) = \frac{\log(T)}{T}lr^2, \quad g_{4,l}(r,T) = \frac{2r(rl+1)}{T}.
    \end{gathered}
\end{equation}
The resulting criteria can be found in Chapter 4.3 of \cite{lutkepohl2005new}.

We now propose a cross-validation strategy tailored to our model. We shall exploit once again the idea of Remark \ref{rem:latent_factor}. Were the factor series $\{Y_t\}_{t\in\mathbb{Z}}$ observed, a natural cross-validation scheme would select $p$ as $l$ minimizing 
\begin{equation*}
    \sum_{b=1}^B\sum_{t} \left\|Y_{t}^{(b)} - \widehat{\Psi}_1^{(-b)}Y_{t-1}^{(b)} - \ldots - \widehat{\Psi}_l^{(-b)}Y_{t-l}^{(b)} \right\|_F^2,
\end{equation*}
where $\widehat{\Psi}_{h}^{(-b)}$ are the VAR transition matrices estimated on the training data, and $Y_{t}^{(b)}$ refer to the testing data. Setting 
\begin{equation*}
    \mathcal{Y}^{(b,l)} = (Y_{t}^{(b)'})_{t},\quad 
    \widehat{\Psi}^{(-b,l)} = \begin{pmatrix}
        \widehat{\Psi}_{1}^{(-b)'} \\ 
        \vdots \\ 
        \widehat{\Psi}_{l}^{(-b)'} \end{pmatrix},\quad 
    \mathcal{X}^{(b,l)} =  \begin{pmatrix}
        Y_{t-1}^{(b)'} & \ldots & Y_{t-l}^{(b)'}
    \end{pmatrix}_t,
\end{equation*}
the function to minimize can be replaced by 
\begin{equation*}
    \sum_{b=1}^B \left\| \mathcal{Y}^{(b,l)} -  \mathcal{X}^{(b,l)}\widehat{\Psi}^{(-b,l)}\right\|_F^2 
    = \sum_{b=1}^B \vecop\left(\mathcal{Y}^{(b,l)} -  \mathcal{X}^{(b,l)}\widehat{\Psi}^{(-b,l)}\right)'\vecop\left(\mathcal{Y}^{(b,l)} -  \mathcal{X}^{(b,l)}\widehat{\Psi}^{(-b,l)}\right)
\end{equation*}
or
\begin{equation}\label{e:mse_p_ell2}
    \sum_{b=1}^B \left( - 2\widehat{\boldsymbol{\psi}}^{(-b,l)'}\widehat{\gamma}_{Y}^{(b,l)} + \widehat{\boldsymbol{\psi}}^{(-b,l)'}\widehat{\Gamma}_{Y}^{(b,l)}\widehat{\boldsymbol{\psi}}^{(-b,l)} \right),
\end{equation}
where $\widehat{\boldsymbol{\psi}}^{(-b,l)} = \vecop(\widehat{\Psi}^{(-b,l)})$ and
\begin{equation}\label{e:gram_matrix}
    \widehat{\gamma}_{Y}^{(b,l)} 
    = \vecop(\mathcal{X}^{(b,l)'}\mathcal{Y}^{(b,l)}), \quad 
    \widehat{\Gamma}_{Y}^{(b,l)} = I_{r} \otimes \mathcal{X}^{(b,l)'}\mathcal{X}^{(b,l)}.
\end{equation}
Following the idea of Remark \ref{rem:latent_factor}, note that the quantities in \eqref{e:gram_matrix} are obtained from the (sample) second-order properties of $(Y_{t}^{(b)})$. For our model, they can be replaced by those implied by the equation \eqref{e:inverse_link_matrix} based on the observed data $(X_{t}^{(b)})$. Similarly, we already have the estimators $\widehat{\Psi}^{(-b,l)}$ calculated from the observed data $(X_{t}^{(-b)})$. In summary, as a cross-validation scheme to select $p$ for our model, we also minimize \eqref{e:mse_p_ell2} but where the quantities involved are computed on the training data $(X_{t}^{(-b)})$ and the testing data $(X_{t}^{(b)})$.

\section{Simulation study for selecting lag order}
\label{se:more_simulation}

The model used for data generation follows the same setting used for $p=1$ in Section \ref{ssse:estimation_main}. However, for considering different true lag orders, we fix $r=2$ but consider $p=1$ and $p=3$. Also, the transition matrix $\Psi$ and covariance matrix $\Sigma_\eta$ are replaced by $\mbox{diag}(\Psi_h)_{ii}=\rho_h$ with $\rho_h=0.2,0.3,0.4$ for $h=1,2,3$, and $\Sigma_{\eta}=0.3448 I_2$, respectively. The four IC methods in \eqref{e:IC_lag_selection} are considered as ad hoc baselines. Our approach is denoted by BCV. In addition, we also apply the cross-validation scheme described in Section \ref{sse:estimation-r} to both latent Gaussian series $\{Z_t\}$ and $\{Y_t\}$, by assuming that those series are observable. We denote them Low dim and Gaussian, respectively.

\begin{figure}[h]
    \centering
    \includegraphics[width=1\textwidth,height=1\textwidth]{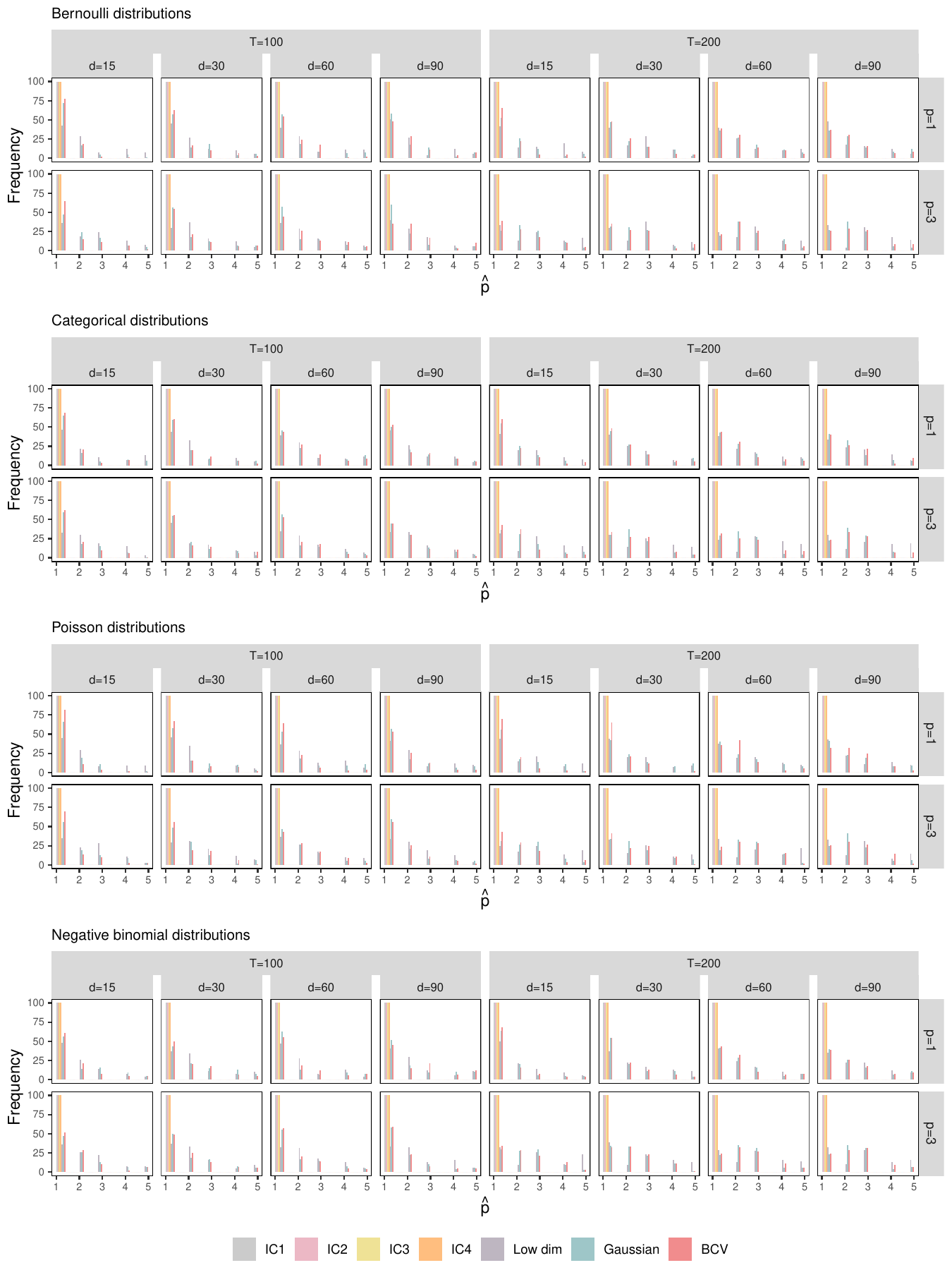}
    \caption{Estimated lag orders from simulated data for various combinations of model parameters including lag orders $p$, dimension $d$, number of time points $T$, and several marginal distributions.}
    \label{fig:Figure_lag}
\end{figure}

Figure \ref{fig:Figure_lag} shows the frequencies of estimated $p$ in 100 replications. The IC methods perform poorly under all conditions. Although the BCV-based approaches also underperform, applying the cross-validation scheme to discrete-valued observed series performs quite similarly to applying it directly to both latent Gaussian series $\{Z_t\}$ and $\{Y_t\}$. Similar to the rank selection in Figure \ref{fig:Figure_rank}, as the dimension $d$ and the sample length $T$ increase, the performance appears to improve. However, this improvement seems relatively minor compared to the rank selection. 
Taking a much larger $T$ (e.g., $T=1,000$) was checked to yield predominantly the true lag order (with the plots omitted). Interestingly, unlike the different distributions of the estimated ranks depending on the marginal distributions, the estimated lag orders appear less sensitive to the marginal distributions of the observed series. Overall, a more accurate lag order selection procedure is still desirable, but note that this is even for the case when observing the $r-$dimensional series $\{Y_t\}$.

\end{document}